\documentclass[superscriptaddress,twocolumn,10pt,nofootinbib,floatfix]{revtex4-2}

\usepackage{amssymb}
\usepackage{amsmath}
\usepackage{appendix}
\usepackage{times}
\usepackage{graphicx}
\usepackage{subeqnarray}
\usepackage{bbold}
\usepackage{enumerate}
\usepackage{color}
\usepackage{bm}
\usepackage[colorlinks=true, letterpaper=true, pdfstartview=FitV, linkcolor=blue, citecolor=blue, urlcolor=blue]{hyperref}
\usepackage{multirow}
\usepackage{babel}
\usepackage{calc}
\usepackage{float}
\usepackage{booktabs}

\usepackage[normalem]{ulem}

\setcitestyle{super,open={},close={}}

\makeatletter
\renewcommand\@biblabel[1]{#1.}
\makeatother

\begin{document}

\title{Layer Construction of Three-Dimensional $\mathbb{Z}_2$ Monopole Charge Nodal Line Semimetals and prediction of the abundant candidate materials}

\author{Yongpan Li}
\affiliation{Centre for Quantum Physics, Key Laboratory of Advanced Optoelectronic Quantum Architecture and Measurement (MOE), School of Physics, Beijing Institute of Technology, Beijing, 100081, China}

\author{Shifeng Qian}
\affiliation{Centre for Quantum Physics, Key Laboratory of Advanced Optoelectronic Quantum Architecture and Measurement (MOE), School of Physics, Beijing Institute of Technology, Beijing, 100081, China}
 
\author{Cheng-Cheng Liu}
\email{ccliu@bit.edu.cn}
\affiliation{Centre for Quantum Physics, Key Laboratory of Advanced Optoelectronic Quantum Architecture and Measurement (MOE), School of Physics, Beijing Institute of Technology, Beijing, 100081, China}

\begin{abstract}
\bf The interplay between symmetry and topology led to the concept of symmetry-protected topological states, including all non-interacting and weakly interacting topological quantum states. Among them, recently proposed nodal line semimetal states with space-time inversion ($\mathcal{PT}$) symmetry which are classified by the Stiefel-Whitney characteristic class associated with real vector bundles and can carry a nontrivial $\mathbb{Z}_2$ monopole charge have attracted widespread attention. However, we know less about such 3D $\mathbb{Z}_2$ nodal line semimetals and do not know how to construct them. In this work, we first extend the layer construction previously used to construct topological insulating states to topological semimetallic systems. We construct 3D $\mathbb{Z}_2$ nodal line semimetals by stacking of 2D $\mathcal{PT}$-symmetric Dirac semimetals via nonsymmorphic symmetries. Based on our construction scheme, effective model and combined with first-principles calculations, we predict two types of candidate electronic materials for $\mathbb{Z}_2$ nodal line semimetals, namely 14 Si and Ge structures and 108 transition metal dichalcogenides $MX_2$ ($M$=Cr, Mo, W, $X$=S, Se, Te). Our theoretical construction scheme can be directly applied to metamaterials and circuit systems. Our work not only greatly enriches the candidate materials and deepens the understanding of $\mathbb{Z}_2$ nodal line semimetal states but also significantly extends the application scope of layer construction.
\end{abstract}
	
\maketitle
\date{\today}
	
\null\cleardoublepage

\noindent

The novel topological states of matter and their material realization have attracted broad interest in the field of condensed-matter research\cite{RevModPhys.82.3045,RevModPhys.83.1057,Gu2009,PhysRevB.99.235125,bradlyn2016beyond,PhysRevB.88.125129,PhysRevB.88.125129,teo2010topological}. In general, there are four characteristic classes: Chern class, Stiefel-Whitney class, Pontrjagin class, and Euler class\cite{nakahara_geometry_2018}. For example, Weyl semimetals belong to the Chern class, originating from the mathematical structure of complex-valued vector bundles associated with complex-valued wave functions. As a result, they exhibit well-defined bulk-boundary correspondence, where a pair of Weyl points are connected by surface Fermi arcs. Recently, a class of systems with negligible spin-orbit coupling and space-time inversion symmetry ($\mathcal{PT}$), where the Hamiltonian and the Bloch wave function can be real-valued, has been causing widespread attention\cite{bzduvsek2017robust,fang2016topological,fang2015topological,ZYX2020realchern}.
The systems are characterized by the Stiefel-Whitney class, which is originated in the mathematical structure of the real-valued vector bundles associated with the real-valued wave functions\cite{PhysRevB.89.235127,PhysRevLett.116.156402,PhysRevLett.118.056401,fang2015topological,bzduvsek2017robust,PhysRevX.8.031069,YangBM2018band}. Among the Stiefel-Whitney class systems, the 3D $\mathcal{PT}$ symmetrical nodal line semimetals, whose valence bands and conduction bands cross each other along closed curves, i.e., nodal lines (NLs)\cite{PhysRevLett.115.036806,PhysRevB.90.205136,PhysRevLett.82.2147,PhysRevB.93.205132,Carbon+allotropes,PhysRevB.94.195104,PhysRevLett.117.096401,PhysRevB.93.201114,hirayama2017topological}, have been extensively researched. Such a NL can be linked by a loop $S^1$ and wrapped by a sphere $S^2$, on which one can define the first Stiefel-Whitney number $w_1$ and the second Steifel-Whitney number $w_2$, respectively, and thus can be doubly charged. If the NL has a nontrivial second Steifel-Whitney number $w_2$, it is called a NL with a nontrivial $\mathbb{Z}_2$ monopole charge and the corresponding NL semimetals are called $\mathbb{Z}_2$ NL semimetals ($\mathbb{Z}_2$NLSMs). Such $\mathbb{Z}_2$NLs belong to the Stiefel-Whitney class and exhibit well-defined higher-order bulk-boundary correspondence, where a pair of $\mathbb{Z}_2$NLs are connected by hinge Fermi arcs.

While $\mathbb{Z}_2$NLSMs have been explored in some theoretic works\cite{fang2015topological,fang2016topological,YangBM2018band,ZYX2020realchern}, the research in this field is still in its infancy. For example, it is not clear how to construct $\mathbb{Z}_2$NLSMs in a traceable way, and give a practical implementation scheme, let alone further predict a series of candidate materials. Furthermore, currently, there are only two electronic material candidates, i.e., ABC-stacking graphdiyne\cite{zyx2022graphdiyne,YangBM2018band,GraphdiyneExp1,GraphdiyneExp2,PhysRevMaterials.2.054204} and $\rm 1T'-MoTe_2$\cite{MoTe2,brown1966crystal}.

Layer construction (LC) is previously introduced in topological crystal insulators (TCIs), as any known TCI can be deformed into a combined state of a  set of decoupled 2D topological nontrivial layers without gap closing and symmetry breaking\cite{PhysRevB.92.081304,PhysRevB.94.125405,PhysRevB.94.155148,PhysRevX.7.011020,PhysRevB.96.205106,song2018quantitative,qian2020layer}. We find that the theory of LC also apply to $\mathbb{Z}_2$NLSMs and can help us in-depth understand and systematically construct $\mathbb{Z}_2$NLSMs. In this paper, for the first time, we generalize LC to the topological semimetals. By using LC, we propose a mechanism to generically construct 3D $\mathbb{Z}_2$NLSMs from usual 2D Dirac semimetals (DSMs). Based on the mechanism, we predict two types of candidate materials for $\mathbb{Z}_2$NLSMs, the 14 3D Si and Ge, and 108 transition metal dichalcogenides (TMDs) $MX_2$ ($M$=Cr,Mo,W, $X$=S,Se,Te), which greatly demonstrates the potential application of LC in semimetals.

\vspace{0.3cm}
\noindent{\bf Results}
\vspace{0.05cm}

\begin{figure*}[t]
\includegraphics[width=1\textwidth]{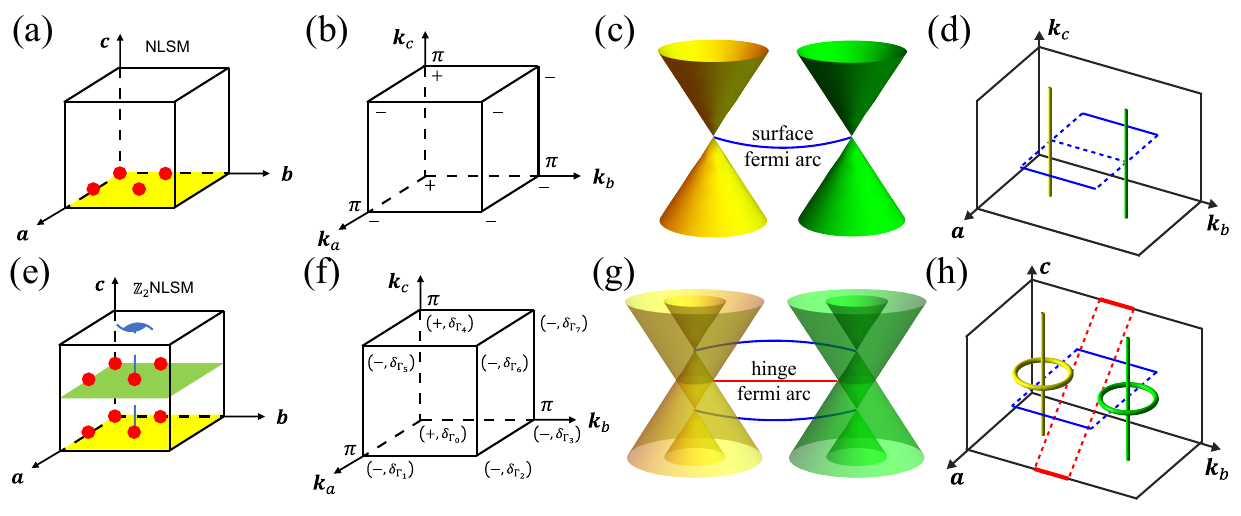}
\caption{\label{figpa} (a),(e) Schematic of the layer construction of 3D $\mathbb{Z}_2$NLSMs. The red dots denote the inversion centers. (a) A 3D NLSM formed by only one layer of 2D DSM in a unit cell. (e) A 3D $\mathbb{Z}_2$NLSM formed by two layers of 2D DSM in a unit cell.  (b),(f) Parity eigenvalues of occupied states at TRIMs and only occupied states that contribute to the NLs are considered. (f) The first parity eigenvalues in the parentheses at eight TRIMs are the same as the parity eigenvalues in (b). If $S_{2z}$ commutates (anticommutates) with $\mathcal{P}$ at $\Gamma_i$, $\delta_{\Gamma_i}$ and the first value in the parenthesis are the same (opposite). (c) Band structure for the NLSM  in (a). The blue arc is the surface Fermi arc. (g) Band structure for the $\mathbb{Z}_2$NLSM in (e). The blue arcs are the surface Fermi arcs while the red arc is the hinge Fermi arc. (d) The NLSM in (a) has a $\mathcal{T}$-reversed pair of straight NLs along $k_c$-direction and the surface Fermi arcs on the side surfaces. (h) The $\mathbb{Z}_2$NLSM in (e) has a $\mathcal{T}$-reversed pair of nodal loops at Fermi energy, which cause the hinge Fermi arcs, and two $\mathcal{T}$-reversed pairs of straight NLs above and below Fermi energy, which contribute to the surface Fermi arcs above and below Fermi energy, respectively. }
\end{figure*}

\begin{figure*}[t]
\includegraphics[width=1\textwidth]{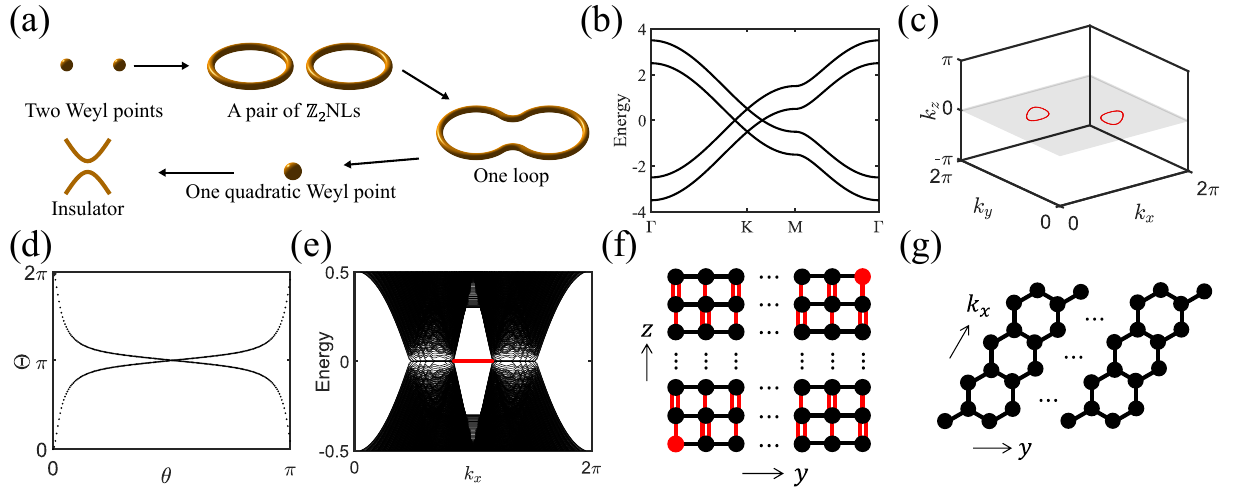}
\caption{\label{fignl} (a) The evolution of the nodes with the increasing of interlayer coupling. (b) The band structure for $t_0=1.0,t_1=0.1,t_2=0.4$. (c) A $\mathcal{T}$-reversed pair of nodal loops of (b). (d) The Wilson loop on a sphere enclosing one of the nodal loops in (b,c), indicating that the nodal loop carries a unit $\mathbb{Z}$ monopole charge (A unit $\mathbb{Z}_2$ monopole charge if the number of occupied bands is larger than 2). (e) The rod bands for parameters in (b) with open boundary in $y,z$-directions. The hinge Fermi arcs are marked by red lines. (f) The dimerization pattern of the hinge states at a pair of opposite hinges. The black dots and red dots are the sites of the orbitals. The red double lines are strong interlayer hoppings (the bigger one of $t_1$ and $t_2$), the red single lines are weak interlayer hoppings (the smaller one of $t_1$ and $t_2$), and black lines are intralayer hoppings ($t_0$). The red dots denote the spatial distribution of the hinge states. (g) The zigzag edge pattern of the edge states at a pair of opposite surfaces. The black dots are the sites of the orbitals and the black lines are intralayer hoppings ($t_0$).}
\end{figure*}

\begin{figure*}[t]
\includegraphics[width=1.0\textwidth]{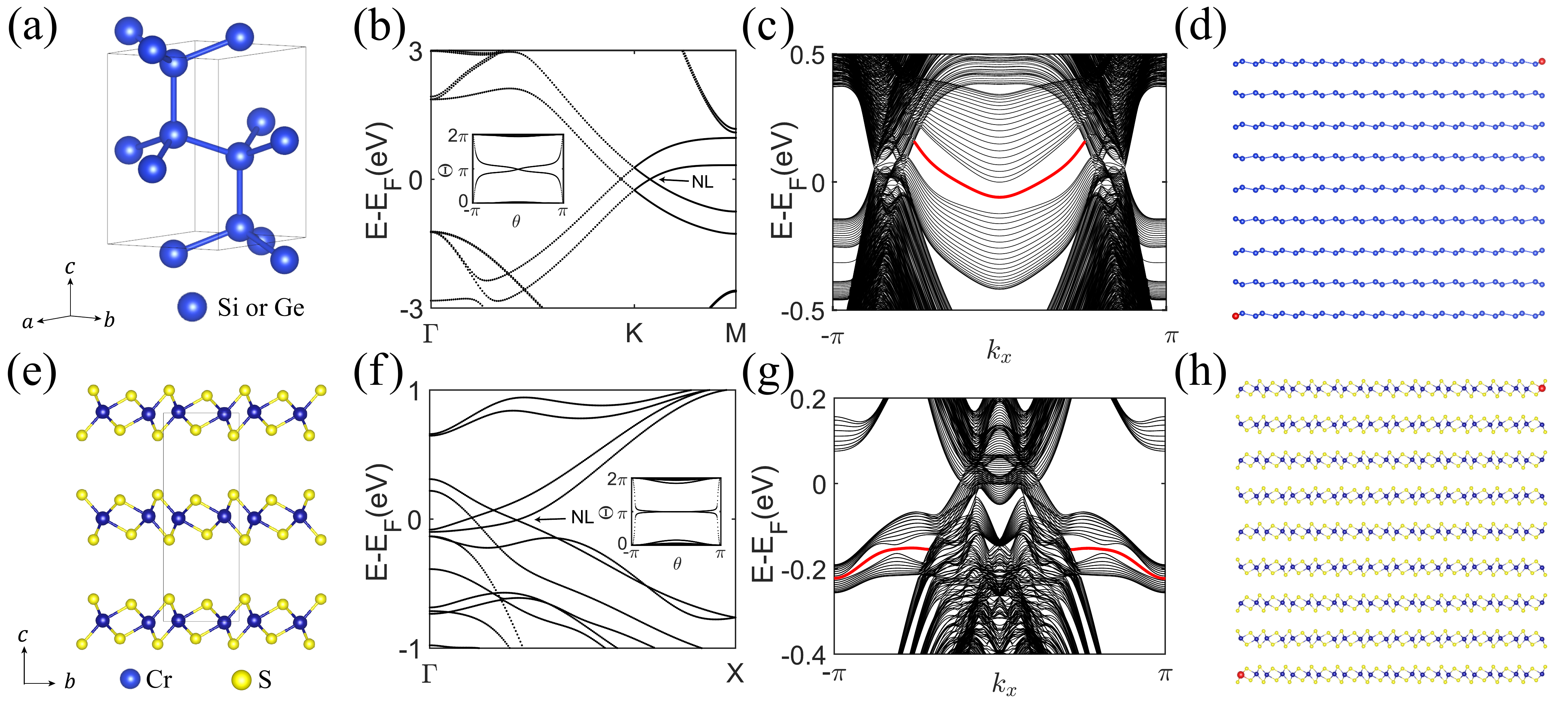}
\caption{\label{figmat}(a)(e) Lattice structures of the wurtzite Si and Ge and lattice structure of T$_\text{d}$-$\mathrm{CrS_2}$, respectively.  (b)(f) Band structures of the wurtzite Si or Ge and Td-$\mathrm{CrS_2}$, respectively. The insets are Wilson loops evaluated on spheres enclosing the NL, which pictorially illustrate the existence of $\mathbb{Z}_2$ monopole charges. (c)(g) Band structures for a rod that is periodic in $x$-direction and the Fermi arcs are colored red. (d)(h) The spatial distribution of the hinge states (red spheres in the opposite hinges.).}
\end{figure*}

\noindent{\bf  Create $\mathbb{Z}_2$ monopole charges by layer construction.} The kernel of our construction mechanism is to make use of the nonsymmorphic symmetry to engineer the parity eigenvalues at time-reversal invariant momentums (TRIMs). The 2D building blocks have time-reversal symmetry $\mathcal{T}$ and inversion symmetry $\mathcal{P}$. Since the twofold screw operator requires minimal symmetries, we use twofold screw symmetry as an example to elaborate our mechanism with the other nonsymmorphic operators considered in Supplementary Material Sec. I. As the engineering of parity eigenvalues is done by the 1/2 translation and the only role of the symmorphic symmetry operator, i.e., the twofold rotation operator, is to add inversion symmetry to the 3D structure, the generalization to other nonsymmorphic symmetry operators is straightforward, as given in Supplementary Material Sec. I.

The existence of $\mathbb{Z}_2$ monopole charge can be detected by parity eigenvalues at TRIMs and we can make use of the commutation or anticommutation relation between twofold screw symmetry $S_{2z}$ and inversion symmetry $\mathcal{P}$ to engineer the parity eigenvalues at TRIMs. As $\mathcal{P}$ acts on real space as $(x,y,z)\rightarrow(-x,-y,-z)$ and $S_{2z}$ acts on real space as $(x,y,z)\rightarrow\left(-x+\mu a/2,-y+\nu b/2,z+c/2\right)$, where $\mu,\nu\in\left[0,2\right)$ and $a,b,c$ are the length of the three lattice vectors, we find in Supplementary Material Sec. II
\begin{equation}
S_{2z}*\mathcal{P}=e^{ik_z-i\mu k_x-i\nu k_y}\mathcal{P}*S_{2z}.\label{eqsp}\\
\end{equation}

Then let us study the construction of 3D $\mathbb{Z}_2$NLSMs from 2D DSMs with individual $\mathcal{P}$ and $\mathcal{T}$. In the rest of the paper, we are referring to 2D Dirac semimetals with $\mathcal{P}$, $\mathcal{T}$ and without $C_{2z}$ when we use the term ``2D DSMs''. The reason why we exclude 2D DSMs with $C_{2z}$ can be found in Supplementary Material Sec. I. The structure of the 3D $\mathbb{Z}_2$NLSM is shown in Fig. \ref{figpa}(e). There are two layers of 2D DSMs in a unit cell. We can obtain another layer by performing $S_{2z}$ to one of the layers. There are only four choices of the twofold screw axis so that the 3D $\mathbb{Z}_2$NLSMs have $\mathcal{P}$, i.e., $\mu,\nu\in\{0,1\}$. Then the value of $e^{ik_c-i\mu k_a-i\nu k_b}$ can only take 1 or -1 at TRIMs, i.e., $S_{2z}$ commutates or anticommutates with $\mathcal{P}$ at any of the eight TRIMs.

In a 3D $\mathbb{Z}_2$NLSM, if we can find two parallel mirror invariant planes that do not intersect with $\mathbb{Z}_2$NLs. The parity eigenvalues in the two mirror invariant planes can be used to detect the $\mathbb{Z}_2$NLs in the Brillouin zone (BZ) as follows
\begin{equation}
e^{i\pi w}=\prod_{i=1}^8(-1)^{\lfloor N_{occ}^-(\Gamma_i)/2\rfloor},\label{eq1}
\end{equation}
where $N_{occ}^-$ is the number of occupied states with $-1$ parity eigenvalue at the TRIM $\Gamma_i$, $\lfloor \rfloor$ is the floor function, and an odd $w$ indicates the existance of $\mathbb{Z}_2$NLs\cite{YangBM2018band}. One should notice that the $\mathbb{Z}_2$ monopole charge $w$ makes sense only if there are NLs that do not cross the entire BZ. To be more intuitive, $w$ can be understood as the number of double band inversions modulo two (Here, we define the double band inversion as two occupied states with both -1 parity eigenvalues.). Then, in order for $w$ to be 1, there should be an odd number of double band inversions.

Now we discuss how the commutation and anticommutation relations between $S_{2z}$ and $\mathcal{P}$ can be used to engineer the parity pattern. For convenience, only bands contribute to the trivial NLs and $\mathbb{Z}_2$NLs are considered, as shown in Figs.\ref{figpa} (c) and (g), respectively. The parity eigenvalues of the occupied states at TRIMs in the 3D NLSM with only a layer of 2D DSM in a unit cell (Fig. \ref{figpa}(a)) are shown in Fig. \ref{figpa}(b). Without loss of generality, we have assumed that the parity eigenvalues of the 2D DSM at three TRIMs are -1 and the rest of one TRIM is +1. In fact, we can choose different inversion centers to change the parity eigenvalues, but the parity of the number of the eigenvalue values of -1/+1 at all TRIMs does not change. Then we consider another 3D NLSM with two layers of 2D DSMs in a unit cell (See Fig. \ref{figpa}(e)) whose parity pattern is shown in Fig. \ref{figpa}(f) and the interlayer coupling is assumed to be weak so that it does not undergo a phase transition. As having been proved in Supplementary Material Sec. III, the bands contributed by the yellow layers in Fig. \ref{figpa}(a) and Fig. \ref{figpa} (e) share the same parity patterns, i.e., the first parity eigenvalues in the parentheses in Fig. \ref{figpa}(f) is just a copy of the parity eigenvalues in Fig. \ref{figpa}(b). At TRIMs where $S_{2z}$ commutates (anticommutates) with $\mathcal{P}$, if there is a state $\left|u_k\right>$ with a certain parity, there exists another state $S_{2z}\left|u_k\right>$ with the same (opposite) parity (See Supplementary Material Sec. II). Thus, one obtains a double band inversion at a TRIM iff (1) the parity eigenvalue in Fig. \ref{figpa}(b) at the TRIM is -1 and (2) $S_{2z}$ commutates with $\mathcal{P}$ at that TRIM according to Eq. \ref{eqsp}.

For our LC of 3D $\mathbb{Z}_2$NLSMs from 2D DSMs, there is a linking structure between the $\mathbb{Z}_2$NL and two NLs above and below the $\mathbb{Z}_2$NL as shown in Fig. \ref{figpa}(h). The 2D DSM has a $\mathcal{T}$-reversed pair of Dirac points. When a 3D NLSM is constructed by 2D DSMs with a 2D DSM in a unit cell (Fig. \ref{figpa}(a)), the pair of Dirac points becomes a $\mathcal{T}$-reversed pair of NLs throughout the BZ (Fig. \ref{figpa}(c), (d)) since we can not remove the Dirac points without symmetry broken. When we construct $\mathbb{Z}_2$NLSMs with two 2D DSMs in a unit cell, a $\mathcal{T}$-reversed pair of $\mathbb{Z}_2$NLs appears at Fermi energy with two $\mathcal{T}$-reversed pairs of NLs throughout the BZ preserved but above and below the $\mathbb{Z}_2$NLs (Fig. \ref{figpa}(g), (h)). While the surface Fermi arcs and hinge Fermi arcs are along the same direction, there should be a considerable energy gap between the upper/bottom surface Fermi arcs so that we are able to distinguish hinge Fermi arcs from surface Fermi arcs.

\noindent{\bf Tight-binding model.} Here we use the 3D structure formed by stacking the 2D honeycomb lattice models to illustrate
 the mechanism. More details about this tight-binding model can be found in Supplementary Material Sec. IV. The three basis lattice vectors of the 3D structure are given by $\bm a=(1,0,0), \bm b=\left(1/2,\sqrt{3}/2,0\right),\bm c=\left(0,0,1\right)$. There are four orbitals in a cell and their reduced coordinates are given by $\left(1/3,1/3,0\right),\left(2/3,2/3,0\right),\left(1/3,1/3,1/2\right),\left(2/3,2/3,1/2\right)$. The Hamiltonian without interlayer coupling is given by
\begin{equation}
H_0(\bm{k})=t_0\sum_j^3\left(\cos(\bm{k}\cdot\bm{\delta_j})\sigma_1\otimes\tau_0-\sin(\bm{k}\cdot\bm{\delta_j})\sigma_2\otimes\tau_0\right),
\end{equation}
where $\bm{\delta}_1=(1/3,1/3,0),\bm{\delta}_2=(1/3,-2/3,0),\bm{\delta}_3=(-2/3,1/3,0)$ and Pauli matrices $\sigma_i$, $\tau_i$ denote the sublattice and layer degrees of freedom, respectively. $H_0(\bm{k})$ has a $\mathcal{T}$-reversed pair of fourfold degeneracy NLs, i.e., $(1/3,2/3,k_z)$ and $(2/3,1/3,k_z)$ in the BZ.

We consider an interlayer coupling term
\begin{equation}
\begin{aligned}
\Delta H(\bm{k})=&(t_1+t_2)*\cos\left(\frac{k_z}{2}\right)\sigma_0\otimes\tau_1\\
&-(t_1-t_2)*\sin\left(\frac{k_z}{2}\right)\sigma_3\otimes\tau_2,
\end{aligned}
\end{equation}
where $t_1\neq t_2$ so that the 2D layer breaks $C_{2z}$. The energy dispersion is given by
\begin{equation}
E=\pm\sqrt{\left(\sqrt{d_1^2+d_2^2}\pm |d_3|\right)^2+d_4^2},
\end{equation}
with $d_1=t_0\sum_j^3\cos(\bm{k}\cdot\bm{\delta_j})$, $d_2=-t_0\sum_j^3\sin(\bm{k}\cdot\bm{\delta_j})$, $d_3=(t_1+t_2)*\cos\left(k_z/2\right)$, and $d_4=-(t_1-t_2)*\sin\left(k_z/2\right)$. The presence of the $\mathbb{Z}_2$NLs can be more intuitively understood by the following steps. Since $\sigma_3\otimes\tau_2$ anticommutates with both $\sigma_0\otimes\tau_1$ and $H_0$, $\sin\left(k_z/2\right)\sigma_3\otimes\tau_2$ gaps all the degeneracy nodes except the pair of Dirac points on $k_z=0$. Then we focus on the term $(t_1+t_2)*\cos\left(k_z/2\right)\sigma_0\otimes\tau_1$, which commutates with $H_0$. By adding $(t_1+t_2)*\cos\left(k_z/2\right)\sigma_0\otimes\tau_1$ to $H_0-(t_1-t_2)*\sin\left(k_z/2\right)\sigma_3\otimes\tau_2$, the pair of Dirac points evolves as shown in Fig. \ref{fignl}(a) as the increasing of the value of $t_1+t_2$. $H_0+\Delta H$ describes a 3D $\mathbb{Z}_2$NLSM when $(t_1+t_2)/t_0<1$, where we can get a $\mathcal{T}$-reversed pair of $\mathbb{Z}_2$NLs on $k_z=0$ shown in Figs. \ref{fignl}(b),(c). The Wilson loop evaluated on a sphere enclosing one of the NL is shown in Fig. \ref{fignl}(d), which indicates that the NL carries a unit $\mathbb{Z}_2$ monopole charge. The band structure of the rod that is only periodic in $x$-direction and the charge spatial distribution of the zero modes are shown in Figs. \ref{fignl}(e), (f), respectively. When $1<(t_1+t_2)/t_0<3$, the pair of $\mathbb{Z}_2$NLs merge into a single NL without carrying a unit monopole charge. Once $(t_1+t_2)/t_0>3$, this Hamiltonian describes an insulator.

In our model, each atom line in $z$-direction can be understood as an analogy of the Su-Schrieffer-Heeger (SSH) model \cite{PhysRevLett.42.1698} and the 2D layers in $x-y$ plane have zigzag edges. As plotted in Fig. \ref{fignl}(g), the rod which is only periodic in $x$-direction has a pair of zigzag edges. As shown in Fig. \ref{fignl}(f), the rod can also be viewed as a series of chains with alternative strong bonds and weak bonds in $z$-direction. The strong bonds in the surface push the surface states into the deep of the conduction bands and valence bands, while the two orbitals that have weak coupling with other orbitals contribute to the hinge states near the Fermi surface.

\noindent{\bf Material candidates.} 
Based on our LC scheme combined with first-principles calculations, we predict two kinds of electronic materials of 3D $\mathbb{Z}_2$NLSMs, which are the 14  Si/Ge 3D structures and 108 TMDs, in terms of their 2D DSM counterparts, i.e., Xene (X=silic, german) and $\rm 1T'$-TMD $\rm MX_2$ ($\rm M$=Cr, Mo, W, $\rm X$=S, Se, Te). (See Supplementary Material Sec. V for the details of all the candidates.)

We take wurtzites Si and Ge, 2 of the 14 Si/Ge 3D candidates with $\mathbb{Z}_2$NLs, as an example.  The wurtzite Si and wurtzite Ge, which have been experimentally prepared nearly half a century ago\cite{silicene,germanene}, crystallize in space group No. 194 with the structures shown in Fig. \ref{figmat}(a). The wurtzite Si and wurtzite Ge can be understood as stacked by their 2D DSMs silicene and germanene through $S_{2z}$ operation, therefore, according to our LC mechanism, one could expect the 3D wurtzite Si and wurtzite Ge to be $\mathbb{Z}_2$NLSMs. Unfortunately, the interlayer coupling in intrinsic wurtzite Si and wurtzite Ge is strong, making the systems in an insulating state, as shown in Fig. \ref{fignl}(a). However, we can reduce the interlayer coupling by inserting an inert layer, such as h-BN, in wurtzite Si or wurtzite Ge, and previous theoretical and experimental works have initially demonstrated the effectiveness of this method\cite{inserLth,inserLex}. For simplicity, we directly doubled the interlayer distances of wurtzite Si and wurtzite Ge to calculate their band structures. They have almost the same band structures as shown in Fig. \ref{figmat}(b). The $\mathcal{T}$-reversed pair of $\mathbb{Z}_2$NLs are located on $k_z=0$ around $K$ and $K'$. The inset shows the Wilson loop evaluated on the sphere enclosing a NL indicating the $\mathbb{Z}_2$ monopole charges on the NLs of the wurtzite Si and the wurtzite Ge. In Fig. \ref{figmat}(c), the energy dispersion for a sample with rod-like geometry is shown, and the hinge Fermi arcs are denoted by red lines. The spatial distribution of the hinge Fermi arcs are shown in Fig. \ref{figmat}(d).

For the 3D counterparts of 2D $\rm 1T'$$-MX_2$ ($M$=Cr,Mo,W, $X$=S,Se,Te), there are 9 possible element combinations and 16 spatial structures for each combination, a total of 144 3D TMDs. For each of the four possible selections of $(\mu,\nu)$, we have one structure for $S_{2z}$ ($\mathrm{T_d}$ phase), i.e., $\boldsymbol{c}$ is perpendicular to both $\boldsymbol{a}$ and $\boldsymbol{b}$, and three structures for $S_{2c}$ ($\mathrm{1T'}$ phase), i.e., $\boldsymbol{c}$ is inclined towards the $\boldsymbol{a}$-axis, the $\boldsymbol{b}$-axis, and both the $\boldsymbol{a}$-axis and $\boldsymbol{b}$-axis. We find all the structures in $\mathrm{T_d}$ phase and two out of three structures in $\rm 1T'$ phase have $\mathbb{Z}_2$NLs, a total of 108 candidate TMDs. We show the results for one structure of 3D $\rm CrS_2$, which has the lightest elements and thus the weakest and most negligible spin-orbit coupling among all TMDs. The other candidates have similar band structures and $\mathbb{Z}_2$NLs. The atomic structure of 3D $\rm CrS_2$ is shown in Fig. \ref{figmat}(e). The band structure with NLs is shown in Fig. \ref{figmat}(f), and the inset presents the Wilson loop evaluated on the sphere enclosing a NL implying the $\mathbb{Z}_2$ monopole charges of the NL. In Fig. \ref{figmat}(g), the energy dispersion for a sample with rod-like geometry is shown, and the hinge Fermi arcs are denoted by red lines. The spatial distribution of the hinge Fermi arcs are shown in Fig. \ref{figmat}(h).  More detailed information about the TMD family with $\mathbb{Z}_2$NLs can be found in Supplementary Material Sec. V.

\vspace{0.3cm}
\noindent{\bf Discussion}
	
\noindent

Similar to the role of inversion symmetry in diagnosing the topological invariant of topological insulators, while we can create a 3D $\mathbb{Z}_2$NLSMs from 2D DSMs making use of $S_{2z}$, $S_{2z}$ is not necessary for the final structure, i.e., the final structure does not need to have $S_{2z}$. In addition, we do not require $\bm c$ to be normal to $\bm a$ and $\bm b$. Let us consider a new symmetry operator $S_{2c}\equiv\left\{C_{2z}|\bm c/2\right\}$. Notice that here $\bm c$ is not normal to $\bm a$ and $\bm b$. Then Eq.\ref{eqsp} becomes
\begin{equation}
S_{2c}*\mathcal{P}=e^{ik_c-i\mu k_a-i\nu k_b}\mathcal{P}*S_{2c},
\end{equation}
where $\mu,\nu\in\{0,1\}$. We now can use the commutation and anticommutation relation between $S_{2\bm c}$ and $\mathcal{P}$ to modify the parity patterns at eight TRIMs.

The analyses for $S_{2z}$ can be easily extended to other screw symmetries. Since $S_{3z}$ neither commutates nor anticommutates with $\mathcal{P}$ at TRIMs, $S_{3z}$ is excluded. For $S_{4z}$ and $S_{6z}$, the layers in a unit cell can always be divided into two groups. In the case of $S_{4z}$ ($S_{6z}$), the first 2 (3) layers are merged into a group while the rest 2 (3) layers are merged into another group. Then the two groups are related by $S_{2z}$. The slide symmetry can not ensure the complete overlap of two Dirac cones from two layers and $\mathbb{Z}_2$NLs are less likely to appear. More discussion on these nonsymmorphic symmetries can be found in Supplementary Material Sec. I.

We generalize the notion of LC from TCIs to $\mathbb{Z}_2$NLSMs and use 2D DSMs with individual $\mathcal{P}$ and $\mathcal{T}$ and without $C_{2z}$  to construct $\mathbb{Z}_2$NLSMs. Our mechanism uses the commutation relationship between $S_{2z}$ and $\mathcal{P}$ to modify the parity eigenvalues at eight TRIMs. 

By our mechanism, on the one hand, we simplify the process of searching 3D $\mathbb{Z}_2$NLSMs to searching 2D DSM with individual $\mathcal{P}$ and $\mathcal{T}$ and without $C_{2z}$ and give two categories of candidate materials, which greatly enriches the real materials of the 3D $\mathbb{Z}_2$NLSMs. On the other hand, our work also deepens the understanding of 3D $\mathbb{Z}_2$NLSMs. The proposed mechanism to construct  3D $\mathbb{Z}_2$NLSM can be directly applied to the classical wave systems with $\mathcal{P}$$\mathcal{T}$ symmetry, including acoustic, photonic, and circuit systems. We believe that our theory can be experimentally verified in the future.

\vspace{0.3cm}
\noindent{\bf Methods}
	
\noindent
The calculations of the structure optimization and band structures of 3D structures of Xene and $\rm 1T'$-TMD are performed using DFT in the Perdew-Becke-Ernzerhof (PBE) generalized gradient approximation (GGA) \cite{PBE} implemented in the Vienna \textit{ab} \textit{initio} simulation package (VASP) \cite{VASP}. The symmetry--adapted Wannier functions are constructed using the WANNIER90 code \cite{wannier1} and VASP. Based on the constructed Wannier functions, we obtain the Wilson loops. 
	
\vspace{0.3cm}
\noindent{\bf Data availability} 
	
\noindent
The data and the code that support the findings of this study are available from the corresponding authors on reasonable request.

	
	\vspace{0.3cm}
	\def\bibsection{\noindent{\bf References}}
	
	\bibliographystyle{naturemag}
	\bibliography{0reference}

\begin{thebibliography}{10}
\expandafter\ifx\csname url\endcsname\relax
  \def\url#1{\texttt{#1}}\fi
\expandafter\ifx\csname urlprefix\endcsname\relax\def\urlprefix{URL }\fi
\providecommand{\bibinfo}[2]{#2}
\providecommand{\eprint}[2][]{\url{#2}}

\bibitem{RevModPhys.82.3045}
\bibinfo{author}{Hasan, M.~Z.} \& \bibinfo{author}{Kane, C.~L.}
\newblock \bibinfo{title}{Colloquium: Topological insulators}.
\newblock \emph{\bibinfo{journal}{Rev. Mod. Phys.}}
  \textbf{\bibinfo{volume}{82}}, \bibinfo{pages}{3045--3067}
  (\bibinfo{year}{2010}).

\bibitem{RevModPhys.83.1057}
\bibinfo{author}{Qi, X.-L.} \& \bibinfo{author}{Zhang, S.-C.}
\newblock \bibinfo{title}{Topological insulators and superconductors}.
\newblock \emph{\bibinfo{journal}{Rev. Mod. Phys.}}
  \textbf{\bibinfo{volume}{83}}, \bibinfo{pages}{1057--1110}
  (\bibinfo{year}{2011}).

\bibitem{Gu2009}
\bibinfo{author}{Gu, Z.-C.} \& \bibinfo{author}{Wen, X.-G.}
\newblock \bibinfo{title}{Tensor-entanglement-filtering renormalization
  approach and symmetry-protected topological order}.
\newblock \emph{\bibinfo{journal}{Phys. Rev. B}} \textbf{\bibinfo{volume}{80}},
  \bibinfo{pages}{155131} (\bibinfo{year}{2009}).

\bibitem{PhysRevB.99.235125}
\bibinfo{author}{Ahn, J.} \& \bibinfo{author}{Yang, B.-J.}
\newblock \bibinfo{title}{Symmetry representation approach to topological
  invariants in ${C}_{2z}t$-symmetric systems}.
\newblock \emph{\bibinfo{journal}{Phys. Rev. B}} \textbf{\bibinfo{volume}{99}},
  \bibinfo{pages}{235125} (\bibinfo{year}{2019}).

\bibitem{bradlyn2016beyond}
\bibinfo{author}{Bradlyn, B.} \emph{et~al.}
\newblock \bibinfo{title}{Beyond dirac and weyl fermions: Unconventional
  quasiparticles in conventional crystals}.
\newblock \emph{\bibinfo{journal}{Science}} \textbf{\bibinfo{volume}{353}},
  \bibinfo{pages}{aaf5037} (\bibinfo{year}{2016}).

\bibitem{PhysRevB.88.125129}
\bibinfo{author}{Morimoto, T.} \& \bibinfo{author}{Furusaki, A.}
\newblock \bibinfo{title}{Topological classification with additional symmetries
  from clifford algebras}.
\newblock \emph{\bibinfo{journal}{Phys. Rev. B}} \textbf{\bibinfo{volume}{88}},
  \bibinfo{pages}{125129} (\bibinfo{year}{2013}).

\bibitem{teo2010topological}
\bibinfo{author}{Teo, J. C.~Y.} \& \bibinfo{author}{Kane, C.~L.}
\newblock \bibinfo{title}{Topological defects and gapless modes in insulators
  and superconductors}.
\newblock \emph{\bibinfo{journal}{Phys. Rev. B}} \textbf{\bibinfo{volume}{82}},
  \bibinfo{pages}{115120} (\bibinfo{year}{2010}).

\bibitem{nakahara_geometry_2018}
\bibinfo{author}{Nakahara, M.}
\newblock \emph{\bibinfo{title}{Geometry, topology and physics}}
  (\bibinfo{publisher}{CRC press}, \bibinfo{year}{2018}).

\bibitem{bzduvsek2017robust}
\bibinfo{author}{Bzdu\ifmmode~\check{s}\else \v{s}\fi{}ek, T. c.~v.} \&
  \bibinfo{author}{Sigrist, M.}
\newblock \bibinfo{title}{Robust doubly charged nodal lines and nodal surfaces
  in centrosymmetric systems}.
\newblock \emph{\bibinfo{journal}{Phys. Rev. B}} \textbf{\bibinfo{volume}{96}},
  \bibinfo{pages}{155105} (\bibinfo{year}{2017}).

\bibitem{fang2016topological}
\bibinfo{author}{Fang, C.}, \bibinfo{author}{Weng, H.}, \bibinfo{author}{Dai,
  X.} \& \bibinfo{author}{Fang, Z.}
\newblock \bibinfo{title}{Topological nodal line semimetals}.
\newblock \emph{\bibinfo{journal}{Chinese Physics B}}
  \textbf{\bibinfo{volume}{25}}, \bibinfo{pages}{117106}
  (\bibinfo{year}{2016}).

\bibitem{fang2015topological}
\bibinfo{author}{Fang, C.}, \bibinfo{author}{Chen, Y.}, \bibinfo{author}{Kee,
  H.-Y.} \& \bibinfo{author}{Fu, L.}
\newblock \bibinfo{title}{Topological nodal line semimetals with and without
  spin-orbital coupling}.
\newblock \emph{\bibinfo{journal}{Phys. Rev. B}} \textbf{\bibinfo{volume}{92}},
  \bibinfo{pages}{081201} (\bibinfo{year}{2015}).

\bibitem{ZYX2020realchern}
\bibinfo{author}{Wang, K.}, \bibinfo{author}{Dai, J.-X.},
  \bibinfo{author}{Shao, L.~B.}, \bibinfo{author}{Yang, S.~A.} \&
  \bibinfo{author}{Zhao, Y.~X.}
\newblock \bibinfo{title}{Boundary criticality of $\mathcal{PT}$-invariant
  topology and second-order nodal-line semimetals}.
\newblock \emph{\bibinfo{journal}{Phys. Rev. Lett.}}
  \textbf{\bibinfo{volume}{125}}, \bibinfo{pages}{126403}
  (\bibinfo{year}{2020}).

\bibitem{PhysRevB.89.235127}
\bibinfo{author}{Morimoto, T.} \& \bibinfo{author}{Furusaki, A.}
\newblock \bibinfo{title}{Weyl and dirac semimetals with ${\mathbb{z}}_{2}$
  topological charge}.
\newblock \emph{\bibinfo{journal}{Phys. Rev. B}} \textbf{\bibinfo{volume}{89}},
  \bibinfo{pages}{235127} (\bibinfo{year}{2014}).

\bibitem{PhysRevLett.116.156402}
\bibinfo{author}{Zhao, Y.~X.}, \bibinfo{author}{Schnyder, A.~P.} \&
  \bibinfo{author}{Wang, Z.~D.}
\newblock \bibinfo{title}{Unified theory of $pt$ and $cp$ invariant topological
  metals and nodal superconductors}.
\newblock \emph{\bibinfo{journal}{Phys. Rev. Lett.}}
  \textbf{\bibinfo{volume}{116}}, \bibinfo{pages}{156402}
  (\bibinfo{year}{2016}).

\bibitem{PhysRevLett.118.056401}
\bibinfo{author}{Zhao, Y.~X.} \& \bibinfo{author}{Lu, Y.}
\newblock \bibinfo{title}{$pt$-symmetric real dirac fermions and semimetals}.
\newblock \emph{\bibinfo{journal}{Phys. Rev. Lett.}}
  \textbf{\bibinfo{volume}{118}}, \bibinfo{pages}{056401}
  (\bibinfo{year}{2017}).

\bibitem{PhysRevX.8.031069}
\bibinfo{author}{Song, Z.}, \bibinfo{author}{Zhang, T.} \&
  \bibinfo{author}{Fang, C.}
\newblock \bibinfo{title}{Diagnosis for nonmagnetic topological semimetals in
  the absence of spin-orbital coupling}.
\newblock \emph{\bibinfo{journal}{Phys. Rev. X}} \textbf{\bibinfo{volume}{8}},
  \bibinfo{pages}{031069} (\bibinfo{year}{2018}).

\bibitem{YangBM2018band}
\bibinfo{author}{Ahn, J.}, \bibinfo{author}{Kim, D.}, \bibinfo{author}{Kim, Y.}
  \& \bibinfo{author}{Yang, B.-J.}
\newblock \bibinfo{title}{Band topology and linking structure of nodal line
  semimetals with ${Z}_{2}$ monopole charges}.
\newblock \emph{\bibinfo{journal}{Phys. Rev. Lett.}}
  \textbf{\bibinfo{volume}{121}}, \bibinfo{pages}{106403}
  (\bibinfo{year}{2018}).

\bibitem{PhysRevLett.115.036806}
\bibinfo{author}{Kim, Y.}, \bibinfo{author}{Wieder, B.~J.},
  \bibinfo{author}{Kane, C.~L.} \& \bibinfo{author}{Rappe, A.~M.}
\newblock \bibinfo{title}{Dirac line nodes in inversion-symmetric crystals}.
\newblock \emph{\bibinfo{journal}{Phys. Rev. Lett.}}
  \textbf{\bibinfo{volume}{115}}, \bibinfo{pages}{036806}
  (\bibinfo{year}{2015}).

\bibitem{PhysRevB.90.205136}
\bibinfo{author}{Chiu, C.-K.} \& \bibinfo{author}{Schnyder, A.~P.}
\newblock \bibinfo{title}{Classification of reflection-symmetry-protected
  topological semimetals and nodal superconductors}.
\newblock \emph{\bibinfo{journal}{Phys. Rev. B}} \textbf{\bibinfo{volume}{90}},
  \bibinfo{pages}{205136} (\bibinfo{year}{2014}).

\bibitem{PhysRevLett.82.2147}
\bibinfo{author}{Mikitik, G.~P.} \& \bibinfo{author}{Sharlai, Y.~V.}
\newblock \bibinfo{title}{Manifestation of berry's phase in metal physics}.
\newblock \emph{\bibinfo{journal}{Phys. Rev. Lett.}}
  \textbf{\bibinfo{volume}{82}}, \bibinfo{pages}{2147--2150}
  (\bibinfo{year}{1999}).

\bibitem{PhysRevB.93.205132}
\bibinfo{author}{Chan, Y.-H.}, \bibinfo{author}{Chiu, C.-K.},
  \bibinfo{author}{Chou, M.~Y.} \& \bibinfo{author}{Schnyder, A.~P.}
\newblock \bibinfo{title}{${\mathrm{ca}}_{3}{\mathrm{p}}_{2}$ and other
  topological semimetals with line nodes and drumhead surface states}.
\newblock \emph{\bibinfo{journal}{Phys. Rev. B}} \textbf{\bibinfo{volume}{93}},
  \bibinfo{pages}{205132} (\bibinfo{year}{2016}).

\bibitem{Carbon+allotropes}
\bibinfo{author}{Chen, Y.} \emph{et~al.}
\newblock \bibinfo{title}{Nanostructured carbon allotropes with weyl-like loops
  and points}.
\newblock \emph{\bibinfo{journal}{Nano Letters}} \textbf{\bibinfo{volume}{15}},
  \bibinfo{pages}{6974--6978} (\bibinfo{year}{2015}).

\bibitem{PhysRevB.94.195104}
\bibinfo{author}{Zhao, J.}, \bibinfo{author}{Yu, R.}, \bibinfo{author}{Weng,
  H.} \& \bibinfo{author}{Fang, Z.}
\newblock \bibinfo{title}{Topological node-line semimetal in compressed black
  phosphorus}.
\newblock \emph{\bibinfo{journal}{Phys. Rev. B}} \textbf{\bibinfo{volume}{94}},
  \bibinfo{pages}{195104} (\bibinfo{year}{2016}).

\bibitem{PhysRevLett.117.096401}
\bibinfo{author}{Li, R.} \emph{et~al.}
\newblock \bibinfo{title}{Dirac node lines in pure alkali earth metals}.
\newblock \emph{\bibinfo{journal}{Phys. Rev. Lett.}}
  \textbf{\bibinfo{volume}{117}}, \bibinfo{pages}{096401}
  (\bibinfo{year}{2016}).

\bibitem{PhysRevB.93.201114}
\bibinfo{author}{Huang, H.}, \bibinfo{author}{Liu, J.},
  \bibinfo{author}{Vanderbilt, D.} \& \bibinfo{author}{Duan, W.}
\newblock \bibinfo{title}{Topological nodal-line semimetals in alkaline-earth
  stannides, germanides, and silicides}.
\newblock \emph{\bibinfo{journal}{Phys. Rev. B}} \textbf{\bibinfo{volume}{93}},
  \bibinfo{pages}{201114} (\bibinfo{year}{2016}).

\bibitem{hirayama2017topological}
\bibinfo{author}{Hirayama, M.}, \bibinfo{author}{Okugawa, R.},
  \bibinfo{author}{Miyake, T.} \& \bibinfo{author}{Murakami, S.}
\newblock \bibinfo{title}{Topological dirac nodal lines and surface charges in
  fcc alkaline earth metals}.
\newblock \emph{\bibinfo{journal}{Nature communications}}
  \textbf{\bibinfo{volume}{8}}, \bibinfo{pages}{1--9} (\bibinfo{year}{2017}).

\bibitem{zyx2022graphdiyne}
\bibinfo{author}{Chen, C.} \emph{et~al.}
\newblock \bibinfo{title}{Second-order real nodal-line semimetal in
  three-dimensional graphdiyne}.
\newblock \emph{\bibinfo{journal}{Phys. Rev. Lett.}}
  \textbf{\bibinfo{volume}{128}}, \bibinfo{pages}{026405}
  (\bibinfo{year}{2022}).

\bibitem{GraphdiyneExp1}
\bibinfo{author}{Matsuoka, R.} \emph{et~al.}
\newblock \bibinfo{title}{Crystalline graphdiyne nanosheets produced at a
  gas/liquid or liquid/liquid interface}.
\newblock \emph{\bibinfo{journal}{Journal of the American Chemical Society}}
  \textbf{\bibinfo{volume}{139}}, \bibinfo{pages}{3145--3152}
  (\bibinfo{year}{2017}).

\bibitem{GraphdiyneExp2}
\bibinfo{author}{Gao, X.} \emph{et~al.}
\newblock \bibinfo{title}{Ultrathin graphdiyne film on graphene through
  solution-phase van der waals epitaxy}.
\newblock \emph{\bibinfo{journal}{Science Advances}}
  \textbf{\bibinfo{volume}{4}}, \bibinfo{pages}{eaat6378}
  (\bibinfo{year}{2018}).

\bibitem{PhysRevMaterials.2.054204}
\bibinfo{author}{Nomura, T.}, \bibinfo{author}{Habe, T.},
  \bibinfo{author}{Sakamoto, R.} \& \bibinfo{author}{Koshino, M.}
\newblock \bibinfo{title}{Three-dimensional graphdiyne as a topological
  nodal-line semimetal}.
\newblock \emph{\bibinfo{journal}{Phys. Rev. Materials}}
  \textbf{\bibinfo{volume}{2}}, \bibinfo{pages}{054204} (\bibinfo{year}{2018}).

\bibitem{MoTe2}
\bibinfo{author}{Wang, Z.}, \bibinfo{author}{Wieder, B.~J.},
  \bibinfo{author}{Li, J.}, \bibinfo{author}{Yan, B.} \&
  \bibinfo{author}{Bernevig, B.~A.}
\newblock \bibinfo{title}{Higher-order topology, monopole nodal lines, and the
  origin of large fermi arcs in transition metal dichalcogenides
  $x{\mathrm{te}}_{2}$ ($x=\mathrm{Mo},\mathrm{W}$)}.
\newblock \emph{\bibinfo{journal}{Phys. Rev. Lett.}}
  \textbf{\bibinfo{volume}{123}}, \bibinfo{pages}{186401}
  (\bibinfo{year}{2019}).

\bibitem{brown1966crystal}
\bibinfo{author}{Brown, B.~E.}
\newblock \bibinfo{title}{The crystal structures of wte2 and high-temperature
  mote2}.
\newblock \emph{\bibinfo{journal}{Acta Crystallographica}}
  \textbf{\bibinfo{volume}{20}}, \bibinfo{pages}{268--274}
  (\bibinfo{year}{1966}).

\bibitem{PhysRevB.92.081304}
\bibinfo{author}{Isobe, H.} \& \bibinfo{author}{Fu, L.}
\newblock \bibinfo{title}{Theory of interacting topological crystalline
  insulators}.
\newblock \emph{\bibinfo{journal}{Phys. Rev. B}} \textbf{\bibinfo{volume}{92}},
  \bibinfo{pages}{081304} (\bibinfo{year}{2015}).

\bibitem{PhysRevB.94.125405}
\bibinfo{author}{Fulga, I.~C.}, \bibinfo{author}{Avraham, N.},
  \bibinfo{author}{Beidenkopf, H.} \& \bibinfo{author}{Stern, A.}
\newblock \bibinfo{title}{Coupled-layer description of topological crystalline
  insulators}.
\newblock \emph{\bibinfo{journal}{Phys. Rev. B}} \textbf{\bibinfo{volume}{94}},
  \bibinfo{pages}{125405} (\bibinfo{year}{2016}).

\bibitem{PhysRevB.94.155148}
\bibinfo{author}{Ezawa, M.}
\newblock \bibinfo{title}{Hourglass fermion surface states in stacked
  topological insulators with nonsymmorphic symmetry}.
\newblock \emph{\bibinfo{journal}{Phys. Rev. B}} \textbf{\bibinfo{volume}{94}},
  \bibinfo{pages}{155148} (\bibinfo{year}{2016}).

\bibitem{PhysRevX.7.011020}
\bibinfo{author}{Song, H.}, \bibinfo{author}{Huang, S.-J.},
  \bibinfo{author}{Fu, L.} \& \bibinfo{author}{Hermele, M.}
\newblock \bibinfo{title}{Topological phases protected by point group
  symmetry}.
\newblock \emph{\bibinfo{journal}{Phys. Rev. X}} \textbf{\bibinfo{volume}{7}},
  \bibinfo{pages}{011020} (\bibinfo{year}{2017}).

\bibitem{PhysRevB.96.205106}
\bibinfo{author}{Huang, S.-J.}, \bibinfo{author}{Song, H.},
  \bibinfo{author}{Huang, Y.-P.} \& \bibinfo{author}{Hermele, M.}
\newblock \bibinfo{title}{Building crystalline topological phases from
  lower-dimensional states}.
\newblock \emph{\bibinfo{journal}{Phys. Rev. B}} \textbf{\bibinfo{volume}{96}},
  \bibinfo{pages}{205106} (\bibinfo{year}{2017}).

\bibitem{song2018quantitative}
\bibinfo{author}{Song, Z.}, \bibinfo{author}{Zhang, T.}, \bibinfo{author}{Fang,
  Z.} \& \bibinfo{author}{Fang, C.}
\newblock \bibinfo{title}{Quantitative mappings between symmetry and topology
  in solids}.
\newblock \emph{\bibinfo{journal}{Nature communications}}
  \textbf{\bibinfo{volume}{9}}, \bibinfo{pages}{1--7} (\bibinfo{year}{2018}).

\bibitem{qian2020layer}
\bibinfo{author}{Qian, Y.} \emph{et~al.}
\newblock \bibinfo{title}{Layer construction of topological crystalline
  insulator lasbte}.
\newblock \emph{\bibinfo{journal}{Science China Physics, Mechanics \&
  Astronomy}} \textbf{\bibinfo{volume}{63}}, \bibinfo{pages}{1--7}
  (\bibinfo{year}{2020}).

\bibitem{PhysRevLett.42.1698}
\bibinfo{author}{Su, W.~P.}, \bibinfo{author}{Schrieffer, J.~R.} \&
  \bibinfo{author}{Heeger, A.~J.}
\newblock \bibinfo{title}{Solitons in polyacetylene}.
\newblock \emph{\bibinfo{journal}{Phys. Rev. Lett.}}
  \textbf{\bibinfo{volume}{42}}, \bibinfo{pages}{1698--1701}
  (\bibinfo{year}{1979}).

\bibitem{silicene}
\bibinfo{author}{Jennings, H.~M.} \& \bibinfo{author}{Richman, M.~H.}
\newblock \bibinfo{title}{A hexagonal (wurtzite) form of silicon}.
\newblock \emph{\bibinfo{journal}{Science}} \textbf{\bibinfo{volume}{193}},
  \bibinfo{pages}{1242--1243} (\bibinfo{year}{1976}).

\bibitem{germanene}
\bibinfo{author}{Eremenko, V.}
\newblock \bibinfo{title}{New crystalline phase of ge}.
\newblock \emph{\bibinfo{journal}{Fiz. Tverd. Tela}}
  \textbf{\bibinfo{volume}{17}}, \bibinfo{pages}{2476--2477}
  (\bibinfo{year}{1975}).

\bibitem{inserLth}
\bibinfo{author}{Wu, Z.-B.}, \bibinfo{author}{Zhang, Y.-Y.},
  \bibinfo{author}{Li, G.}, \bibinfo{author}{Du, S.} \& \bibinfo{author}{Gao,
  H.-J.}
\newblock \bibinfo{title}{Electronic properties of silicene in bn/silicene van
  der waals heterostructures}.
\newblock \emph{\bibinfo{journal}{Chinese Physics B}}
  \textbf{\bibinfo{volume}{27}}, \bibinfo{pages}{077302}
  (\bibinfo{year}{2018}).

\bibitem{inserLex}
\bibinfo{author}{Wiggers, F.~B.} \emph{et~al.}
\newblock \bibinfo{title}{Van der waals integration of silicene and hexagonal
  boron nitride}.
\newblock \emph{\bibinfo{journal}{2D Materials}} \textbf{\bibinfo{volume}{6}},
  \bibinfo{pages}{035001} (\bibinfo{year}{2019}).

\bibitem{PBE}
\bibinfo{author}{Perdew, J.~P.}, \bibinfo{author}{Burke, K.} \&
  \bibinfo{author}{Ernzerhof, M.}
\newblock \bibinfo{title}{Generalized gradient approximation made simple}.
\newblock \emph{\bibinfo{journal}{Phys. Rev. Lett.}}
  \textbf{\bibinfo{volume}{77}}, \bibinfo{pages}{3865} (\bibinfo{year}{1996}).

\bibitem{VASP}
\bibinfo{author}{Kresse, G.} \& \bibinfo{author}{Furthm{\"u}ller, J.}
\newblock \bibinfo{title}{Efficient iterative schemes for $ab$ initio
  total-energy calculations using a plane-wave basis set}.
\newblock \emph{\bibinfo{journal}{Phys. Rev. B}} \textbf{\bibinfo{volume}{54}},
  \bibinfo{pages}{11169} (\bibinfo{year}{1996}).

\bibitem{wannier1}
\bibinfo{author}{Mostofi, A.~A.} \emph{et~al.}
\newblock \bibinfo{title}{wannier90: A tool for obtaining maximally-localised
  wannier functions}.
\newblock \emph{\bibinfo{journal}{Comput. Phys. Commun.}}
  \textbf{\bibinfo{volume}{178}}, \bibinfo{pages}{685--699}
  (\bibinfo{year}{2008}).

\end{thebibliography}


\begin{thebibliography}{5}%
\makeatletter
\providecommand \@ifxundefined [1]{%
 \@ifx{#1\undefined}
}%
\providecommand \@ifnum [1]{%
 \ifnum #1\expandafter \@firstoftwo
 \else \expandafter \@secondoftwo
 \fi
}%
\providecommand \@ifx [1]{%
 \ifx #1\expandafter \@firstoftwo
 \else \expandafter \@secondoftwo
 \fi
}%
\providecommand \natexlab [1]{#1}%
\providecommand \enquote  [1]{``#1''}%
\providecommand \bibnamefont  [1]{#1}%
\providecommand \bibfnamefont [1]{#1}%
\providecommand \citenamefont [1]{#1}%
\providecommand \href@noop [0]{\@secondoftwo}%
\providecommand \href [0]{\begingroup \@sanitize@url \@href}%
\providecommand \@href[1]{\@@startlink{#1}\@@href}%
\providecommand \@@href[1]{\endgroup#1\@@endlink}%
\providecommand \@sanitize@url [0]{\catcode `\\12\catcode `\$12\catcode
  `\&12\catcode `\#12\catcode `\^12\catcode `\_12\catcode `\%12\relax}%
\providecommand \@@startlink[1]{}%
\providecommand \@@endlink[0]{}%
\providecommand \url  [0]{\begingroup\@sanitize@url \@url }%
\providecommand \@url [1]{\endgroup\@href {#1}{\urlprefix }}%
\providecommand \urlprefix  [0]{URL }%
\providecommand \Eprint [0]{\href }%
\providecommand \doibase [0]{https://doi.org/}%
\providecommand \selectlanguage [0]{\@gobble}%
\providecommand \bibinfo  [0]{\@secondoftwo}%
\providecommand \bibfield  [0]{\@secondoftwo}%
\providecommand \translation [1]{[#1]}%
\providecommand \BibitemOpen [0]{}%
\providecommand \bibitemStop [0]{}%
\providecommand \bibitemNoStop [0]{.\EOS\space}%
\providecommand \EOS [0]{\spacefactor3000\relax}%
\providecommand \BibitemShut  [1]{\csname bibitem#1\endcsname}%
\let\auto@bib@innerbib\@empty
\bibitem [{\citenamefont {Evarestov}\ and\ \citenamefont
  {Smirnov}(2012)}]{evarestov2012site}%
  \BibitemOpen
  \bibfield  {author} {\bibinfo {author} {\bibfnamefont {R.~A.}\ \bibnamefont
  {Evarestov}}\ and\ \bibinfo {author} {\bibfnamefont {V.~P.}\ \bibnamefont
  {Smirnov}},\ }\href@noop {} {\emph {\bibinfo {title} {Site symmetry in
  crystals: theory and applications}}},\ Vol.\ \bibinfo {volume} {108}\
  (\bibinfo  {publisher} {Springer Science \& Business Media},\ \bibinfo {year}
  {2012})\BibitemShut {NoStop}%
\bibitem [{\citenamefont {Cano}\ \emph {et~al.}(2018)\citenamefont {Cano},
  \citenamefont {Bradlyn}, \citenamefont {Wang}, \citenamefont {Elcoro},
  \citenamefont {Vergniory}, \citenamefont {Felser}, \citenamefont {Aroyo},\
  and\ \citenamefont {Bernevig}}]{PhysRevB.97.035139}%
  \BibitemOpen
  \bibfield  {author} {\bibinfo {author} {\bibfnamefont {J.}~\bibnamefont
  {Cano}}, \bibinfo {author} {\bibfnamefont {B.}~\bibnamefont {Bradlyn}},
  \bibinfo {author} {\bibfnamefont {Z.}~\bibnamefont {Wang}}, \bibinfo {author}
  {\bibfnamefont {L.}~\bibnamefont {Elcoro}}, \bibinfo {author} {\bibfnamefont
  {M.~G.}\ \bibnamefont {Vergniory}}, \bibinfo {author} {\bibfnamefont
  {C.}~\bibnamefont {Felser}}, \bibinfo {author} {\bibfnamefont {M.~I.}\
  \bibnamefont {Aroyo}},\ and\ \bibinfo {author} {\bibfnamefont {B.~A.}\
  \bibnamefont {Bernevig}},\ }\bibfield  {title} {\bibinfo {title} {Building
  blocks of topological quantum chemistry: Elementary band representations},\
  }\href {https://doi.org/10.1103/PhysRevB.97.035139} {\bibfield  {journal}
  {\bibinfo  {journal} {Phys. Rev. B}\ }\textbf {\bibinfo {volume} {97}},\
  \bibinfo {pages} {035139} (\bibinfo {year} {2018})}\BibitemShut {NoStop}%
\bibitem [{\citenamefont {Liu}\ \emph {et~al.}(2011)\citenamefont {Liu},
  \citenamefont {Jiang},\ and\ \citenamefont {Yao}}]{Liu2011}%
  \BibitemOpen
  \bibfield  {author} {\bibinfo {author} {\bibfnamefont {C.-C.}\ \bibnamefont
  {Liu}}, \bibinfo {author} {\bibfnamefont {H.}~\bibnamefont {Jiang}},\ and\
  \bibinfo {author} {\bibfnamefont {Y.}~\bibnamefont {Yao}},\ }\bibfield
  {title} {\bibinfo {title} {Low-energy effective hamiltonian involving
  spin-orbit coupling in silicene and two-dimensional germanium and tin},\
  }\href {https://doi.org/10.1103/PhysRevB.84.195430} {\bibfield  {journal}
  {\bibinfo  {journal} {Phys. Rev. B}\ }\textbf {\bibinfo {volume} {84}},\
  \bibinfo {pages} {195430} (\bibinfo {year} {2011})}\BibitemShut {NoStop}%
\bibitem [{\citenamefont {Kresse}\ and\ \citenamefont
  {Furthm{\"u}ller}(1996)}]{vasp}%
  \BibitemOpen
  \bibfield  {author} {\bibinfo {author} {\bibfnamefont {G.}~\bibnamefont
  {Kresse}}\ and\ \bibinfo {author} {\bibfnamefont {J.}~\bibnamefont
  {Furthm{\"u}ller}},\ }\bibfield  {title} {\bibinfo {title} {Efficient
  iterative schemes for $ab$ initio total-energy calculations using a
  plane-wave basis set},\ }\href {https://doi.org/10.1103/physrevb.54.11169}
  {\bibfield  {journal} {\bibinfo  {journal} {Phys. Rev. B}\ }\textbf {\bibinfo
  {volume} {54}},\ \bibinfo {pages} {11169} (\bibinfo {year}
  {1996})}\BibitemShut {NoStop}%
\bibitem [{\citenamefont {Mostofi}\ \emph {et~al.}(2008)\citenamefont
  {Mostofi}, \citenamefont {Yates}, \citenamefont {Lee}, \citenamefont {Souza},
  \citenamefont {Vanderbilt},\ and\ \citenamefont {Marzari}}]{wannier90}%
  \BibitemOpen
  \bibfield  {author} {\bibinfo {author} {\bibfnamefont {A.~A.}\ \bibnamefont
  {Mostofi}}, \bibinfo {author} {\bibfnamefont {J.~R.}\ \bibnamefont {Yates}},
  \bibinfo {author} {\bibfnamefont {Y.-S.}\ \bibnamefont {Lee}}, \bibinfo
  {author} {\bibfnamefont {I.}~\bibnamefont {Souza}}, \bibinfo {author}
  {\bibfnamefont {D.}~\bibnamefont {Vanderbilt}},\ and\ \bibinfo {author}
  {\bibfnamefont {N.}~\bibnamefont {Marzari}},\ }\bibfield  {title} {\bibinfo
  {title} {wannier90: A tool for obtaining maximally-localised wannier
  functions},\ }\href {https://doi.org/10.1016/j.cpc.2007.11.016} {\bibfield
  {journal} {\bibinfo  {journal} {Comput. Phys. Commun.}\ }\textbf {\bibinfo
  {volume} {178}},\ \bibinfo {pages} {685} (\bibinfo {year}
  {2008})}\BibitemShut {NoStop}%
\end{thebibliography}%
	
	\vspace{0.3cm}
	\noindent{\bf Acknowledgements}
	
	 \noindent
	The work is supported by the National Key R\&D Program of China (Grant No. 2020YFA0308800) and the NSF of China (Grants No. 11922401).

	\vspace{0.3cm}
	\noindent{\bf Author contributions} 
	
	 \noindent
	C.-C. L. conceived the research. Y. L., S. Q., and C.-C. L. performed research, analyzed data, and wrote the manuscript.
	
	\vspace{0.3cm}
	\noindent{\bf Additional information}	
	
	\noindent{\bf Supplementary information} The online version contains supplementary material available at xxx.
	
	\noindent{\bf Competing interests:} The authors declare no competing financial interests.

\end{document}


\title{Supplementary Material for ``Layer Construction of Three-Dimensional $\mathbb{Z}_2$ Monopole Charge Nodal Line Semimetals and Prediction of Abundant Candidate Materials"}
\maketitle

\tableofcontents

\section{Symmetry consideration}
In our layer construction scheme, it is important to utilize the nonsymmorphic symmetry operators, i.e. screw symmetry and slide symmetry, to engineer the parity pattern at time-reversal invariant momentums (TRIMs). In the following, we will consider these nonsymmorphic symmetry operators one by one.
\subsection{Twofold screw symmetry}
 We have claimed in the text that the twofold screw symmetry $S_{2z}$ can be replaced by another symmetry $S_{2c}\equiv\left\{C_{2z}|\bm c/2\right\}$, i.e., a twofold rotation whose rotation axis is perpendicular to $\bm a$ and $\bm b$ following by a translation $\bm c/2$. Here we consider the case of $S_{2z}$ and the result for $S_{2c}$ can be immediately obtained by removing the limitation that $\bm c$ is normal to $\bm a$ and $\bm b$.

\begin{figure}[h]
\includegraphics[width=0.6\textwidth]{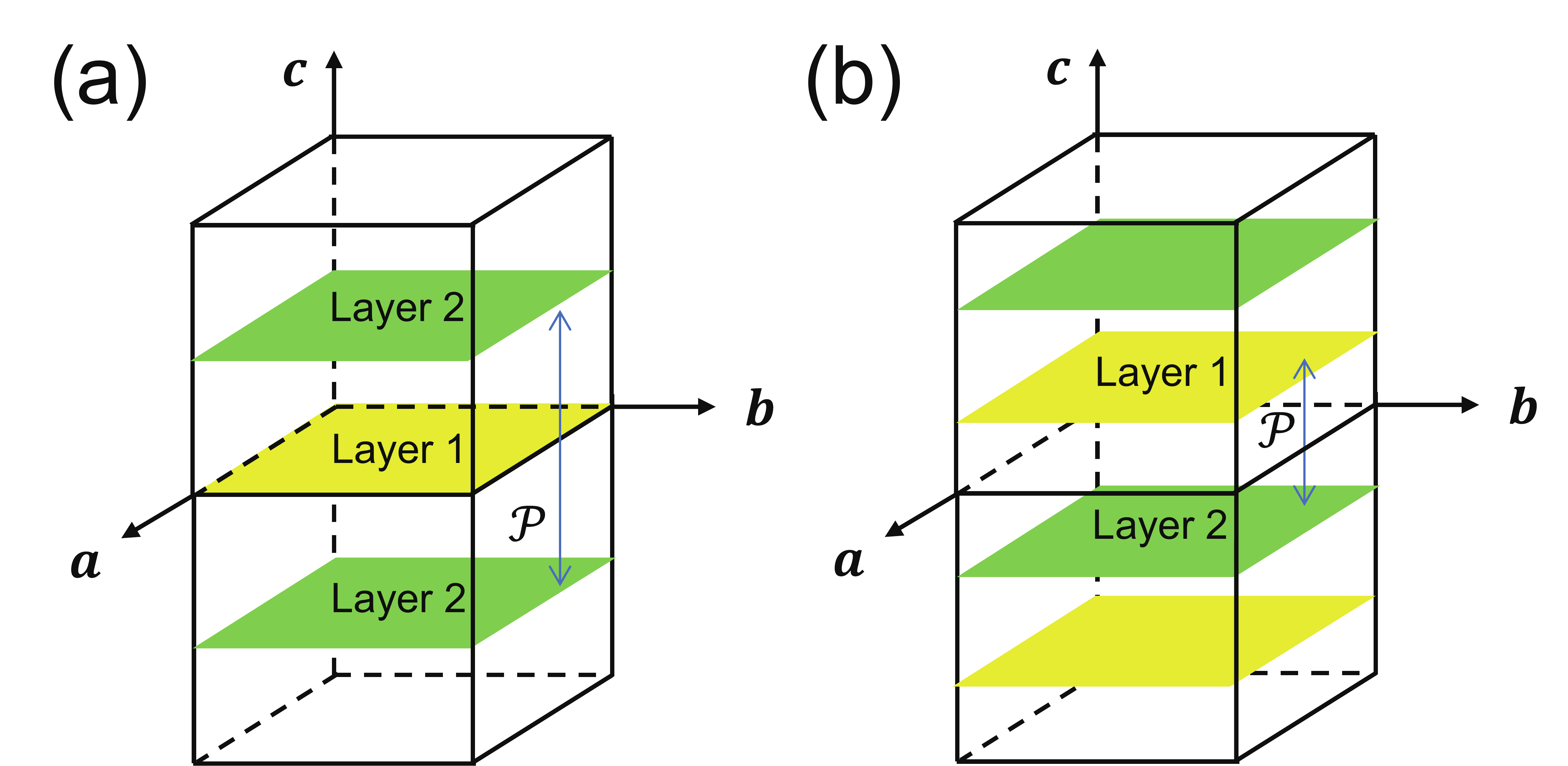}
\caption{\label{s2z} Schematics of the 3D structures with $\mathcal{P}$ and two layers of 2D DSMs per unit cell. (a) 3D structures formed by 2D DSMs without $C_{2z}$ and $\mathcal{P}$ leaves the 2D layers invariant. (b) 3D structure formed by 2D DSMs with $C_{2z}$ and $\mathcal{P}$ exchanges two 2D layers.}
\end{figure}

There are two layers of 2D DSMs per unit cell. The 3D structure should also have $\mathcal{P}$ and there are two ways to restore $\mathcal{P}$ as shown in Figs. \ref{s2z} (a) and (b), i.e., two layers are invariant under $\mathcal{P}$ and $\mathcal{P}$ exchanges two layers, respectively. And we discuss them separately.

\subsubsection{2D layers invariant under $\mathcal{P}$}\label{s2z1}
We first consider the case that two layers are invariant under $\mathcal{P}$ as shown in Fig. \ref{s2z}(a). $S_{2z}$ acts in real space as
\begin{equation}
(x,y,z)\rightarrow\left(-x+\frac{\mu a}{2},-y+\frac{\nu b}{2},z+\frac{c}{2}\right),
\end{equation}
where $\mu,\nu\in[0,2)$ and $a,b,c$ are the length of three lattice vectors.

Firstly, we prove that the values of $\mu$ and $\nu$ can only be 0 and 1. Let us consider the restriction of the inversion symmetry of the 3D structure on the positions of inversion centers of each 2D layer. Suppose one of the inversion centers of layer 1 is given by (0,0,0). Performing $S_{2z}$ to (0,0,0), we obtain the corresponding inversion center of layer 2 at the top: $(\mu a/2,\nu b/2,c/2)$. Performing $S_{2z}^{-1}$ to (0,0,0), we obtain the corresponding inversion center of layer 2 at the bottom: $(\mu a/2,\nu b/2,-c/2)$. In the most general case, $\mathcal{P}$ can only exchange these two inversion centers, i.e.,
\begin{equation}
\left(\frac{\mu a}{2},\frac{\nu b}{2},-\frac{c}{2}\right)\stackrel{\mathcal{P}}{\longrightarrow}\left(-\frac{\mu a}{2},-\frac{\nu b}{2},\frac{c}{2}\right)\equiv\left(\frac{(2-\mu) a}{2},\frac{(2-\nu) b}{2},\frac{c}{2}\right)\approx\left(\frac{\mu a}{2},\frac{\nu b}{2},\frac{c}{2}\right),\label{eqiv}
\end{equation}
where $\approx$ denotes that the two sides are equal modulo a lattice vector. Therefore, the values of $\mu,\nu$ are restricted to $\{0,1\}$. We get the same result by choosing other inversion centers of layer 1, i.e., $(a/2,0,0)$, $(0,b/2,0)$, and $(a/2,b/2,0)$. 

Next, we prove that the atoms in the 3D structure also have inversion symmetry. Here, $(0,0,0)$ is still one of the inversion centers of the 3D structure and the 2D layer 1. Consider an arbitrary atom in the layer 2 located at $(x+\mu a/2,y+\nu b/2,z+c/2)$ where $(\mu a/2,\nu b/2,c/2)$ is the coordinate of an inversion center given by Eq. \ref{eqiv}. Since the values of $\mu,\nu$ are limited to $\{0,1\}$, $(\mu a/2,\nu b/2,c/2)$ and $(-\mu a/2,-\nu b/2,-c/2)$ differ only by a lattice vector and $(-\mu a/2,-\nu b/2,-c/2)$ is also an inversion center of layer 2. By the translation invariance, there is an equivalent atom at $(x-\mu a/2,y-\nu b/2,z-c/2)$. By the inversion symmetry of the 2D layer 2, the atom of the layer 2 at $(x-\mu a/2,y-\nu b/2,z-c/2)$ has an equivalent atom of the layer 2 at $(-x-\mu a/2,-y-\nu b/2,-z-c/2)$. (Recall that $(-\mu a/2,-\nu b/2,-c/2)$ is also an inversion center.) Therefore, for every atom in layer 2 on the bottom, there exists an atom in layer 2 on the top that the two atoms ($(x+\mu a/2,y+\nu b/2,z+c/2)$ and $(-x-\mu a/2,-y-\nu b/2,-z-c/2)$) form an $\mathcal{P}$-reversed pair.

However, this is true only when the 2D DSMs do not have $C_{2z}$. Let us consider 2D DSMs with $C_{2z}$. In this case, two-fold rotation/screw axes may be $(0,0,z)$, $(0,0.25,z)$, $(0.25,0,z)$, or $(0.25,0.25,z)$. As a result, $C_{2z}$ followed by $S_{2z}$ acts in real space as
\begin{equation}
(x,y,z)\xrightarrow{S_{2z}*C_{2z}}\left(x,y,z+\frac{c}{2}\right),
\end{equation}
i.e., the two layers only differ by a pure translation $\bm c/2$ and thus the primitive cell actually contains only one layer of the 2D DSM.

\subsubsection{2D layers exchanged under $\mathcal{P}$}
We now consider the case that $\mathcal{P}$ exchanges two 2D layers as shown in Fig. \ref{s2z}(b). In this section, we choose a gauge that $S_{2z}$ acts in real space as
\begin{equation}
(x,y,z)\rightarrow\left(-x,-y,z+\frac{c}{2}\right),
\end{equation}
i.e., the twofold screw axis  becomes $(0,0,z)$.

Here, we prove that the 3D structure can not have inversion symmetry in the most general case. Assume there is an atom of layer 1 at $(x+i_x,y+i_y,z+c/4)$ where $(i_x,i_y,c/4)$ is the coordinate of an inversion center of layer 1. Again, there should be an equivalent atom at $(-x+i_x,-y+i_y,-z+c/4)$. Performing $S_{2z}$ on $(x+i_x,y+i_y,z+c/4)$, we have
\begin{equation}
\left(x+i_x,y+i_y,z+\frac{c}{4}\right)\xrightarrow{S_{2z}}\left(-x-i_x,-y-i_y,z+\frac{3c}{4}\right).
\end{equation}
For the most general case, we can not find a $\mathcal{P}$-reversed pair for every two atoms. However, things become different when the 2D DSMs have $C_{2z}$. In that case, $z$ can only be 0 due to the presence of $C_{2z}$ and $\mathcal{P}$ in a single layer. Therefore, for an atom at $(x+i_x,y+i_y,c/4)$, there is always an atom at $(-x-i_x,-y-i_y,-c/4)$. But, in this case, $\mathcal{P}$ does not commutate or anticommutate with $S_{2z}$ at TRIMs, which will be clear in Sec. \ref{S2zP}.

\subsection{Slide symmetry}
\begin{figure}[h]
\includegraphics[width=0.7\textwidth]{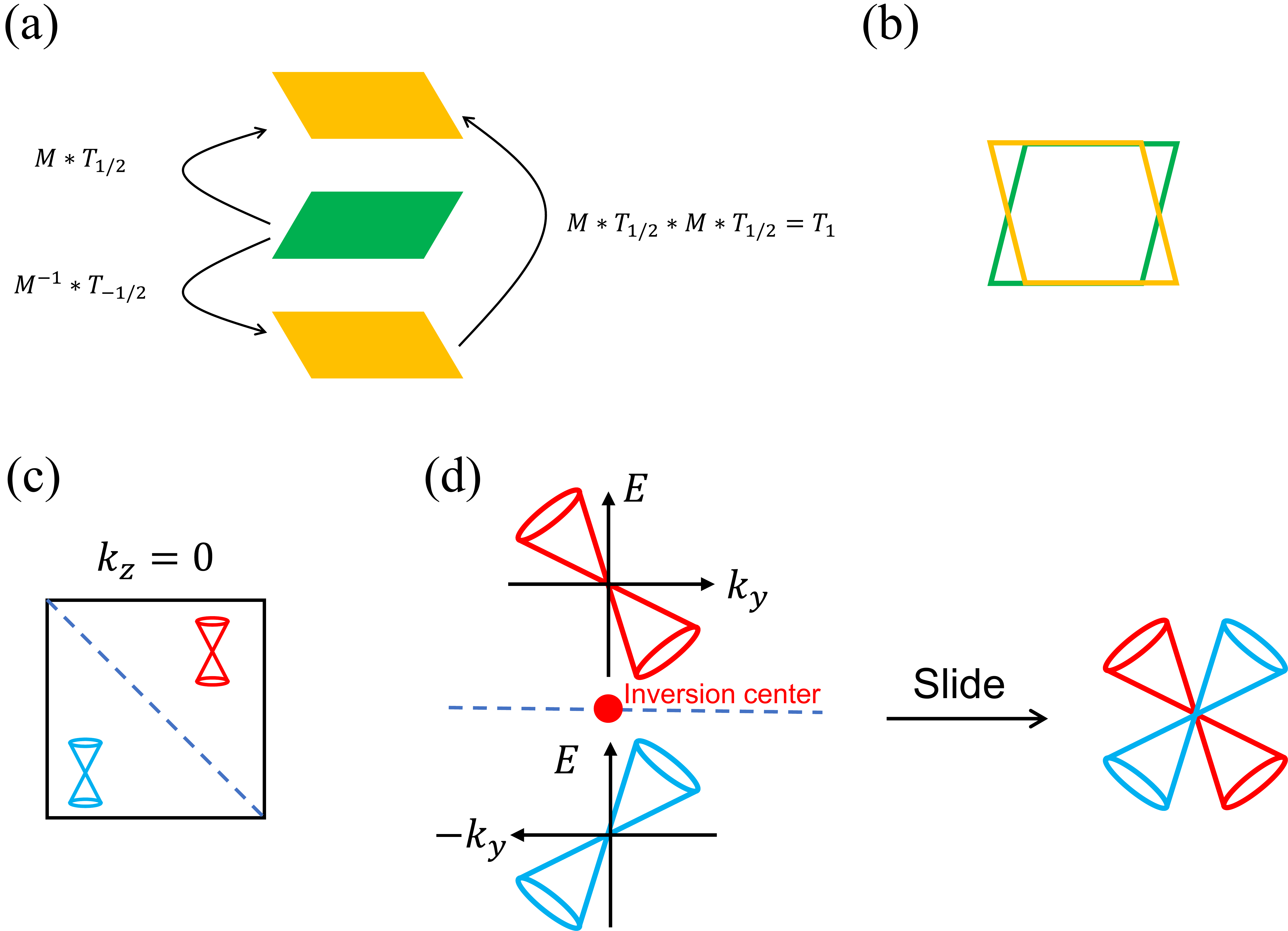}
\caption{\label{m1} (a) The action of the mirror operator of slide symmetry on a 2D layer and the fractional translation is 1/2 of the lattice constant. (b) When the slide symmetry is used, we need to consider the commensurability between two layers and the intervalley scattering which gaps the Dirac cones. (c) If the two layers are commensurate and intervalley scattering is not involved, the selection of the mirror operator also needs to ensure that two Dirac cones can coincide. (d) The combination of two type-II Dirac cones may not result in NLs.}
\end{figure}

In the case of the slide symmetry, while the fundamental symmetries of the 2D DSMs are still time reversal $\mathcal{T}$ and inversion $\mathcal{P}$, additional symmetries are needed. First of all, we need to deal with the commensurability between two layers as shown in Fig. \ref{m1}(b). Even if the two layers are commensurate, we also need to consider the possible intervalley scattering, which may gap the Dirac cones.

If the above problem does not exist, we still consider two cases as shown in Fig. \ref{s2z}. Without loss of generality, we assume that Dirac cones are located at $k_x$ axis and use $\{M_x|c/2\}$ as an example in this section. Again, one should notice that the $c$ axis does not need to be perpendicular to $a$ and $b$ as in the case of $S_{2c}$.

\subsubsection{2D layers invariant under $\mathcal{P}$}
Consider the case that two layers are invariant under $\mathcal{P}$ as shown in Fig. \ref{s2z}(a). $\{M_x|c/2\}$ acts in real space as
\begin{equation}
(x,y,z)\rightarrow\left(2\mu-x,y,z+\frac{c}{2}\right),
\end{equation}
where $x=\mu$ is the mirror plane of the slide symmetry operator and $a,b,c$ are the length of three lattice vectors.

First, we identify the constraints imposed by the positions of Dirac cones. Consider a pair of Dirac cones at $(d,0,0)$ and $(-d,0,0)$, then the allowed mirror planes are $k_x=0,\pi$ in momentum space and $x=0,0.5a$ in real space.

Next, we identify the constraints imposed by the inversion symmetry. Suppose one of the inversion centers of layer 1 is given by (0,0,0). Performing $\{M_x|c/2\}$ to (0,0,0), we obtain the corresponding inversion center of layer 2 at the top: $(2\mu,0,c/2)$. Performing $\{M_x|c/2\}^{-1}$ to (0,0,0), we obtain the corresponding inversion center of layer 2 at the bottom: $(2\mu,0,-c/2)$. In the most general case, $\mathcal{P}$ can only exchange these two inversion centers, i.e.,
\begin{equation}
\left(2\mu,0,\frac{c}{2}\right)\stackrel{\mathcal{P}}{\longrightarrow}\left(-2\mu,0,-\frac{c}{2}\right)\approx\left(2\mu,0,-\frac{c}{2}\right)
\end{equation}
where $\approx$ denotes that the two sides are equal modulo a lattice vector. Therefore, the values of $\mu$ are restricted to $\{0,0.5\}$. We get the same result by choosing other inversion centers of layer 1, i.e., $(a/2,0,0)$, $(0,b/2,0)$, or $(a/2,b/2,0)$. By similar steps in Sec.\ref{s2z1}, we can also prove that the atoms in the 3D structure also have inversion symmetry.

However, we need to take the shape of Dirac cones into account. In the case of $S_{2z}$, there is no need to care about the shape of the Dirac cones for the following reason. At $k_z=0$ plane, the inversion symmetry is equivalent to $C_{2z}$ and the inversion symmetry can exchange the $\mathcal{T}$-reversed pair of Dirac cones without changing the band structure. As a result, $C_{2z}$ can also exchange the $\mathcal{T}$-reversed pair of Dirac cones without changing the band structure at $k_z=0$ plane. As $S_{2z}$ is just $C_{2z}$ on $k_z=0$ plane, $S_{2z}$ exchanges the $\mathcal{T}$-reversed pair of Dirac cones without changing the band structure at $k_z=0$ plane. When the interlayer coupling is not considered, there is no energy dispersion in $k_z$-direction and the band structures of the two layers overlap completely. Since strong interlayer coupling will destroy the NLs, we only consider weak interlayer coupling in the whole momentum space, which will not drastically change the band structure. However, we can not find such a relation between inversion symmetry and mirror symmetry. Thus, slide symmetry can not ensure the complete overlapping of the $\mathcal{T}$-reversed pair of Dirac cones and $\mathbb{Z}_2$NLs are less likely to appear.

Finally, let us consider 2D DSMs with mirror symmetry. In this case, $M_x$ followed by $\{M_x|c/2\}$ acts in real space as
\begin{equation}
(x,y,z)\xrightarrow{\{M_x|c/2\}*M_x}\left(x,y,z+\frac{c}{2}\right),
\end{equation}
i.e., the two layers only differ by a pure translation $\bm c/2$, and thus the primitive cell contains only one layer of 2D DSM.

\subsubsection{2D layers exchanged under $\mathcal{P}$}
Consider the case that $\mathcal{P}$ exchanges two 2D layers as shown in Fig. \ref{s2z}(b). In this section, we choose a gauge that $\{M_x|c/2\}$ acts in real space as
\begin{equation}
(x,y,z)\rightarrow\left(-x,y,z+\frac{c}{2}\right),
\end{equation}
i.e., the mirror plane becomes $x=0$.

Here, we prove that the 3D structure can not have inversion symmetry in the most general case. Assume there is an atom of layer 1 at $(x+i_x,y+i_y,z+c/4)$ where $(i_x,i_y,c/4)$ is the coordinate of an inversion center of layer 1. Performing $\{M_x|c/2\}$ on $(x+i_x,y+i_y,z+c/4)$, we have
\begin{equation}
\left(x+i_x,y+i_y,z+\frac{c}{4}\right)\xrightarrow{\{M_x|c/2\}}\left(-x-i_x,y+i_y,z+\frac{3c}{4}\right).
\end{equation}
In general, we can not find a $\mathcal{P}$-reversed pair for every two atoms.

\subsection{Other screw symmetries}

\subsubsection{$S_{3z}$}
Since, generally, $S_{3z}$ does not commutate or anticommutate with $\mathcal{P}$ at TRIMs, we exclude this symmetry.

\subsubsection{$S_{4z}$ and $S_{6z}$}
\begin{figure}[h]
\includegraphics[width=0.4\textwidth]{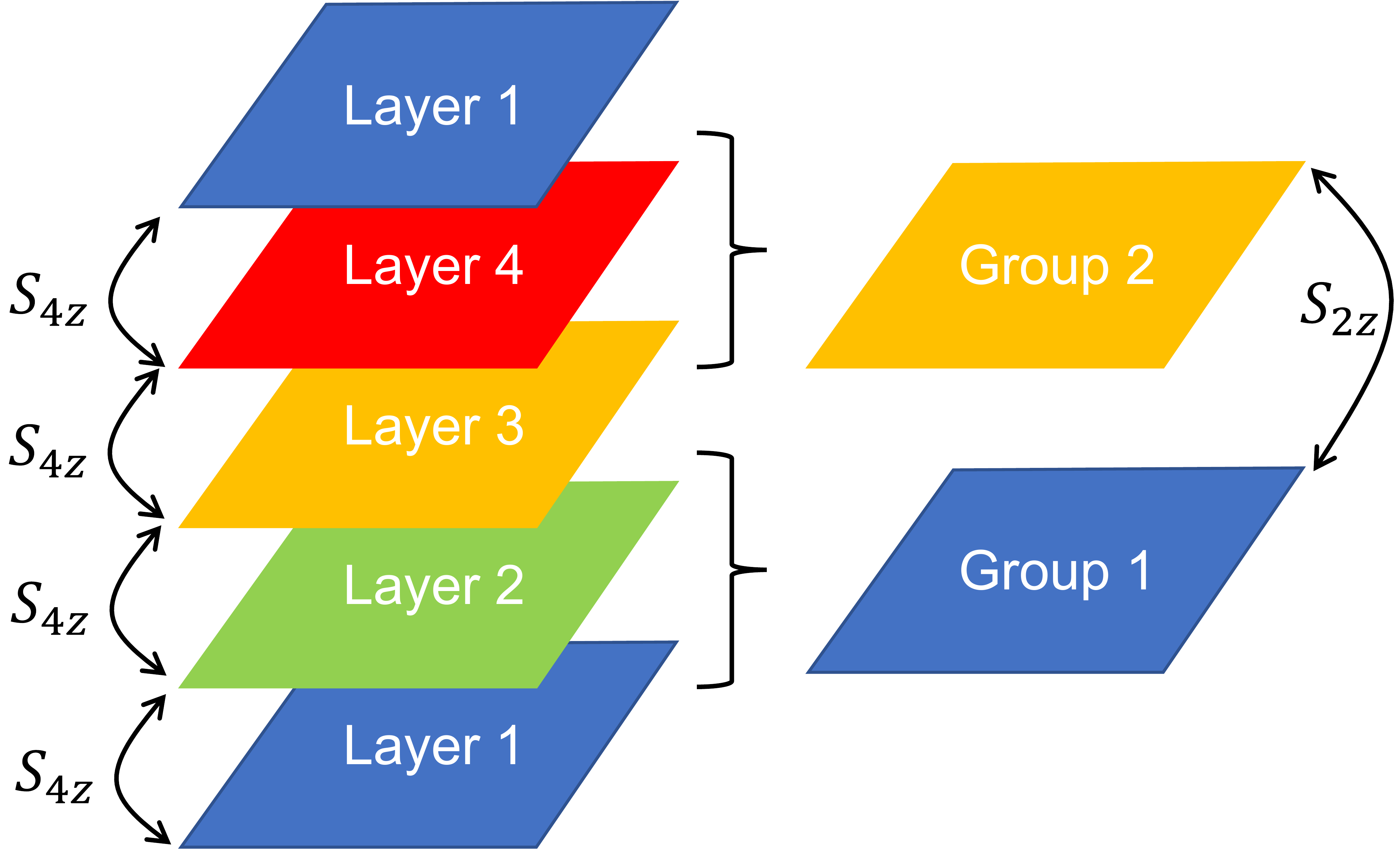}
\caption{\label{s4z} $S_{4z}$ can be understood as $S_{2z}$. We can group the four layers in the case of $S_{4z}$ into two groups, i.e., group 1 contains layer 1 and layer 2 and group 2 contains layer 3 and layer 4. Then the following analyses are the same as in the case of $S_{2z}$.}
\end{figure}

We can group the layers into 2 groups, i.e., two layers in a group for $S_{4z}$ as shown in Fig. \ref{s4z} and three layers in a group for $S_{6z}$.  What is special in these two cases is that we need to consider the commensurability between layers in a group and the corresponding intervalley scatterings. After that, a group can be regarded as the 2D layer in $S_{2z}$, and this 2D layer should have inversion symmetry and time reversal symmetry but not $C_{2z}$. Then the analyses of $S_{2z}$ can be performed similarly.

\section{Derivation of commutation or anticommutation relation between $S_{2z}$ and $\mathcal{P}$ in the Wannier representation}\label{S2zP}

\begin{figure}[h]
\includegraphics[width=1\textwidth]{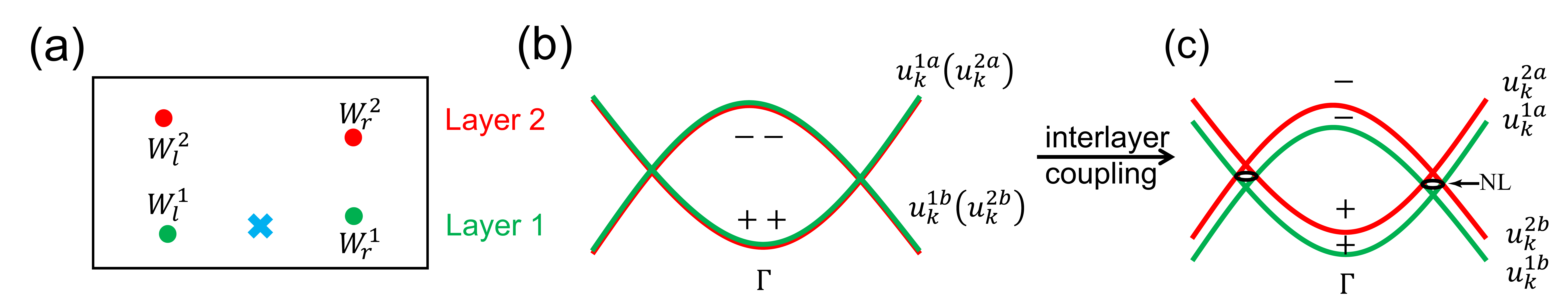}
\caption{\label{figs2} (a) Schematic of the 3D structure. The black rectangles denote the boundary of the primitive cell. The blue cross is the inversion center of the 3D structure and layer 1. The red and green circles are the position of the orbitals. $\mathcal{P}$ exchanges orbitals in the same layer. $W_{\mu}^\nu$ are wannier functions. (b) The double band inversion and the bands doubly degenerate when the interlayer coupling is absent. The $\Gamma$ is a TRIM. (c) The $\mathcal{T}$-reversed pair of $\mathbb{Z}_2$NL when the interlayer coupling is present.}
\end{figure}

In the derivation of the relation between $S_{2z}$ and $\mathcal{P}$, we only consider bands that contribute to the $\mathbb{Z}_2$NL, i.e., four bands as shown in Figs. \ref{figs2}(b),(c). And we also limit ourselves to only four Wannier functions in a unit cell as shown in Fig. \ref{figs2}(a). These limitations are not essential and can be easily removed.

Firstly, we derive the transformation of the wannier functions $W_\nu^\mu$ under $S_{2z}$ and $\mathcal{P}$ using the notations of Refs.\cite{evarestov2012site,PhysRevB.97.035139}. For $S_{2z}$,

\begin{equation}
\begin{aligned}
S_{2z}*W_l^1(\bm r- \bm r_\mu)=&\left\{C_{2z}\left|\left(-\frac{\mu a}{2},-\frac{\nu b}{2},\frac{c}{2}\right)\right.\right\}\{E|\bm r_\mu\}W_l^1(\bm r)\\
=&\left\{E\left|C_{2z}\bm r_\mu+\left(-\frac{\mu a}{2},-\frac{\nu b}{2},\frac{c}{2}\right)\right.\right\}S_{2z}*W_l^1(\bm r)\\
=&\left\{E\left|-\bm r_\mu+\left(-\frac{\mu a}{2},-\frac{\nu b}{2},\frac{c}{2}\right)\right.\right\}W_r^2(\bm r)\\
=&W_r^2\left(\bm r+\bm r_\mu-\left(-\frac{\mu a}{2},-\frac{\nu b}{2},\frac{c}{2}\right)\right).
\end{aligned}
\end{equation}
Similarly,
\begin{equation}
S_{2z}*W_r^1(\bm r- \bm r_\mu)=W_l^2\left(\bm r+\bm r_\mu-\left(-\frac{\mu a}{2},-\frac{\nu b}{2},\frac{c}{2}\right)\right).
\end{equation}

For $\mathcal{P}$,

\begin{equation}
\begin{aligned}
\mathcal{P}*W_l^1(\bm r-\bm r_\mu)=&\{\mathcal{P}|0\}\{E|\bm r_\mu\}W_l^1(\bm r)\\
=&\{E|-\bm r_\mu\}\{\mathcal{P}|0\}W_l^1(\bm r)\\
=&\{E|-\bm r_\mu\}W_r^1(\bm r)\\
=&W_r^1(\bm r+\bm r_\mu).
\end{aligned}
\end{equation}
Similarly,
\begin{equation}
\begin{aligned}
\mathcal{P}*W_l^2(\bm r-\bm r_\mu)=&\{\mathcal{P}|0\}\{E|\bm r_\mu\}W_l^2(\bm r)\\
=&\{E|-\bm r_\mu\}\{\mathcal{P}|0\}W_l^2(\bm r)\\
=&\{E|-\bm r_\mu\}\{E|(-\mu a,-\nu b,c)\}W_r^2(\bm r)\\
=&W_r^2(\bm r+\bm r_\mu-\bm r_B),
\end{aligned}
\end{equation}
where $\bm r_B=(-\mu a,-\nu b,c)$. Because we obtain a $W_r^2$ in a different unit cell by performing $\mathcal{P}$ to a $W_l^2$, $\bm r_B$ appears in the equation to move $W_r^2$ back to the same unit cell with $W_l^2$.

Next, we derive the eigenvalues of the four wavefunctions. Consider the 3D structure when the interlayer coupling is absent and the band structure is shown in Fig. \ref{figs2}(b). Suppose $u_{\bm k}^{1a}$ and $u_{\bm k}^{1b}$ are induced only by obitals in layer 1, and $u_{\bm k}^{2a}$ and $u_{\bm k}^{2b}$ are induced only by obitals in layer 2. Then we have

\begin{equation}
\begin{aligned}
&u_{\bm k}^{1a}(\bm r)=\sum_{\bm r_\mu}e^{i\bm k\cdot \bm r_\mu}(W_{r}^1(\bm r-\bm r_\mu)+W_{l}^1(\bm r-\bm r_\mu)),\\
&u_{\bm k}^{1b}(\bm r)=\sum_{\bm r_\mu}e^{i\bm k\cdot \bm r_\mu}(W_{r}^1(\bm r-\bm r_\mu)-W_{l}^1(\bm r-\bm r_\mu)),\\
&u_{\bm k}^{2a}(\bm r)=\sum_{\bm r_\mu}e^{i\bm k\cdot \bm r_\mu}(W_{r}^2(\bm r-\bm r_\mu)+W_{l}^2(\bm r-\bm r_\mu)),\\
&u_{\bm k}^{2b}(\bm r)=\sum_{\bm r_\mu}e^{i\bm k\cdot \bm r_\mu}(W_{r}^2(\bm r-\bm r_\mu)-W_{l}^2(\bm r-\bm r_\mu)),\label{eqbp}\\
\end{aligned}
\end{equation}
where now $\bm r_\mu$ is a Bravais lattice vector. We find the parity eigenvalues of these wavefunctions by performing $\mathcal{P}$ on them. Since we only care about the parity eigenvalues at TRIMs, in the rest of this section and the next section, the momentum $\bm k$ is restricted to TRIMs without explicitly specifying.

\begin{equation}
\begin{aligned}
\mathcal{P}*u_{\bm k}^{1a}(\bm r)=&\mathcal{P}*\sum_{\bm r_\mu}e^{i\bm k\cdot \bm r_\mu}(W_{r}^1(\bm r-\bm r_\mu)+W_{l}^1(\bm r-\bm r_\mu))\\
=&\sum_{\bm r_\mu}e^{i\bm k\cdot \bm r_\mu}(W_{l}^1(\bm r+\bm r_\mu)+W_{r}^1(\bm r+\bm r_\mu))\\
=&\sum_{\bm r_\mu}e^{2i\bm k\cdot \bm r_\mu}e^{-i\bm k\cdot \bm r_\mu}\left[W_l^1\left(\bm r+\bm r_\mu\right)+W_r^1\left(\bm r+\bm r_\mu\right)\right]\\
=&\sum_{\bm r_\mu}e^{-i\bm k\cdot \bm r_\mu}\left[W_l^1\left(\bm r+\bm r_\mu\right)+W_r^1\left(\bm r+\bm r_\mu\right)\right]\\
=&u_{\bm k}^{1a}(\bm r).\label{eqpe}
\end{aligned}
\end{equation}
When moving into the fourth line, we have used $e^{2i\bm k\cdot \bm r_\mu}=1$. Similarly,
\begin{equation}
\mathcal{P}*u_{\bm k}^{1b}(\bm r)=-u_{\bm k}^{1b}(\bm r).\label{eqpe2}
\end{equation}
Therefore, at all TRIMs, the parity eigenvalue of $u_{\bm k}^{1a}(\bm r)$ is 1 and the parity eigenvalue of $u_{\bm k}^{1b}(\bm r)$ is -1.

\begin{equation}
\begin{aligned}
\mathcal{P}*u_{\bm k}^{2a}(\bm r)=&\mathcal{P}*\sum_{\bm r_\mu}e^{i\bm k\cdot \bm r_\mu}(W_{r}^2(\bm r-\bm r_\mu)+W_{l}^2(\bm r-\bm r_\mu))\\
=&\sum_{\bm r_\mu}e^{i\bm k\cdot \bm r_\mu}(W_{l}^2(\bm r+\bm r_\mu-\bm r_B)+W_{r}^2(\bm r+\bm r_\mu-\bm r_B))\\
=&e^{-i\bm k\cdot \bm r_B}\sum_{\bm r_\mu}e^{2i\bm k\cdot \bm r_\mu}e^{-i\bm k\cdot \bm r_\mu+i\bm k\cdot \bm r_B}\left[W_l^2\left(\bm r+\bm r_\mu-\bm r_B\right)+W_r^2\left(\bm r+\bm r_\mu-\bm r_B\right)\right]\\
=&e^{-i\bm k\cdot \bm r_B}u_{\bm k}^{2a}(\bm r).\\
\end{aligned}
\end{equation}
Similarly,
\begin{equation}
\mathcal{P}*u_{\bm k}^{2b}(\bm r)=-e^{-i\bm k\cdot \bm r_B}u_{\bm k}^{2b}(\bm r).
\end{equation}

Then, we derive the relationship between the wavefunctions. Perform $S_{2z}$ on $u_{\bm k}^{1a}(\bm r)$,
\begin{equation}
\begin{aligned}
S_{2z}*u_{\bm k}^{1a}(\bm r)=&\sum_{\bm r_\mu}e^{i\bm k\cdot \bm r_\mu}(S_{2z}*W_{r}^1(\bm r-\bm r_\mu)+S_{2z}*W_{l}^1(\bm r-\bm r_\mu))\\
=&\sum_{\bm r_\mu}e^{i\bm k\cdot \bm r_\mu}\left[W_l^2\left(\bm r+\bm r_\mu-\bm r_{S_{2z}}\right)+W_r^2\left(\bm r+\bm r_\mu-\bm r_{S_{2z}}\right)\right]\\
=&\sum_{\bm r_\mu}e^{2i\bm k\cdot \bm r_\mu}e^{-i\bm k\cdot \bm r_\mu}\left[W_l^2\left(\bm r+\bm r_\mu-\bm r_{S_{2z}}\right)+W_r^2\left(\bm r+\bm r_\mu-\bm r_{S_{2z}}\right)\right]\\
=&e^{-i\bm k\cdot\bm r_{S_{2z}}}\sum_{\bm r_\mu}e^{2i\bm k\cdot \bm r_\mu}e^{-i\bm k\cdot \bm r_\mu+i\bm k\cdot\bm r_{S_{2z}}}\left[W_l^2\left(\bm r+\bm r_\mu-\bm r_{S_{2z}}\right)+W_r^2\left(\bm r+\bm r_\mu-\bm r_{S_{2z}}\right)\right]\\
=&e^{-i\bm k\cdot\bm r_{S_{2z}}}\sum_{\bm r_\mu}e^{-i\bm k\cdot \bm r_\mu+i\bm k\cdot\bm r_{S_{2z}}}\left[W_l^2\left(\bm r+\bm r_\mu-\bm r_{S_{2c}}\right)+W_r^2\left(\bm r+\bm r_\mu-\bm r_{S_{2c}}\right)\right],\\
=&e^{-i\bm k\cdot\bm r_{S_{2z}}}u_{\bm k}^{2a}(\bm r).\label{eqsp}
\end{aligned}
\end{equation}
where $\bm r_{S_{2z}}=(\mu a/2,\nu b/2,c/2)$. When moving into the fifth line, we have used $e^{2i\bm k\cdot \bm r_\mu}=1$. Similarly,
\begin{equation}
S_{2z}*u_{\bm k}^{1b}(\bm r)=e^{-i\bm k\cdot\bm r_{S_{2z}}}u_{\bm k}^{2b}(\bm r).\label{eqsp2}
\end{equation}
For this reason, we claimed in the main text that $\left|u_k\right>$ and $S_{2z}\left|u_k\right>$ are two different states.

Finally, we derive the commutation or anticommutation relation between $S_{2z}$ and $\mathcal{P}$. For $u_{\bm k}^{1a}(\bm r)$ and $u_{\bm k}^{2a}(\bm r)$,
\begin{equation}
\begin{aligned}
S_{2z}*\mathcal{P}*u_{\bm k}^{1a}(\bm r)=&e^{-i\bm k\cdot\bm r_{S_{2z}}}u_{\bm k}^{2a}(\bm r),\\
\mathcal{P}*S_{2z}*u_{\bm k}^{1a}(\bm r)=&e^{-i\bm k\cdot \bm r_B}e^{-i\bm k\cdot\bm r_{S_{2z}}}u_{\bm k}^{2a}(\bm r).\\
\end{aligned}
\end{equation}
For $u_{\bm k}^{1b}(\bm r)$ and $u_{\bm k}^{2b}(\bm r)$,
\begin{equation}
\begin{aligned}
S_{2z}*\mathcal{P}*u_{\bm k}^{1b}(\bm r)=&-e^{-i\bm k\cdot\bm r_{S_{2z}}}u_{\bm k}^{2b}(\bm r),\\
\mathcal{P}*S_{2z}*u_{\bm k}^{1b}(\bm r)=&-e^{-i\bm k\cdot \bm r_B}e^{-i\bm k\cdot\bm r_{S_{2z}}}u_{\bm k}^{2b}(\bm r).\\
\end{aligned}
\end{equation}
In both cases, we have
\begin{equation}
S_{2z}*\mathcal{P}=e^{i\bm k\cdot \bm r_B}\mathcal{P}*S_{2z},
\end{equation}
or,
\begin{equation}
S_{2z}*\mathcal{P}=e^{ik_z-i\mu k_x-i\nu k_y}\mathcal{P}*S_{2z},
\end{equation}
which is just Eq. 1 in the main text.

This relation can also be understood in a more general way as in the main text. The $\mathcal{P}$ acts in real space as
\begin{equation}
(x,y,z)\rightarrow\left(-x,-y,-z\right),
\end{equation}
and $S_{2z}$ acts in real space as
\begin{equation}
(x,y,z)\rightarrow\left(-x+\frac{\mu a}{2},-y+\frac{\nu b}{2},z+\frac{c}{2}\right).
\end{equation}
Consider the commutation or anticommutation relation between $\mathcal{P}$ and $S_{2z}$
\begin{equation}
\mathcal{P}*S_{2z}:(x,y,z)\rightarrow\left(x-\frac{\mu a}{2},y-\frac{\nu b}{2},-z-\frac{c}{2}\right),
\end{equation}
\begin{equation}
S_{2z}*\mathcal{P}:(x,y,z)\rightarrow\left(x+\frac{\mu a}{2},y+\frac{\nu b}{2},-z+\frac{c}{2}\right),
\end{equation}
from which we obtain
\begin{equation}
\begin{aligned}
S_{2z}*\mathcal{P}=&T_{(\mu a,\nu b,-c)}\mathcal{P}*S_{2z}\\
=&e^{ik_z-i\mu k_x-i\nu k_y}\mathcal{P}*S_{2z}.\\
\end{aligned}\label{S2zPeq}
\end{equation}

At TRIMs where $S_{2z}$ commutates with $\mathcal{P}$, i.e., $e^{ik_z-i\mu k_x-i\nu k_y}=1$. By applying $S_{2z}$ to $u_k^{1a}$ and $u_k^{1b}$, we can obtain $u_k^{2a}$ and $u_k^{2b}$, respectively. That is, if there is an occupied state $\left|u_k^{1\sigma}\right>$ with a certain parity, there exists another occupied state $\left|u_k^{2\sigma}\right>=S_{2z}\left|u_k^{1\sigma}\right>$ with the same parity. At TRIMs where $S_{2z}$ anticommutates with $\mathcal{P}$, i.e., $e^{ik_z-i\mu k_x-i\nu k_y}=-1$. If there is an occupied state $\left|u_k^{1\sigma}\right>$ with a certain parity, there exists another occupied state $\left|u_k^{2\sigma}\right>=S_{2z}\left|u_k^{1\sigma}\right>$ with a different parity.

By adding proper interlayer coupling terms, the fourfold degeneracy nodal lines in Fig. \ref{figs2}(b) (They are nodal lines instead of Dirac point since there is no dispersion in $k_z$-direction) can evolve into a pair of $\mathbb{Z}_2$NLs in Fig. \ref{figs2}(c).

\section{All possible parity patterns of the 3D $\mathbb{Z}_2$NLSMs from the 2D Dirac SMs}
In this section, we explicitly show all possible parity patterns for which the appearance of $\mathbb{Z}_2$NLs is allowed. 

\begin{figure}[t]
\includegraphics[width=0.8\textwidth]{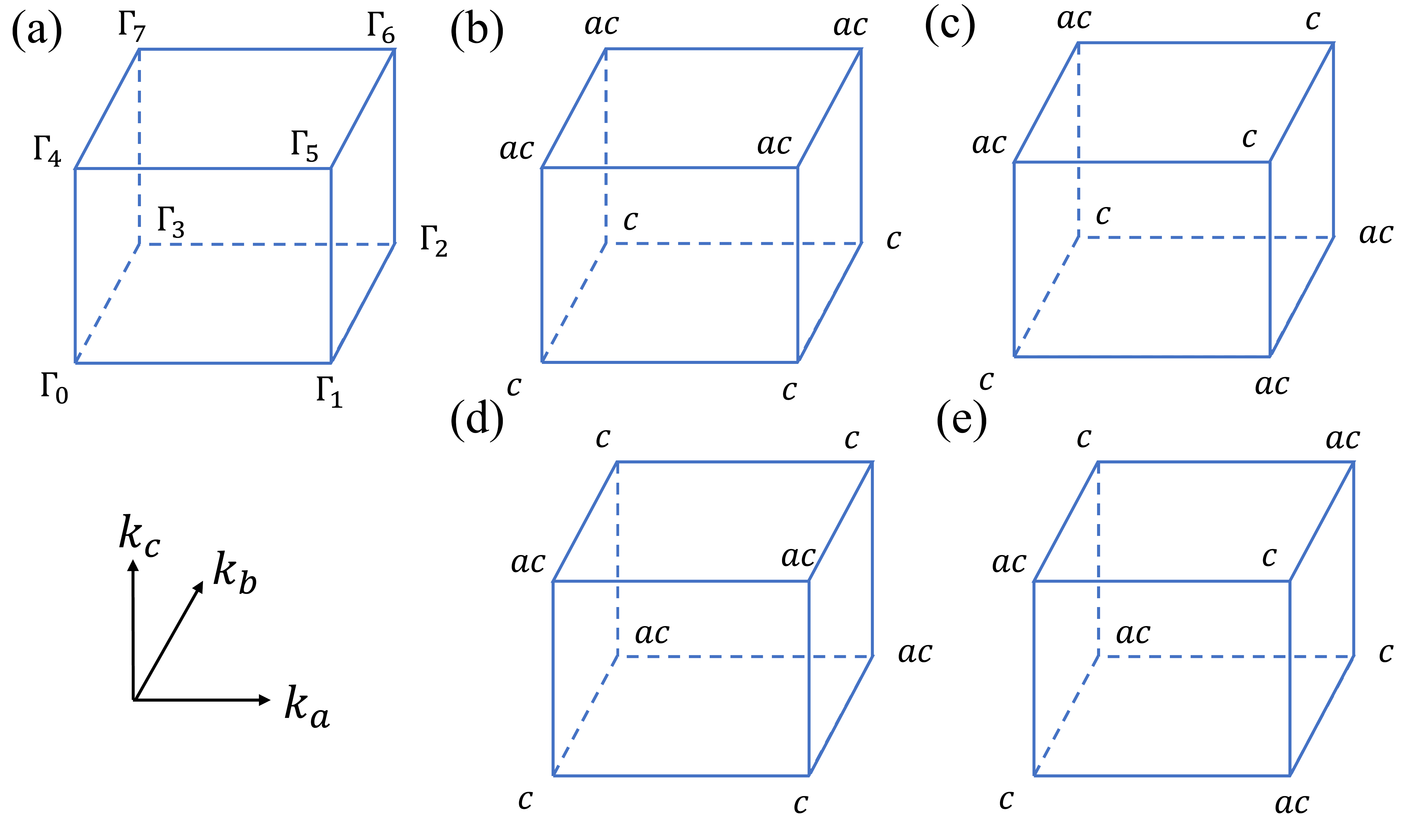}
\caption{\label{figs3}(a) The coordinates of the eight TRIMs $\Gamma_i$ and $\Gamma_0=(0,0,0)$. (b-e) The commutation relations between $S_{2c}$ and $\mathcal{P}$ at TRIMs for four different cases of $(\mu,\nu)$. $c$ means $S_{2c}$ commutates with $\mathcal{P}$ and $ac$ means $S_{2c}$ anticommutates with $\mathcal{P}$. (b) $(\mu,\nu)=(0,0)$. (c) $(\mu,\nu)=(1,0)$. (d) $(\mu,\nu)=(0,1)$. (e) $(\mu,\nu)=(1,1)$.}
\end{figure}

The eight TRIMs are named by $\Gamma_i,i=0,1,...,7$, and the coordinates of the $\Gamma_i$ are shown in Fig. \ref{figs3}(a) with $\Gamma_0=(0,0,0)$. The value of $(\mu,\nu)$ when the 2D DSMs does not have $C_{2z}$ are restricted to four cases, i.e. $(0,0),(0,1),(1,0),(1,1)$. Correspondingly, according to
\begin{equation}
S_{2c}*\mathcal{P}=e^{ik_c-i\mu k_a-i\nu k_b}\mathcal{P}*S_{2c},
\end{equation}
the commutation or anticommutation relations between $S_{2c}$ and $\mathcal{P}$ at TRIMs are shown in Figs. \ref{figs3}(b-e) for the four cases.

\begin{figure}[h]
\includegraphics[width=1.0\textwidth]{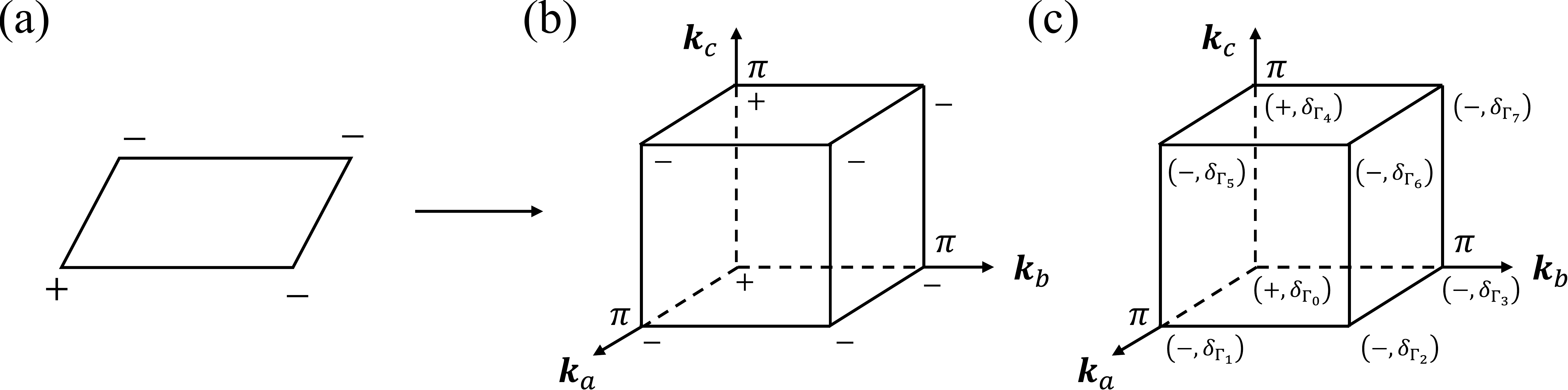}
\caption{\label{figs4} (a) The parity eigenvalues of occupied states for the 2D Dirac semimetal. (b) The parity eigenvalues of occupied states of a 3D NLSM with only one layer per unit cell (c) The parity eigenvalues of occupied states for a 3D structure with two 2D layers per unit cell.}
\end{figure}

Then we consider a 2D DSM with a parity pattern as shown in Fig. \ref{figs4}(a). We stack the 2D DSM into a 3D NLSM with only one layer per unit cell whose parity pattern (The occupied one of $u_k^{1a}$ or $u_k^{1b}$.) is shown in Fig. \ref{figs4}(b). Here we assume the absence of interlayer coupling. The parity eigenvalues can be obtained from Eq. \ref{eqpe} and Eq. \ref{eqpe2}. Then we need to find the parity eigenvalues of bands from another layer, i.e., the parity eigenvalues of the occupied one among $u_k^{2a}$ and $u_k^{2b}$.

To this end, we first find out the commutation or anticommutation relation between $S_{2c}$ and $\mathcal{P}$ at all eight TRIMs as shown in Fig. \ref{figs3}. The two parity eigenvalues of the occupied bands are the same if $S_{2c}$ commutates with $\mathcal{P}$ and are opposite if $S_{2c}$ anticommutates with $\mathcal{P}$, as explained under Eq. \ref{S2zPeq}. As shown in Fig. \ref{figs4}(c), the first parity eigenvalues in the parentheses are the parity eigenvalues of the occupied one of $u_k^{1a}$ or $u_k^{1b}$ and the second parity eigenvalues in the parentheses are the parity eigenvalues of the occupied one of $u_k^{2a}$ or $u_k^{2b}$. We then diagnose $\mathbb{Z}_2$NLs  from the parity pattern by
\begin{equation}
e^{i\pi w}=\prod_{i=1}^8(-1)^{\lfloor N_{occ}^-(\Gamma_i)/2\rfloor}.
\end{equation}
If $w=1$, then the appearing of $\mathbb{Z}_2$NLs is allowed. Thus, there should be an odd number of double band inversions
(two occupied states with both -1 parity eigenvalues). A double band inversion appears when the first parity eigenvalue in the parentheses is -1 and $S_{2c}$ commutates with $\mathcal{P}$ at that TRIM.

\section{More details on the tight-binding model}
\begin{figure}[h]
\includegraphics[width=1\textwidth]{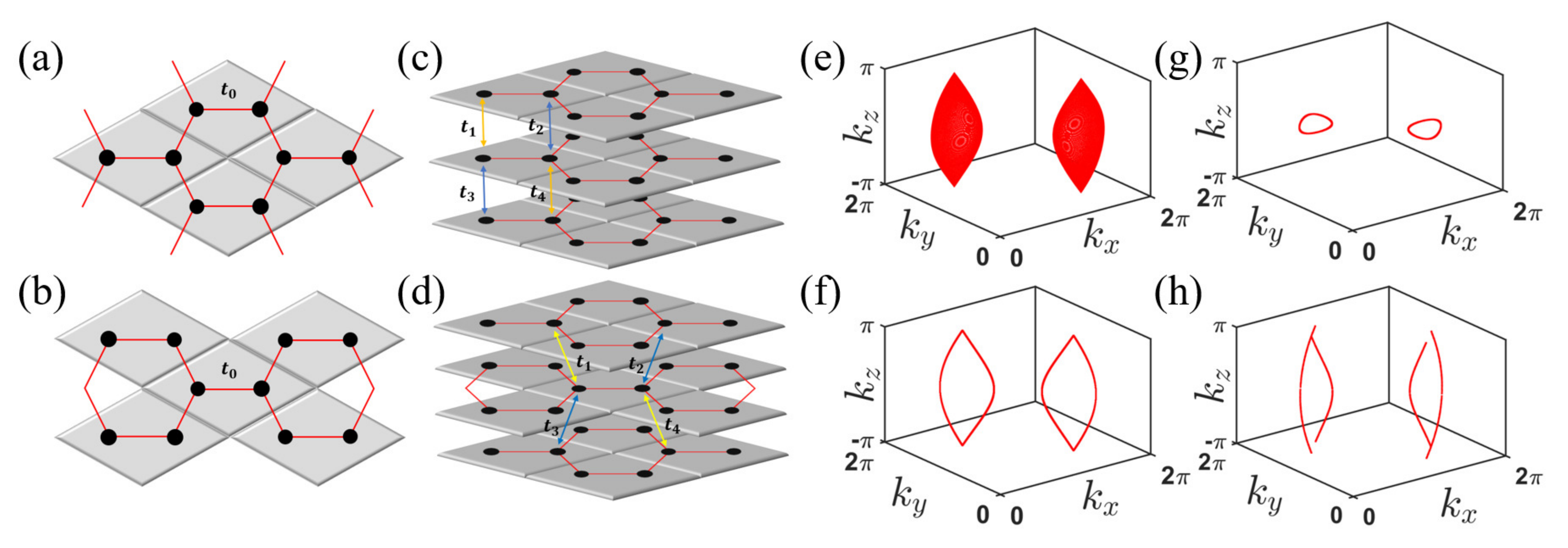}
\caption{\label{fig_model}(a)(b) Schematics of the crystal structure of 2D Honeycomb lattice model. And (b) can be obtained by a shift $1/2(\bm{r}_1+\bm{r}_2)$ or $1/2(\bm{r}_1-\bm{r}_2)$ from (a). (c)(d) Schematics of the two kinds of 3D crystal structures. The hoppings $t_1,t_2,t_3,t_4$ are interlayer hoppings and $t_1=t_4,t_2=t_3$ so that the 3D crystal has $\mathcal{P}$. (c) The 3D structure formed by stacking of 2D structures in (a). (d) The 3D structure formed by stacking of 2D structures in (a) and (b). (e) The spindle-shaped nodal surfaces for $t_0=1.0,t_1=t_2=t_3=t_4=0.4$ in (c). (f) A pair of NLs winding around the whole Brillouin zone (BZ) for $t_0=1.0,t_1=t_2=t_3=t_4=0.4$ in (d). (g) The NL carrying a unit $\mathbb{Z}_2$ monopole charge for $t_0=1.0,t_1=t_4=0.1,t_2=t_3=0.4$ in (c). (h) Four NLs winding around the whole BZ for $t_0=1.0,t_1=t_4=0.1,t_2=t_3=0.4$.}
\end{figure}

The tight-binding model of the 2D Dirac semimetal we use in the text originates from the tight-binding model of the silicene, i.e., buckled honeycomb lattice, with only $p_z$ orbitals\cite{Liu2011}. In this section, we give four models with their nodal surfaces or NLs shown in Figs. \ref{fig_model}(e-h). The first model which has spindle-shaped nodal surfaces (Fig. \ref{fig_model}(e)) is just due to the supercell effect. The rest models with the same parity patterns give three types of NLs (Fig.\ref{fig_model}(f-h)), while only one of them has nontrivial $\mathbb{Z}_2$ monopole charge (Fig.\ref{fig_model}(g)).

As shown in Fig .\ref{fig_model}(c), we choose a screw axis so that it is coincident with the inversion center at $(0,0,0)$ or $(1/2,1/2,0)$, i.e., $\mu,\nu=0$ or $\mu,\nu=1$. The three basis lattice vectors are given by
\begin{equation}
\bm a=(1,0,0),
\bm b=\left(\frac{1}{2},\frac{\sqrt{3}}{2},0\right),
\bm c=\left(0,0,1\right).\label{lat}
\end{equation}
There are four orbitals in the unit cell and their reduced coordinates are given by
\begin{equation}
\begin{aligned}
&\left(\frac{1}{3},\frac{1}{3},0\right),\left(\frac{2}{3},\frac{2}{3},0\right),\\
&\left(\frac{1}{3},\frac{1}{3},\frac{1}{2}\right),\left(\frac{2}{3},\frac{2}{3},\frac{1}{2}\right).
\end{aligned}
\end{equation}

The Hamiltonian without interlayer coupling reads
\begin{equation}
H_0(\bm{k})=t_0\sum_j^3\left(\cos(\bm{k}\cdot\bm{\delta_j})\sigma_1\otimes\tau_0-\sin(\bm{k}\cdot\bm{\delta_j})\sigma_2\otimes\tau_0\right),\label{h0}
\end{equation}
where $\bm{\delta}_1=(1/3,1/3,0),\bm{\delta}_2=(1/3,-2/3,0),\bm{\delta}_3=(-2/3,1/3,0)$ and $\sigma_i$, $\tau_i$ are Pauli matrices and denote sublattice and interlayer degrees of freedom, respectively.

We first assume $t_1=t_2=t_3=t_4$. As shown in Fig. \ref{fig_model}(e), we obtain a pair of spindle-shaped nodal surfaces related by $\mathcal{P}$ when the interlayer coupling is weak. This can be understood as follows. First, there is a pair of Dirac points in 2D DSM which are related by $\mathcal{P}$. The pure translation operator does not change the position of the two Dirac points in momentum space while the twofold rotation operator exchanges the position of the two Dirac points in momentum space. So, now we have a pair of 3D four-fold degenerate Dirac points. When there is no interlayer coupling, the bands have no dispersion in the $k_z$ direction, i.e., the pair of Dirac points become a pair of Dirac NLs in the $k_z$ direction. Then we consider the interlayer coupling term
\begin{equation}
\Delta H_1(\bm{k})=\cos\left(\frac{k_z}{2}\right)\sigma_0\otimes\tau_1,
\end{equation}
which commutates with $H_0$. The energy dispersion of the Hamiltonian $H_1(\bm{k})=H_0(\bm{k})+\Delta H_1(\bm{k})$  is given by
\begin{equation}
E_1=\pm\left(\sqrt{\left(t_0\sum_j^3\cos(\bm{k}\cdot\bm{\delta_j})\right)^2+\left(t_0\sum_j^3\sin(\bm{k}\cdot\bm{\delta_j})\right)^2}\pm \left|(t_1+t_3)\cos\left(\frac{k_z}{2}\right)\right|\right).
\end{equation}
The Dirac nodal line above can be regarded as a series of Dirac points and the series of Dirac points become a series of Weyl rings except at $k_z=\pi$ where $\cos(k_z/2)=0$ with $2t_1*\Delta H_1$ added. As a result, we get a pair of spindle-shaped nodal surfaces related by $\mathcal{P}$, as plotted in Fig. \ref{fig_model}(e).

Then we set $t_1=t_4\neq t_2=t_3$ and consider the interlayer coupling term
\begin{equation}
\Delta H_2(\bm{k})=-\sin\left(\frac{k_z}{2}\right)\sigma_3\otimes\tau_2,
\end{equation}
which anticommutates with $H_0$. The Hamiltonian reads
\begin{equation}
H_2(\bm{k})=H_0+(t_1+t_3)*\Delta H_1+(t_1-t_3)*\Delta H_2.
\end{equation}
The energy dispersion is
\begin{equation}
E_2=\pm\sqrt{\left(\sqrt{\left(t_0\sum_j^3\cos(\bm{k}\cdot\bm{\delta_j})\right)^2+\left(t_0\sum_j^3\sin(\bm{k}\cdot\bm{\delta_j})\right)^2}\pm \left|(t_1+t_3)\cos\left(\frac{k_z}{2}\right)\right|\right)^2+\left((t_1-t_3)\sin\left(\frac{k_z}{2}\right)\right)^2}.
\end{equation}
When the values of $t_1,t_2,t_3,t_4$ are small enough, adding $(t_1-t_3)*\Delta H_2$ to $H_0+(t_1+t_3)*\Delta H_1$ would gap all the degenerate points of the nodal surfaces in Fig. \ref{fig_model}(e) except the degenerate points on $k_z=0$, i.e., only a pair of NLs on $k_z=0$ is left as shown in Fig. \ref{fig_model}(g).

Next, we consider the case the screw axis coincides with the inversion center at $(0,1/2,0)$ or $(1/2,0,0)$, i.e., $(\mu,\nu)=(0,1)$ or $(1,0)$. While the lattice vectors are the same as Eq. \ref{lat}, the reduced coordinates of the four orbitals are given by
\begin{equation}
\begin{aligned}
&\left(\frac{1}{3},\frac{1}{3},0\right),\left(\frac{2}{3},\frac{2}{3},0\right),\\
&\left(\frac{1}{6},\frac{1}{6},\frac{1}{2}\right),\left(\frac{5}{6},\frac{5}{6},\frac{1}{2}\right),
\end{aligned}
\end{equation}
as shown in Fig .\ref{fig_model}(d).

The Hamiltonian without interlayer coupling is still $H_0$ in Eq. \ref{h0}. We assume $t_1=t_2=t_3=t_4$ and consider the interlayer coupling term
\begin{equation}
\Delta H_3(\bm{k})=\cos\left(\frac{k_x}{6}+\frac{k_y}{6}\right)\cos\left(\frac{k_z}{2}\right)\sigma_0\otimes\tau_1+\sin\left(\frac{k_x}{6}+\frac{k_y}{6}\right)\cos\left(\frac{k_z}{2}\right)\sigma_3\otimes\tau_2,
\end{equation}
where $\sigma_0\otimes\tau_1$ anticommutates with $H_0$ and $\sigma_3\otimes\tau_2$ commutates with $H_0$. The Hamiltonian is
\begin{equation}
H_3(\bm{k})=H_0+2t_1*\Delta H_3.
\end{equation}
Adding $2t_1*\Delta H_3$ to $H_0$, we obtain a pair of NLs as plotted in Fig. \ref{fig_model}(f) since only degenerate points on the plane $k_x+k_y=0 \mod 2\pi$ in a spindle-shaped nodal surface similar to Fig. \ref{fig_model}(e) still exist. Since we can not find a 2D closed manifold, like a sphere, enclosing the NL, the $\mathbb{Z}_2$ monopole charge is ill-defined for such NLs.

Finally, we set $t_1=t_4\neq t_2=t_3$ and consider the interlayer coupling term
\begin{equation}
\Delta H_4(\bm{k})=-\cos\left(\frac{k_x}{6}+\frac{k_y}{6}\right)\sin\left(\frac{k_z}{2}\right)\sigma_0\otimes\tau_2+\sin\left(\frac{k_x}{6}+\frac{k_y}{6}\right)\sin\left(\frac{k_z}{2}\right)\sigma_3\otimes\tau_0,\\
\end{equation}
and the Hamiltonian is
\begin{equation}
H_4(\bm{k})=(t_1+t_3)\Delta H_3+(t_1-t_3)\Delta H_4.
\end{equation}
We can obtain two pairs of NLs as shown in Fig. \ref{fig_model}(h) and this kind of NLs do not carry $\mathbb{Z}_2$ monopole charges, either.

\section{Detailed results and calculation details about material candidates}
Our prediction of material candidates is based on the DFT methods using the PBE form for the GGA as implemented in the Vienna ab initio simulation package (VASP)\cite{vasp}. The Gamma scheme $k$-point mesh size is set to $8\times10\times6$ for transition metal dichalcogenides (TMDs) and $10\times10\times6$ for wurtzite Si/Ge and the plane-wave cutoff energy is set to 500 eV. The lattice structure is optimized until the forces on the atoms are less than 0.002 eV/$\rm \mathring{A}$. Then a model is constructed using wannier90\cite{wannier90}. We use both parity analysis and the Wilson loop method to verify the nontrivial monopole charges of the nodal lines.

\subsection{TMDs}
As $\rm{1T'-MoTe_2}$ has been grown in experiments, we first perform structure optimization on it to verify our calculation parameters. We find that the vdW correction is not needed and the lattice structure of the experiments and lattice structure after structure optimization is shown in Table. \ref{tab_MoTe2}. Thus, in the following calculations, we do not consider the vdW corrections.

\begin{table}[b]
\caption{\label{tab_MoTe2}The lattice structure of $\rm{1T'-MoTe_2}$ of experiments and the structure optimization. The lattice parameters of experiments are $a=6.37$ $\rm \mathring{A}$, $b=3.49$ $\rm \mathring{A}$, $c=15.55$ $\rm \mathring{A}$, $\alpha=90\degree$, $\beta=92.39\degree$, $\gamma=90\degree$. The lattice parameters of the structure optimization are $a=6.39$ $\rm \mathring{A}$, $b=3.44$ $\rm \mathring{A}$, $c=15.74$ $\rm \mathring{A}$, $\alpha=90\degree$, $\beta=91.78\degree$, $\gamma=90\degree$.}
\begin{ruledtabular}
\begin{tabular}{ccccccc}
 &\multicolumn{3}{c}{Experiments}&\multicolumn{3}{c}{Structure optimization}\\
Element&x&y&z&x&y&z\\ \hline
Te&0.581&0.750&0.595&0.576&0.750&0.594\\
Te&0.419&0.250&0.405&0.424&0.250&0.406\\
Te&0.091&0.250&0.633&0.088&0.250&0.631\\
Te&0.909&0.750&0.367&0.912&0.750&0.369\\
Te&0.564&0.250&0.867&0.568&0.250&0.869\\
Te&0.436&0.750&0.133&0.432&0.750&0.131\\
Te&0.062&0.750&0.905&0.062&0.750&0.906\\
Te&0.938&0.250&0.095&0.938&0.250&0.094\\
Mo&0.182&0.750&0.506&0.182&0.750&0.506\\
Mo&0.818&0.250&0.493&0.818&0.250&0.494\\
Mo&0.320&0.250&0.007&0.319&0.250&0.006\\
Mo&0.680&0.750&0.993&0.681&0.750&0.994\\
\end{tabular}
\end{ruledtabular}
\end{table}

For the 3D counterparts of 2D $\rm 1T'$$-MX_2$ ($M$=Cr,Mo,W, $X$=S,Se,Te), there are 9 possible element combinations and 16 spatial structures for each combination, a total of 144 3D TMDs. For each of the four combinations of $(\mu,\nu)$, we have one structure for $S_{2z}$ ($T_d$ phase), i.e., $\boldsymbol{c}$ is perpendicular to both $\boldsymbol{a}$ and $\boldsymbol{b}$, and three structures for $S_{2c}$ ($\rm 1T'$ phase), i.e., $\boldsymbol{c}$ is inclined towards the $\boldsymbol{a}$-axis, $\boldsymbol{c}$ is inclined towards the $\boldsymbol{b}$-axis, and $\boldsymbol{c}$ is inclined towards both the $\boldsymbol{a}$-axis and $\boldsymbol{b}$-axis. We find all the structures in the $T_d$ phase and two out of three structures in the $\rm 1T'$ phase have $\mathbb{Z}_2$NLs, as shown in Table. \ref{table_rel}, a total of 108 3D TMDs. Although the 12 spatial structures with $\mathbb{Z}_2$NLs are structurally different, they share the same parity feature as in Table. \ref{tab_parity}.

\begin{figure}[h]
\includegraphics[width=1.0\textwidth]{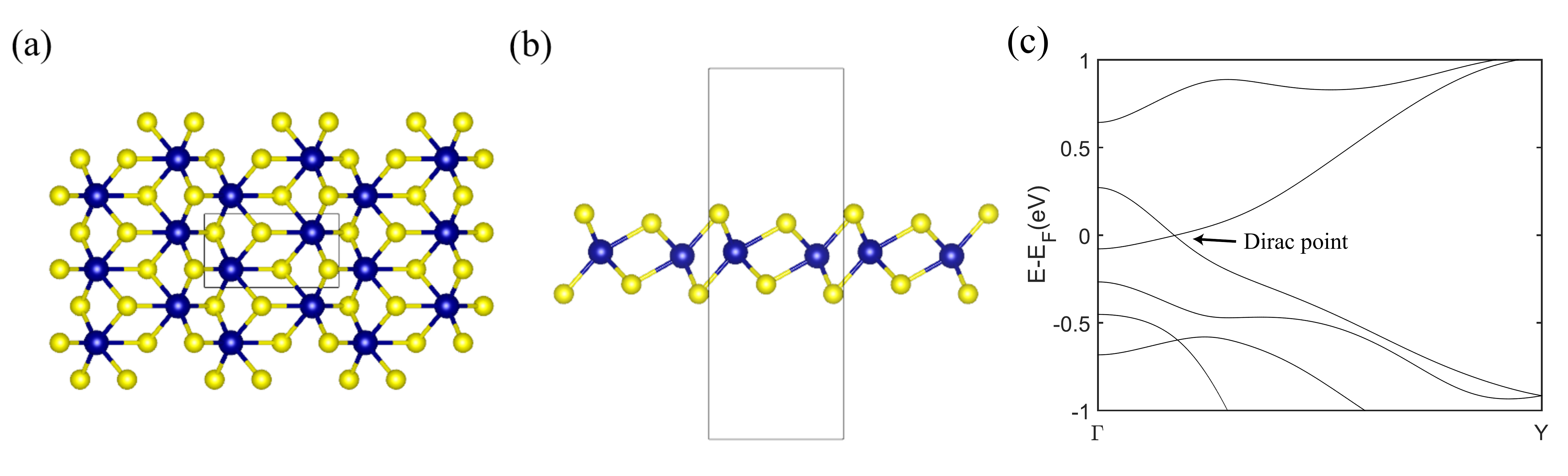}
\caption{\label{fig1}The top view (a) and front view (b) of monolayer $\rm 1T'-CrS_2$. (c) Band structure of monolayer $\rm 1T'-CrS_2$.}
\end{figure}

Here we just show the results for $\rm CrS_2$ and other materials that have similar band structures. The lattice structure of monolayer $\rm 1T'-CrS_2$ is shown in Figs .\ref{fig1}(a)(b). The monolayer $\rm 1T'-CrS_2$ has a $\mathcal{T}$-reversed pair of Dirac points along $\Gamma\mathrm{Y}$.

\begin{table}[h]
\caption{\label{table_rel} The relationship between the 16 structures. The 4 structures with strikethroughs do not have $\mathbb{Z}_2$NLs, and the remaining 12 structures have $\mathbb{Z}_2$NLs.}
\begin{ruledtabular}
\begin{tabular}{crrrr}
\diagbox{$(\mu,\nu)$}{Axes}&$\boldsymbol{c}\perp\boldsymbol{a}$ \& $\boldsymbol{c}\perp\boldsymbol{b}$
&{$\boldsymbol{c}\perp\boldsymbol{a}$} \& $\boldsymbol{c}\not\perp\boldsymbol{b}$
&$\boldsymbol{c}\not\perp\boldsymbol{a}$ \& $\boldsymbol{c}\perp\boldsymbol{b}$
&$\boldsymbol{c}\not\perp\boldsymbol{a}$ \& $\boldsymbol{c}\not\perp\boldsymbol{b}$\\
\hline
(0,0)&crystal 1 (SG.59)&crystal 2 (SG.13)&\sout{(SG.11)}&crystal 3 (SG.2)\\
\hline
(0,1)&crystal 4 (SG.57)&crystal 5 (SG.13)&\sout{(SG.11)}&crystal 6 (SG.2)\\
\hline
(1,0)&crystal 7 (SG.62)&\sout{(SG.14)}&crystal 8 (SG.11)&crystal 9 (SG.2)\\
\hline
(1,1)&crystal 10  (SG.62)&crystal 11  (SG.14)&\sout{(SG.11)}&crystal 12  (SG.2)\\
\end{tabular}
\end{ruledtabular}
\end{table}

\begin{table}[h]
\caption{\label{tab_parity}The number of states with -1 parity eigenvalues $n_-(\vec{k})$ at eight TRIMs for all 8 crystal structures and there are 48 occupied states in total. The $w_2$ obtained from parity criterion $(-1)^{w_2}=\prod_{i=1}^4(-1)^{\lfloor N_{occ}^-(\Gamma_i)/2\rfloor}$ is 1 on $k_z=0$ and 0 on $k_z=\pi$.}
\begin{ruledtabular}
\begin{tabular}{ccccccccc}
 &\multicolumn{4}{c}{$k_z$=0}&\multicolumn{4}{c}{$k_z=\pi$}\\
\cline{2-5}\cline{6-9}
TRIM&$\Gamma$&X&Y&Z&S&T&U&R\\ \hline
$n_-(\vec{k})$&22&24&24&24&24&24&24&24\\
\end{tabular}
\end{ruledtabular}
\end{table}

\clearpage
\subsubsection{$(\mu,\nu)=(0,0)$}
Crystal 1 is the Td phase, and the rest two structures are the $\mathrm{1T'}$ phase. Crystal 2 can be gotten by changing the angle between $c$ and $b$ axes in crystal 1, i.e., the two layers have undergone a relative displacement in the $b$-direction. Crystal 3 can be gotten by changing the angles between both $c$ and $b$ axes and $c$ and $a$ axes in crystal 1, i.e., the two layers have undergone a relative displacement in both $a$ and $b$-directions.

\begin{table}[b]
\caption{\label{tab00}The 3D lattice structures from 2D $\rm{1T'-CrS_2}$ with $(\mu,\nu)=(0,0)$. The lattice parameters of crystal 1 are $a=5.473$ $\rm \mathring{A}$, $b=2.996$ $\rm \mathring{A}$, $c=15.081$ $\rm \mathring{A}$, $\alpha=90\degree$, $\beta=90\degree$, $\gamma=90\degree$. The lattice parameters of crystal 2 are $a=5.473$ $\rm \mathring{A}$, $b=2.996$ $\rm \mathring{A}$, $c=15.345$ $\rm \mathring{A}$, $\alpha=100.11\degree$, $\beta=90\degree$, $\gamma=90\degree$. The lattice parameters of crystal 3 are $a=5.469$ $\rm \mathring{A}$, $b=2.999$ $\rm \mathring{A}$, $c=14.794$ $\rm \mathring{A}$, $\alpha=88.30\degree$, $\beta=80.22\degree$, $\gamma=90\degree$.}
\begin{ruledtabular}
\begin{tabular}{cccccccccc}
 &\multicolumn{3}{c}{crystal 1}&\multicolumn{3}{c}{crystal 2}&\multicolumn{3}{c}{crystal 3}\\
Element&x&y&z&x&y&z&x&y&z\\ \hline
Cr&0.197&0.250&0.495&0.197&0.755&0.005&0.200&0.251&0.995\\
Cr&0.803&0.750&0.505&0.803&0.245&0.995&0.800&0.749&0.005\\
Cr&0.803&0.750&0.995&0.803&0.755&0.505&0.305&0.251&0.495\\
Cr&0.197&0.250&0.005&0.197&0.245&0.495&0.695&0.749&0.505\\
S&0.924&0.250&0.892&0.924&0.347&0.608&0.474&0.766&0.388\\
S&0.076&0.750&0.108&0.076&0.653&0.392&0.526&0.234&0.611\\
S&0.425&0.750&0.918&0.425&0.824&0.582&0.964&0.262&0.414\\
S&0.575&0.250&0.082&0.575&0.176&0.418&0.036&0.738&0.585\\
S&0.076&0.750&0.392&0.076&0.347&0.108&0.128&0.766&0.888\\
S&0.924&0.250&0.608&0.924&0.653&0.892&0.872&0.234&0.112\\
S&0.575&0.250&0.418&0.575&0.824&0.082&0.615&0.262&0.915\\
S&0.425&0.750&0.582&0.425&0.176&0.918&0.385&0.738&0.085\\
\end{tabular}
\end{ruledtabular}
\end{table}

\begin{figure}[h]
\includegraphics[width=1\textwidth]{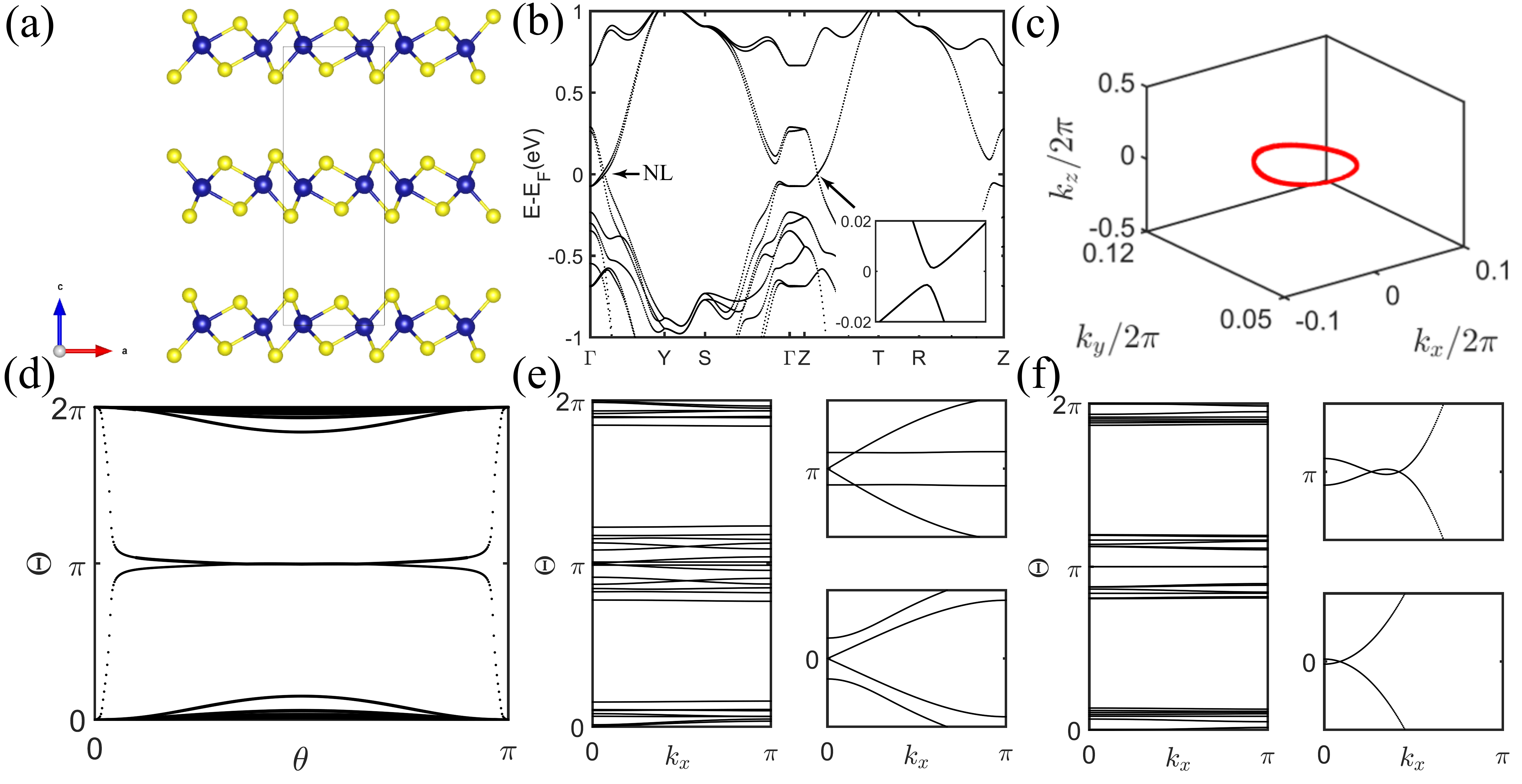}
\caption{\label{c1n}(a) The lattice structure of crystal 1 (SG. 59). (b) The band structure. (c) The position of one of the nodal lines. (d) Wilson loop on the sphere enclosing one of the nodal lines, indicating that the nodal line carries a unit $\mathbb{Z}_2$ monopole charge. (e) Wilson loop on the plane defined by $k_y=0$. (f) The Wilson loop on the plane defined on $k_y=\pi$.}
\end{figure}
\begin{figure}[h]
\includegraphics[width=1\textwidth]{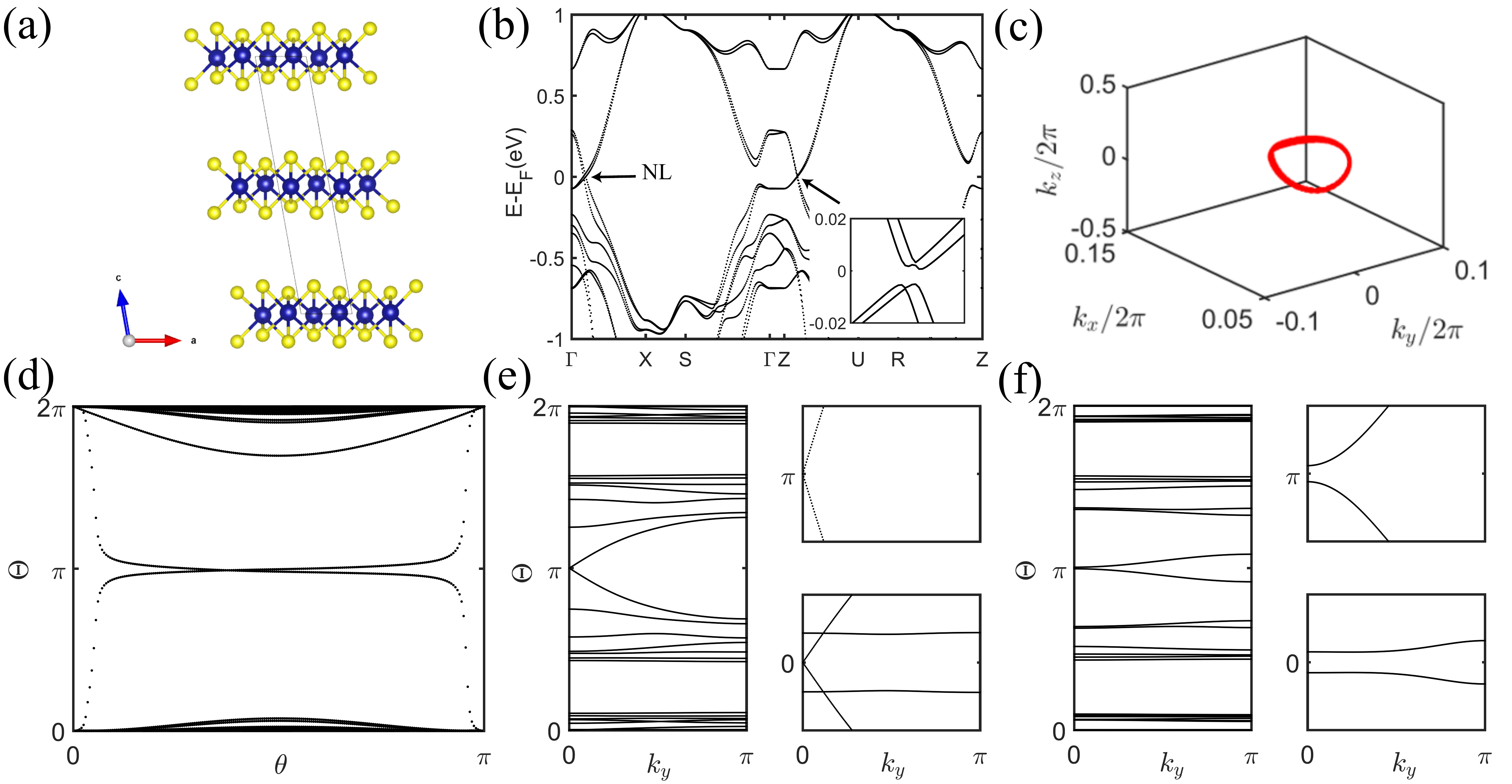}
\caption{\label{c1y} (a) The lattice structure of crystal 2 (SG. 13). (b) The band structure. (b) The position of one of the irregular nodal lines. (c) Wilson loop on the sphere enclosing one of the nodal lines, indicating that the nodal line carries a unit $\mathbb{Z}_2$ monopole charge. (d) Wilson loop on the plane defined by $k_x=0$. (f) The Wilson loop on the plane defined on $k_x=\pi$.}
\end{figure}
\begin{figure}[h]
\includegraphics[width=1\textwidth]{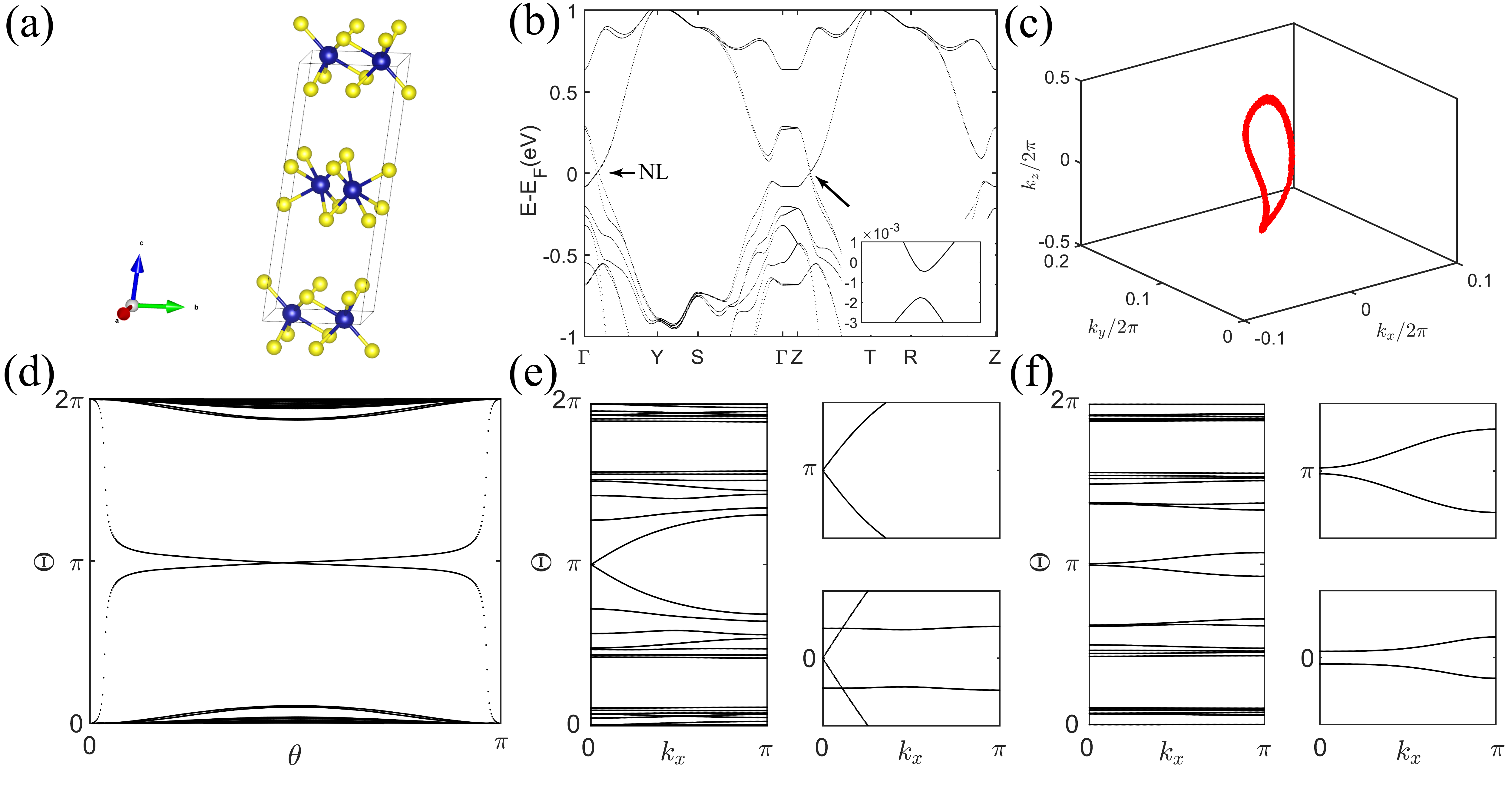}
\caption{\label{c1xy} (a) The lattice structure of crystal 3 (SG. 2). (b) The band structure. (b) The position of one of the irregular nodal lines. (c) Wilson loop on the sphere enclosing one of the nodal lines, indicating that the nodal line carries a unit $\mathbb{Z}_2$ monopole charge. (d) Wilson loop on the plane defined by $k_x=0$. (f) The Wilson loop on the plane defined on $k_x=\pi$.}
\end{figure}
\clearpage

\subsubsection{$(\mu,\nu)=(0,1)$}
Crystal 4 is the Td phase, and the rest two structures are the $\mathrm{1T'}$ phase. Crystal 5 can be gotten by changing the angle between $c$ and $b$ axes in crystal 4, i.e., the two layers have undergone a relative displacement in the $b$-direction. Crystal 6 can be gotten by changing the angles between both $c$ and $b$ axes and $c$ and $a$ axes in crystal 4, i.e., the two layers have undergone a relative displacement in both $a$ and $b$-directions.

\begin{table}[b]
\caption{\label{tab3}The 3D lattice structure from 2D $\rm{1T'-CrS_2}$ with $(\mu,\nu)=(0,1)$. The lattice parameters of crystal 4 are $a=5.474$ $\rm \mathring{A}$, $b=2.997$ $\rm \mathring{A}$, $c=14.569$ $\rm \mathring{A}$, $\alpha=90\degree$, $\beta=90\degree$, $\gamma=90\degree$. The lattice parameters of crystal 5 are $a=5.473$ $\rm \mathring{A}$, $b=2.997$ $\rm \mathring{A}$, $c=14.574$ $\rm \mathring{A}$, $\alpha=90.390\degree$, $\beta=90\degree$, $\gamma=90\degree$. The lattice parameters of crystal 6 are $a=5.450$ $\rm \mathring{A}$, $b=2.998$ $\rm \mathring{A}$, $c=14.571$ $\rm \mathring{A}$, $\alpha=89.304\degree$, $\beta=95.994\degree$, $\gamma=90\degree$.}
\begin{ruledtabular}
\begin{tabular}{cccccccccc}
 &\multicolumn{3}{c}{crystal 4}&\multicolumn{3}{c}{crystal 5}&\multicolumn{3}{c}{crystal 6}\\
Element&x&y&z&x&y&z&x&y&z\\ \hline
Cr&0.803&0.250&0.995&0.803&0.750&0.505&0.804&0.250&0.00510\\
Cr&0.197&0.750&0.005&0.197&0.250&0.495&0.196&0.750&0.99490\\
Cr&0.197&0.250&0.495&0.197&0.750&0.005&0.199&0.250&0.50507\\
Cr&0.803&0.750&0.505&0.803&0.250&0.995&0.801&0.751&0.49493\\
S&0.076&0.750&0.388&0.076&0.254&0.112&0.109&0.743&0.61225\\
S&0.924&0.250&0.612&0.924&0.746&0.888&0.891&0.257&0.38775\\
S&0.575&0.250&0.415&0.575&0.753&0.085&0.599&0.245&0.58569\\
S&0.425&0.750&0.585&0.425&0.247&0.915&0.401&0.755&0.41431\\
S&0.924&0.750&0.888&0.924&0.254&0.612&0.953&0.743&0.11223\\
S&0.076&0.250&0.112&0.076&0.746&0.388&0.047&0.257&0.88777\\
S&0.425&0.250&0.915&0.425&0.753&0.585&0.448&0.245&0.08572\\
S&0.575&0.750&0.085&0.575&0.247&0.415&0.552&0.755&0.91428\\
\end{tabular}
\end{ruledtabular}
\end{table}

\begin{figure}[h]
\includegraphics[width=1\textwidth]{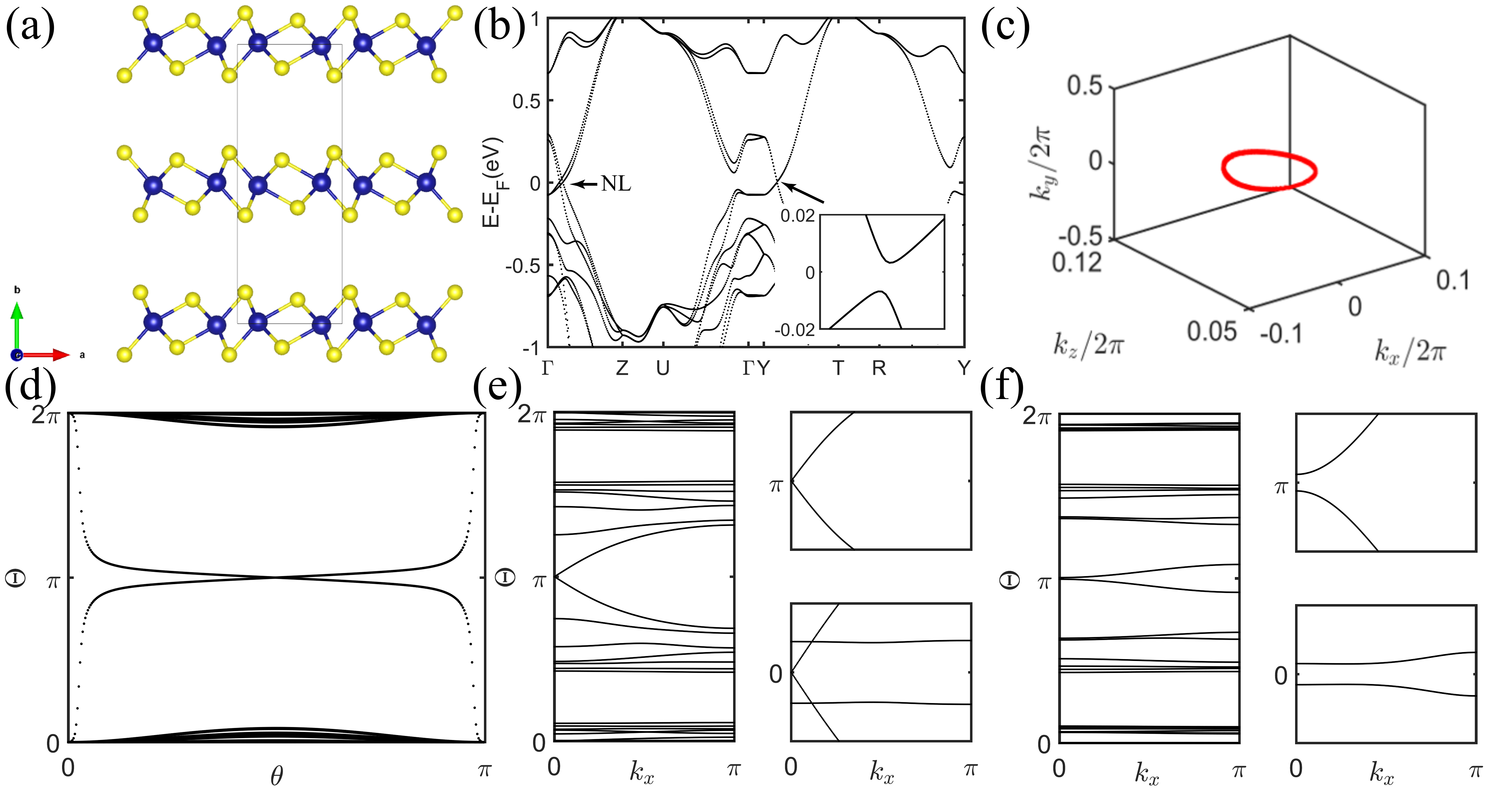}
\caption{\label{c2n}(a) The lattice structure of crystal 4 (SG. 57). (b) The band structure. (c) The position of one of the nodal lines. (d) Wilson loop on the sphere enclosing one of the nodal lines, indicating that the nodal line carries a unit $\mathbb{Z}_2$ monopole charge. (e) Wilson loop on the plane defined by $k_y=0$. (f) The Wilson loop on the plane defined on $k_y=\pi$.}
\end{figure}
\begin{figure}[h]
\includegraphics[width=1\textwidth]{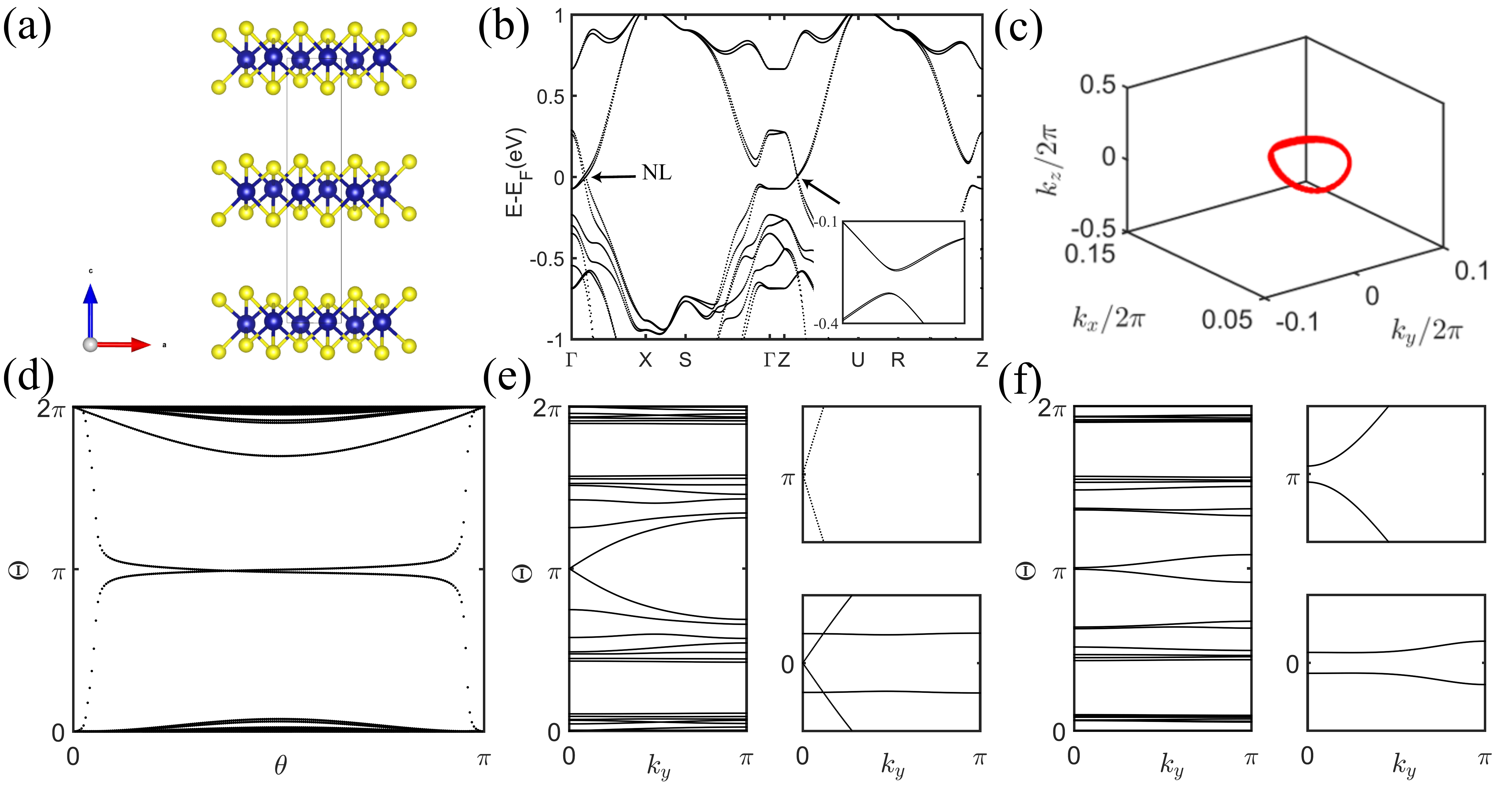}
\caption{\label{c2y} (a) The lattice structure of crystal 5 (SG.13). (b) The band structure. (b) The position of one of the irregular nodal lines. (c) Wilson loop on the sphere enclosing one of the nodal lines, indicating that the nodal line carries a unit $\mathbb{Z}_2$ monopole charge. (d) Wilson loop on the plane defined by $k_x=0$. (f) The Wilson loop on the plane defined on $k_x=\pi$.}
\end{figure}
\begin{figure}[h]
\includegraphics[width=1\textwidth]{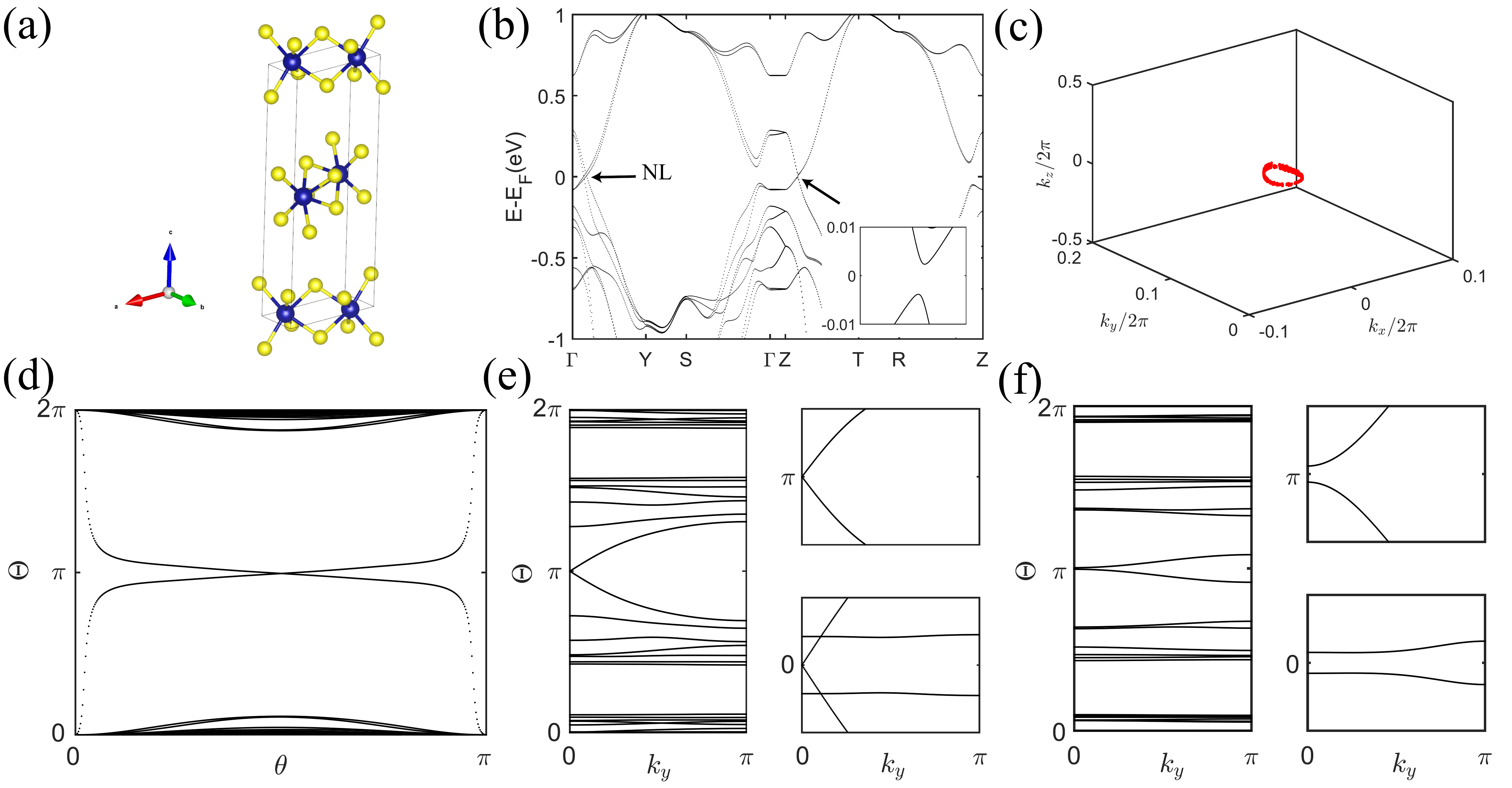}
\caption{\label{c2xy} (a) The lattice structure of crystal 6 (SG. 2). (b) The band structure. (b) The position of one of the irregular nodal lines. (c) Wilson loop on the sphere enclosing one of the nodal lines, indicating that the nodal line carries a unit $\mathbb{Z}_2$ monopole charge. (d) Wilson loop on the plane defined by $k_x=0$. (f) The Wilson loop on the plane defined on $k_x=\pi$.}
\end{figure}
\clearpage

\subsubsection{$(\mu,\nu)=(1,0)$}
Crystal 7 is the Td phase, and the rest two structures are the $\mathrm{1T'}$ phase. Crystal 8 can be gotten by changing the angle between $c$ and $a$ axes in crystal 8, i.e., the two layers have undergone a relative displacement in the $a$-direction. Crystal 9 can be gotten by changing the angles between both $c$ and $b$ axes and $c$ and $a$ axes in crystal 7, i.e., the two layers have undergone a relative displacement in both $a$ and $b$-directions.

\begin{table}[b]
\caption{\label{tab4}The 3D lattice structure from 2D $\rm{1T'-CrS_2}$ with $(\mu,\nu)=(1,0)$. The lattice parameters of crystal 7 are $a=5.475$ $\rm \mathring{A}$, $b=2.998$ $\rm \mathring{A}$, $c=13.984$ $\rm \mathring{A}$, $\alpha=90\degree$, $\beta=90\degree$, $\gamma=90\degree$. The lattice parameters of crystal 8 are $a=5.475$ $\rm \mathring{A}$, $b=2.998$ $\rm \mathring{A}$, $c=13.997$ $\rm \mathring{A}$, $\alpha=90\degree$, $\beta=92.368\degree$, $\gamma=90\degree$. The lattice parameters of crystal 9 are $a=5.470$ $\rm \mathring{A}$, $b=3.001$ $\rm \mathring{A}$, $c=14.0484$ $\rm \mathring{A}$, $\alpha=90.432\degree$, $\beta=91.941\degree$, $\gamma=90\degree$.}
\begin{ruledtabular}
\begin{tabular}{cccccccccc}
 &\multicolumn{3}{c}{crystal 7}&\multicolumn{3}{c}{crystal 8}&\multicolumn{3}{c}{crystal 9}\\
Element&x&y&z&x&y&z&x&y&z\\ \hline
Cr&0.697&0.750&0.006&0.698&0.250&0.006&0.697&0.750&0.995\\
Cr&0.303&0.250&0.994&0.302&0.750&0.994&0.303&0.250&0.005\\
Cr&0.803&0.250&0.506&0.803&0.750&0.506&0.802&0.250&0.495\\
Cr&0.197&0.750&0.494&0.197&0.250&0.494&0.198&0.750&0.505\\
S&0.924&0.750&0.616&0.936&0.250&0.616&0.913&0.746&0.384\\
S&0.076&0.250&0.384&0.064&0.750&0.384&0.087&0.254&0.616\\
S&0.425&0.250&0.589&0.434&0.750&0.589&0.417&0.247&0.412\\
S&0.575&0.750&0.411&0.566&0.250&0.411&0.583&0.753&0.588\\
S&0.576&0.250&0.116&0.588&0.750&0.116&0.567&0.246&0.884\\
S&0.424&0.750&0.884&0.412&0.250&0.884&0.433&0.754&0.116\\
S&0.075&0.750&0.089&0.084&0.250&0.089&0.068&0.747&0.912\\
S&0.925&0.250&0.911&0.916&0.750&0.911&0.932&0.253&0.088\\
\end{tabular}
\end{ruledtabular}
\end{table}

\begin{figure}[h]
\includegraphics[width=1\textwidth]{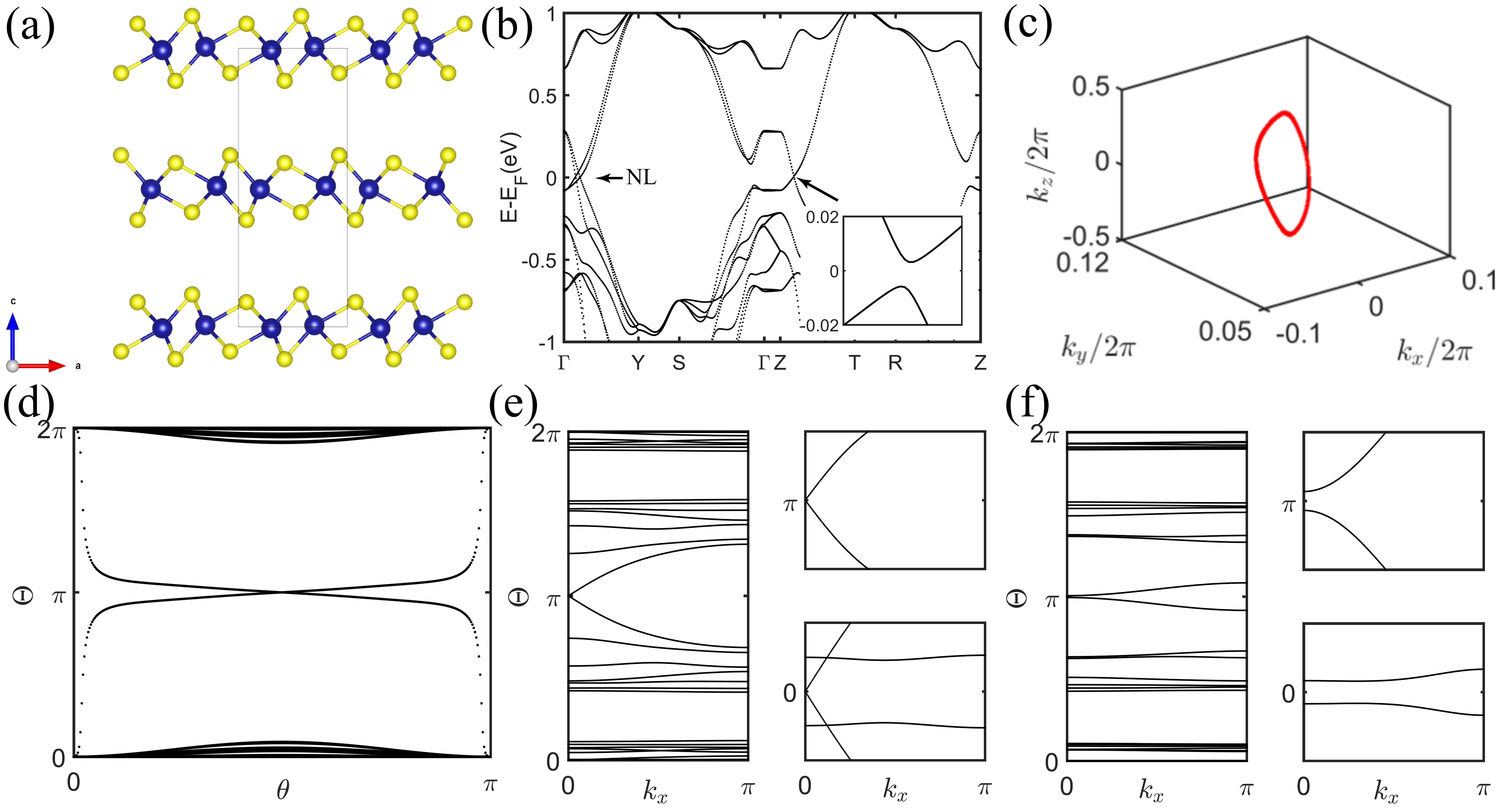}
\caption{\label{c3n}(a) The lattice structure of crystal 7 (SG. 62). (b) The band structure. (c) The position of one of the nodal lines. (d) Wilson loop on the sphere enclosing one of the nodal lines, indicating that the nodal line carries a unit $\mathbb{Z}_2$ monopole charge. (e) Wilson loop on the plane defined by $k_y=0$. (f) The Wilson loop on the plane defined on $k_y=\pi$.}
\end{figure}
\begin{figure}[h]
\includegraphics[width=1\textwidth]{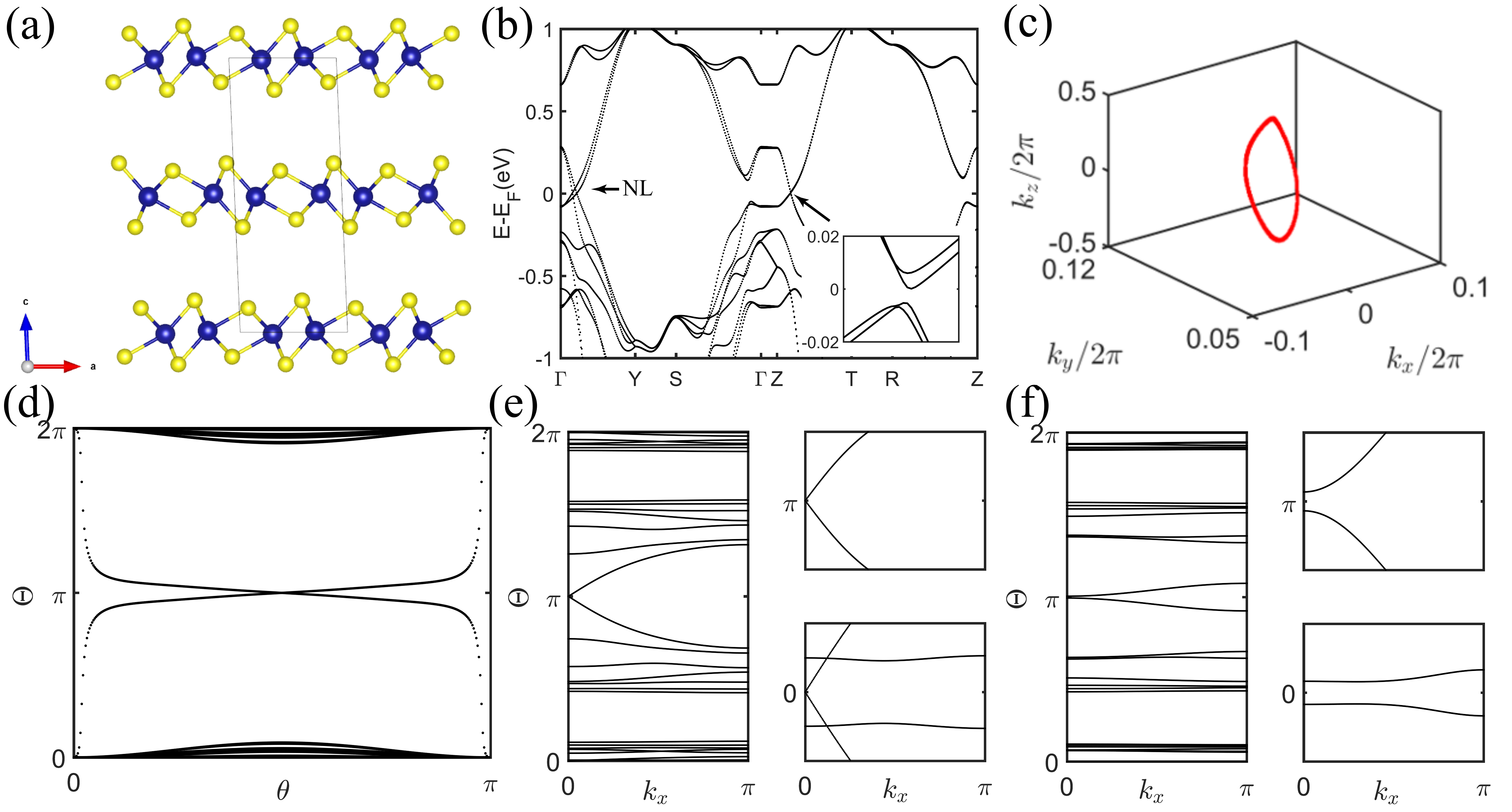}
\caption{\label{c3x} (a) The lattice structure of crystal 8 (SG. 11). (b) The band structure. (b) The position of one of the irregular nodal lines. (c) Wilson loop on the sphere enclosing one of the nodal lines, indicating that the nodal line carries a unit $\mathbb{Z}_2$ monopole charge. (d) Wilson loop on the plane defined by $k_x=0$. (f) The Wilson loop on the plane defined on $k_x=\pi$.}
\end{figure}
\begin{figure}[h]
\includegraphics[width=1\textwidth]{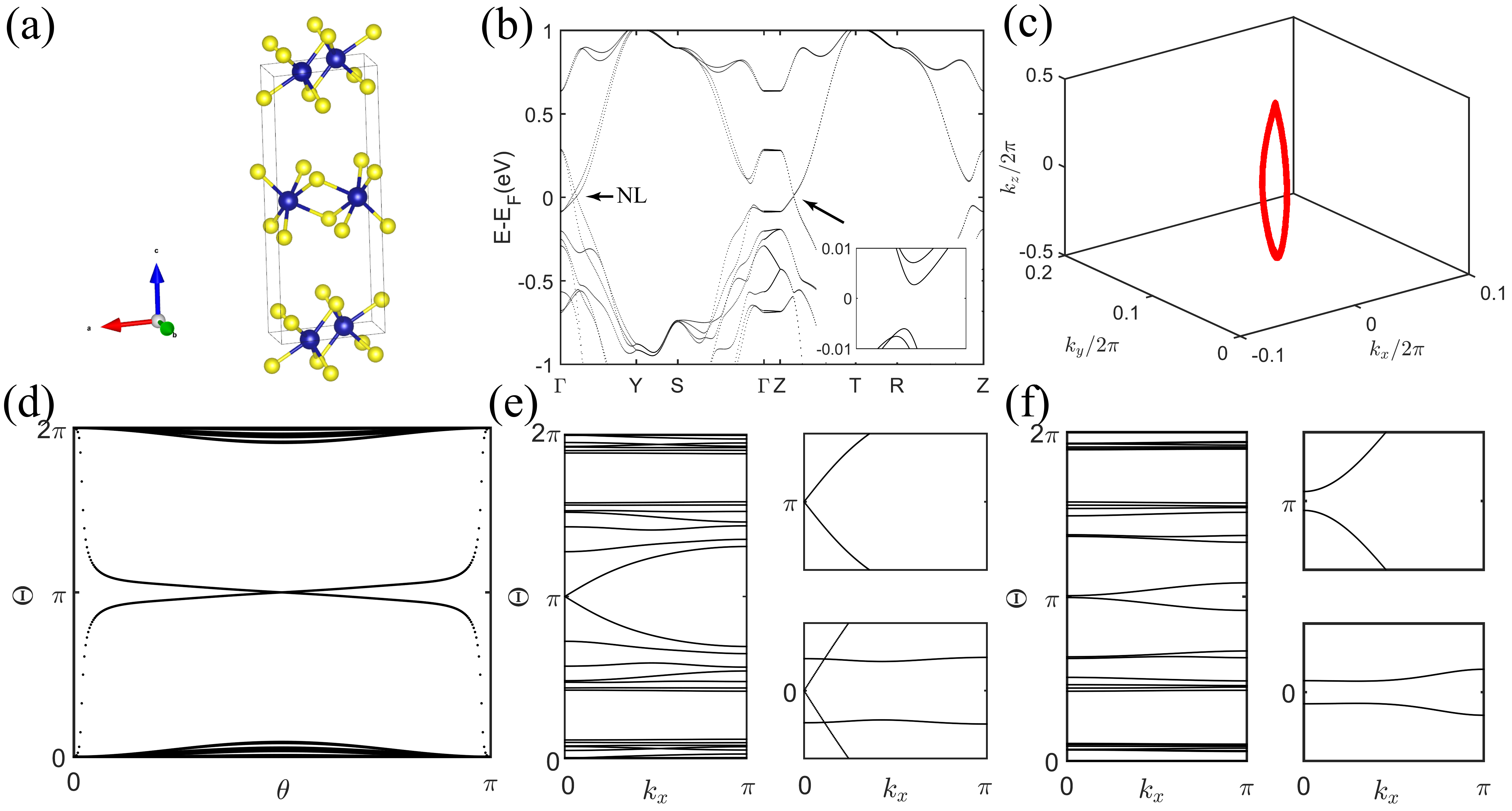}
\caption{\label{c3xy} (a) The lattice structure of crystal 9 (SG. 2). (b) The band structure. (b) The position of one of the irregular nodal lines. (c) Wilson loop on the sphere enclosing one of the nodal lines, indicating that the nodal line carries a unit $\mathbb{Z}_2$ monopole charge. (d) Wilson loop on the plane defined by $k_x=0$. (f) The Wilson loop on the plane defined on $k_x=\pi$.}
\end{figure}
\clearpage

\subsubsection{$(\mu,\nu)=(1,1)$}
Crystal 10 is the Td phase, and the rest two structures are the $\mathrm{1T'}$ phase. Crystal 11 can be gotten by changing the angle between $c$ and $b$ axes in crystal 10, i.e., the two layers have undergone a relative displacement in the $b$-direction. Crystal 9 can be gotten by changing the angles between both $c$ and $b$ axes and $c$ and $a$ axes in crystal 7, i.e., the two layers have undergone a relative displacement in both $a$ and $b$-directions.

\begin{table}[b]
\caption{\label{tab5}The 3D lattice structure from 2D $\rm{1T'-CrS_2}$ with $(\mu,\nu)=(1,1)$. The lattice parameters of crystal 10 are $a=5.474$ $\rm \mathring{A}$, $b=2.997$ $\rm \mathring{A}$, $c=14.542$ $\rm \mathring{A}$, $\alpha=90\degree$, $\beta=90\degree$, $\gamma=90\degree$. The lattice parameters of crystal 11 are $a=5.477$ $\rm \mathring{A}$, $b=2.995$ $\rm \mathring{A}$, $c=14.526$ $\rm \mathring{A}$, $\alpha=90.971\degree$, $\beta=90\degree$, $\gamma=90\degree$. The lattice parameters of crystal 12 are $a=5.469$ $\rm \mathring{A}$, $b=2.999$ $\rm \mathring{A}$, $c=14.566$ $\rm \mathring{A}$, $\alpha=87.611\degree$, $\beta=87.837\degree$, $\gamma=90\degree$.}
\begin{ruledtabular}
\begin{tabular}{cccccccccc}
 &\multicolumn{3}{c}{crystal 10}&\multicolumn{3}{c}{crystal 11}&\multicolumn{3}{c}{crystal 12}\\
Element&x&y&z&x&y&z&x&y&z\\ \hline
Cr&0.197&0.250&0.494&0.697&0.750&0.005&0.802&0.749&0.505\\
Cr&0.803&0.750&0.506&0.303&0.250&0.995&0.198&0.251&0.495\\
Cr&0.303&0.250&0.994&0.803&0.750&0.505&0.697&0.749&0.005\\
Cr&0.697&0.750&0.006&0.197&0.250&0.495&0.303&0.251&0.995\\
S&0.424&0.750&0.888&0.924&0.259&0.612&0.566&0.227&0.112\\
S&0.576&0.250&0.112&0.076&0.741&0.388&0.434&0.773&0.888\\
S&0.925&0.250&0.915&0.425&0.757&0.585&0.067&0.735&0.085\\
S&0.075&0.750&0.085&0.575&0.243&0.415&0.933&0.267&0.915\\
S&0.076&0.750&0.388&0.576&0.259&0.112&0.912&0.227&0.612\\
S&0.924&0.250&0.612&0.424&0.741&0.888&0.088&0.773&0.388\\
S&0.575&0.250&0.415&0.075&0.757&0.085&0.416&0.733&0.585\\
S&0.425&0.750&0.585&0.925&0.243&0.915&0.584&0.267&0.415\\
\end{tabular}
\end{ruledtabular}
\end{table}

\begin{figure}[h]
\includegraphics[width=1\textwidth]{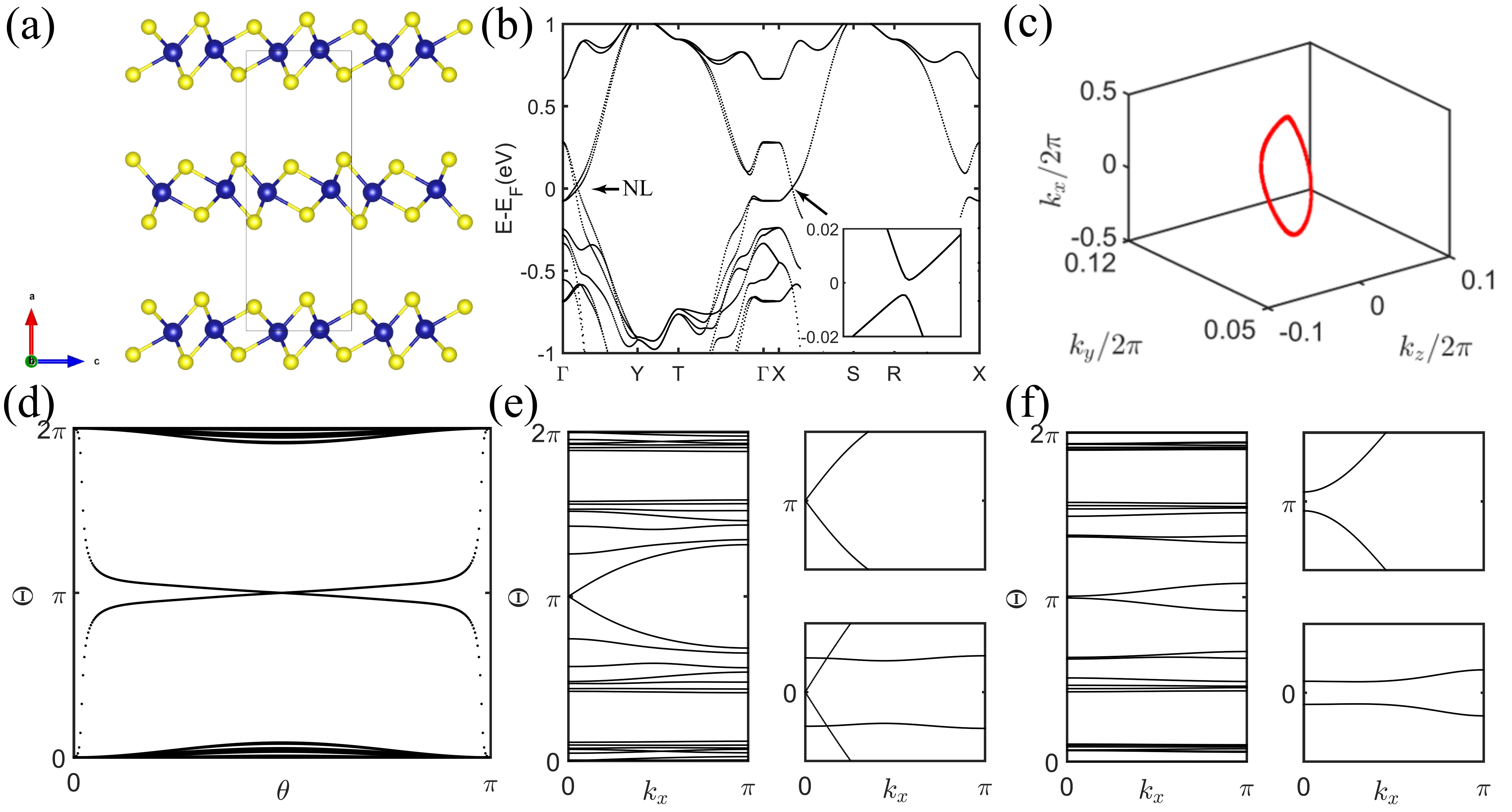}
\caption{\label{c4n}(a) The lattice structure of crystal 10 (SG. 62). (b) The band structure. (c) The position of one of the nodal lines. (d) Wilson loop on the sphere enclosing one of the nodal lines, indicating that the nodal line carries a unit $\mathbb{Z}_2$ monopole charge. (e) Wilson loop on the plane defined by $k_y=0$. (f) The Wilson loop on the plane defined on $k_y=\pi$.}
\end{figure}
\begin{figure}[h]
\includegraphics[width=1\textwidth]{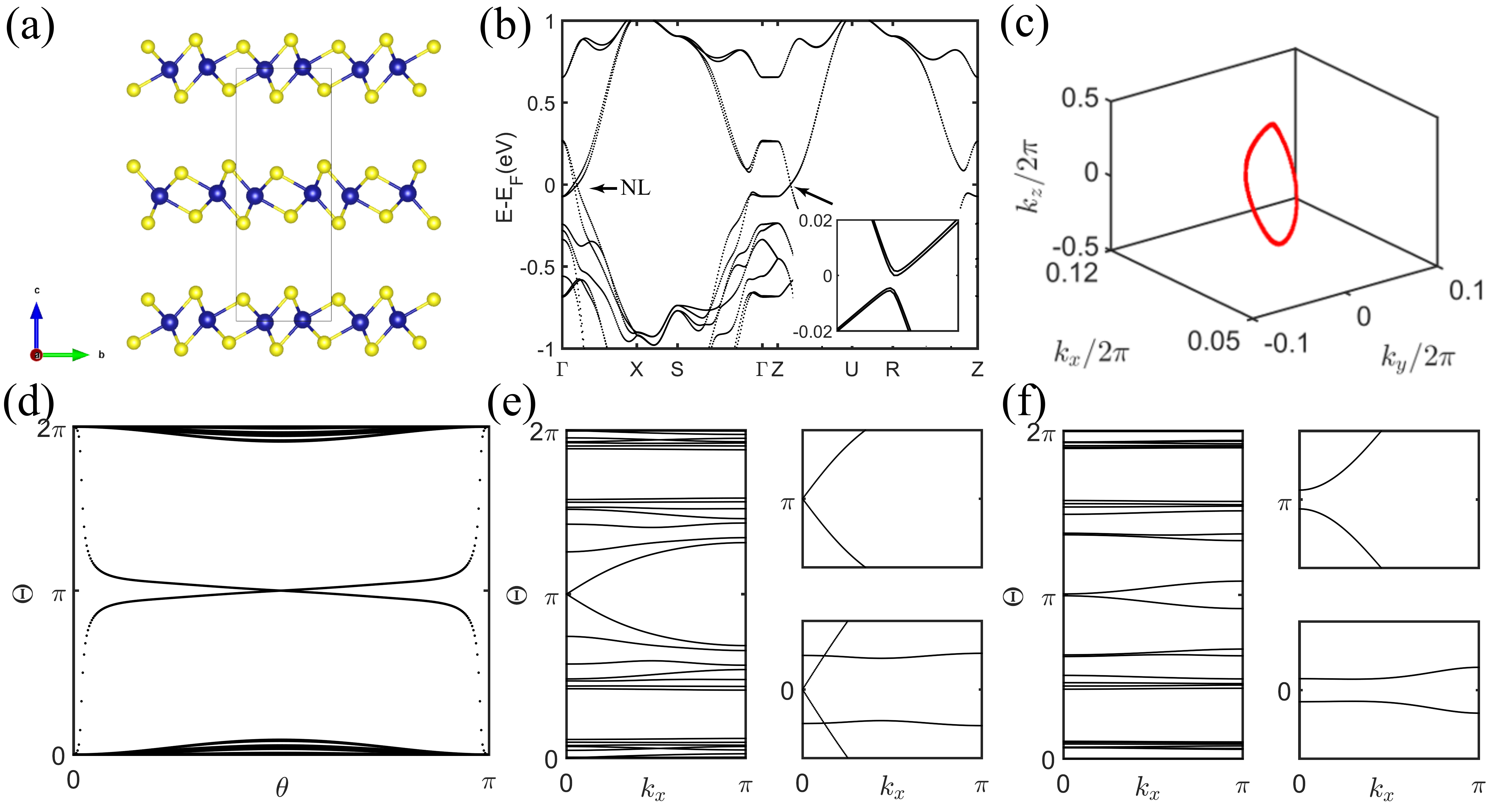}
\caption{\label{c4y} (a) The lattice structure of crystal 11 (SG. 14). (b) The band structure. (b) The position of one of the irregular nodal lines. (c) Wilson loop on the sphere enclosing one of the nodal lines, indicating that the nodal line carries a unit $\mathbb{Z}_2$ monopole charge. (d) Wilson loop on the plane defined by $k_x=0$. (f) The Wilson loop on the plane defined on $k_x=\pi$.}
\end{figure}
\begin{figure}[h]
\includegraphics[width=1\textwidth]{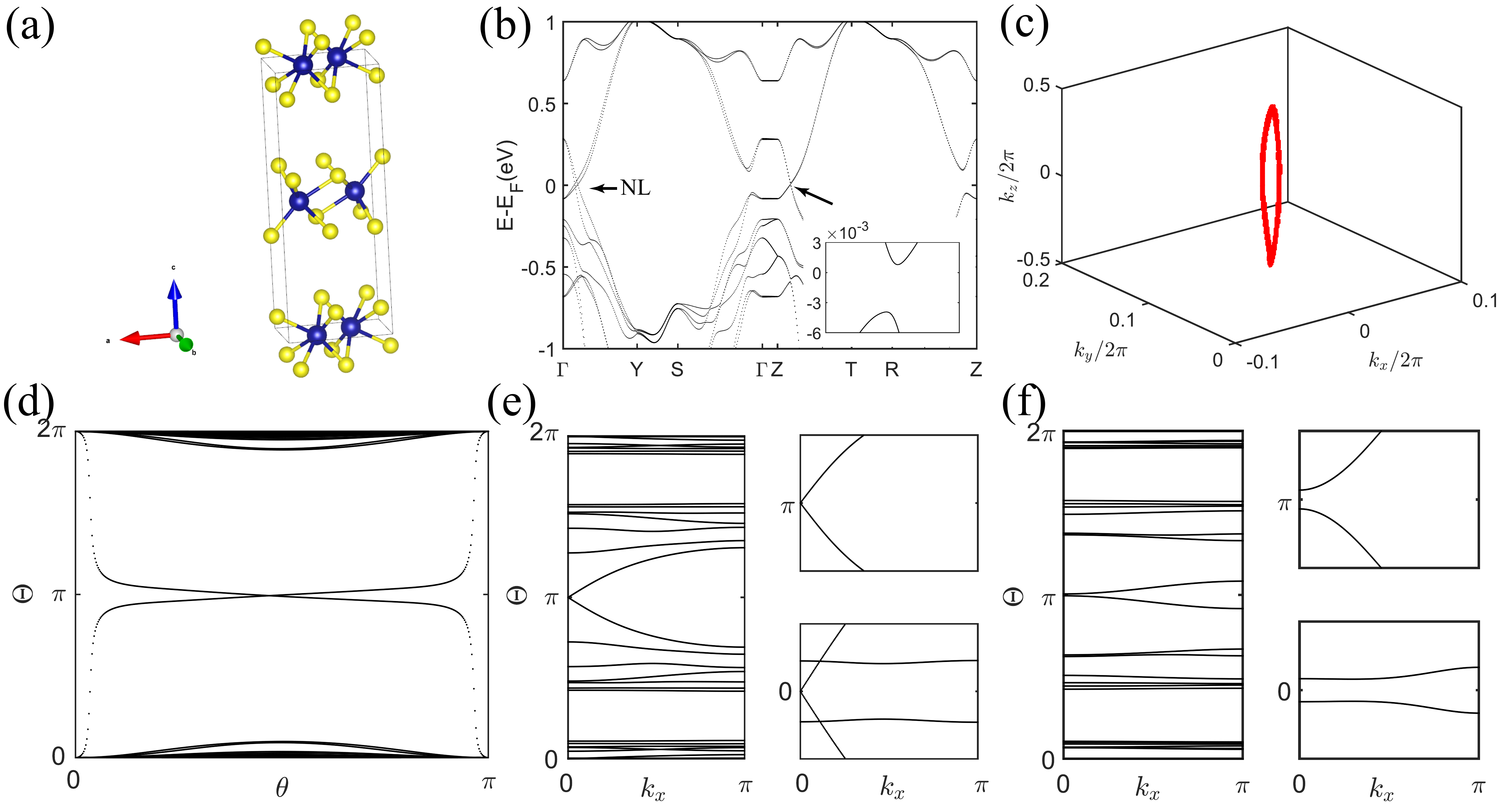}
\caption{\label{c4xy} (a) The lattice structure of crystal 12 (SG. 2). (b) The band structure. (b) The position of one of the irregular nodal lines. (c) Wilson loop on the sphere enclosing one of the nodal lines, indicating that the nodal line carries a unit $\mathbb{Z}_2$ monopole charge. (d) Wilson loop on the plane defined by $k_x=0$. (f) The Wilson loop on the plane defined on $k_x=\pi$.}
\end{figure}
\clearpage

\subsection{Xene}

\begin{figure}[h]
\includegraphics[width=1\textwidth]{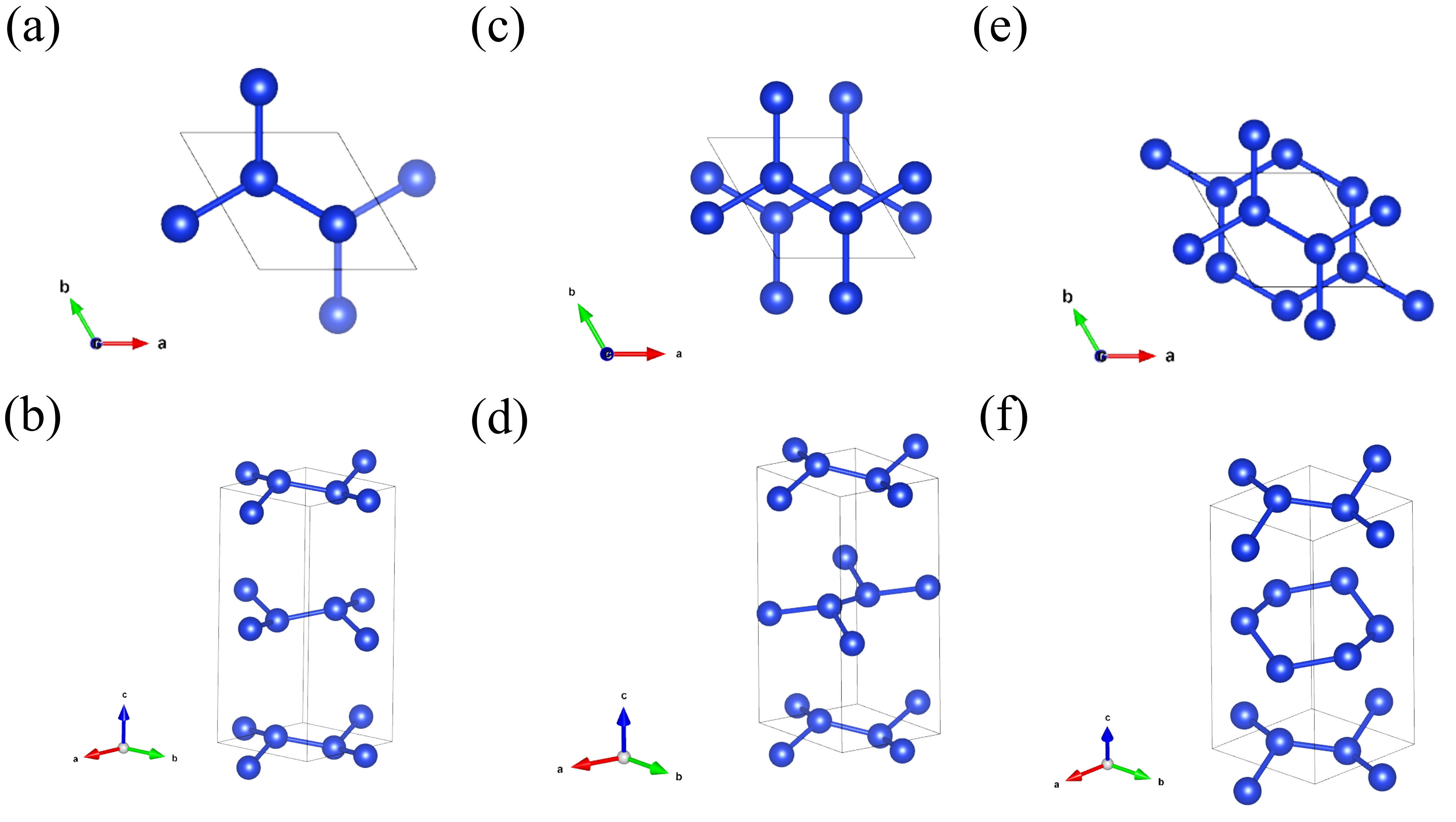}
\caption{\label{sistru}The three starting structures with $\boldsymbol{c}\perp\boldsymbol{a}$ \& $\boldsymbol{c}\perp\boldsymbol{b}$. (a),(b) $(\mu,\nu)=(0,0)$. (c),(d) $(\mu,\nu)=(0,1)$ or $(\mu,\nu)=(1,0)$. (e),(f) $(\mu,\nu)=(1,1)$ }
\end{figure}

For 3D structures of Xene (X=Si, Ge, Sn, Pb), since stanene and plumbene tend to undergo dimerization when stacked into 3D structures, we only consider silicene and germanene. The $a$-axis and $b$-axis are equivalent in Xene, i.e., (0,1) and (1,0) are equivalent, and thus we have 3 combinations of $(\mu,\nu)$. For each combination of $(\mu,\nu)$, we have 3 structures, i.e., $c$ is perpendicular to both $a$ and $b$, $c$ is inclined toward $a$, and $c$ is inclined toward both $a$ and $b$. The three starting structures with $\boldsymbol{c}\perp\boldsymbol{a}$ \& $\boldsymbol{c}\perp\boldsymbol{b}$ $(\boldsymbol{c'}=\boldsymbol{c})$ are given in Fig. \ref{sistru}, with $\boldsymbol{c'}$ defined in Table. \ref{table_rel_si}. We find 7 of the 9 structures, a total of 14 3D structures of Xene, with $\mathbb{Z}_2$NLs. The relationship between the 9 structures is given in Table. \ref{table_rel_si}. The wurtzite structure in the text corresponds to $(\mu,\nu)=(0,0)$.

\begin{table}[h]
\caption{\label{table_rel_si} The relationship between the 9 structures. The three lattice vectors of the wurtzite structure are taken to be $\boldsymbol{a},\boldsymbol{b},\boldsymbol{c}$, while the lattice vectors of other structures are taken to be $\boldsymbol{a},\boldsymbol{b},\boldsymbol{c'}$, i.e., they differ only in $c$-axis. $c_i$ are arbitrary nonzero coefficients. The 2 structures with strikethroughs do not have $\mathbb{Z}_2$NLs, and thus the remaining 7 structures have $\mathbb{Z}_2$NLs.}
\begin{ruledtabular}
\begin{tabular}{crrr}
\diagbox{$(\mu,\nu)$}{Axes}&$\boldsymbol{c'}=\boldsymbol{c}$
&$\boldsymbol{c'}=c_1\boldsymbol{a}+c_2\boldsymbol{c}$
&$\boldsymbol{c'}=c_1\boldsymbol{a}+c_2\boldsymbol{c}+c_3\boldsymbol{b}$\\
\hline
(0,0)&crystal 1 (SG.194) &crystal 2 (SG.15)&crystal 3 (SG.2)\\
\hline
(0,1)&\sout{(SG.64)}&crystal 4 (SG.15)&crystal 5 (SG.2)\\
\hline
(1,1)&\sout{(SG.64)}&crystal 6  (SG.2)&crystal 7  (SG.2)\\
\end{tabular}
\end{ruledtabular}
\end{table}

\begin{figure}[h]
\includegraphics[width=1\textwidth]{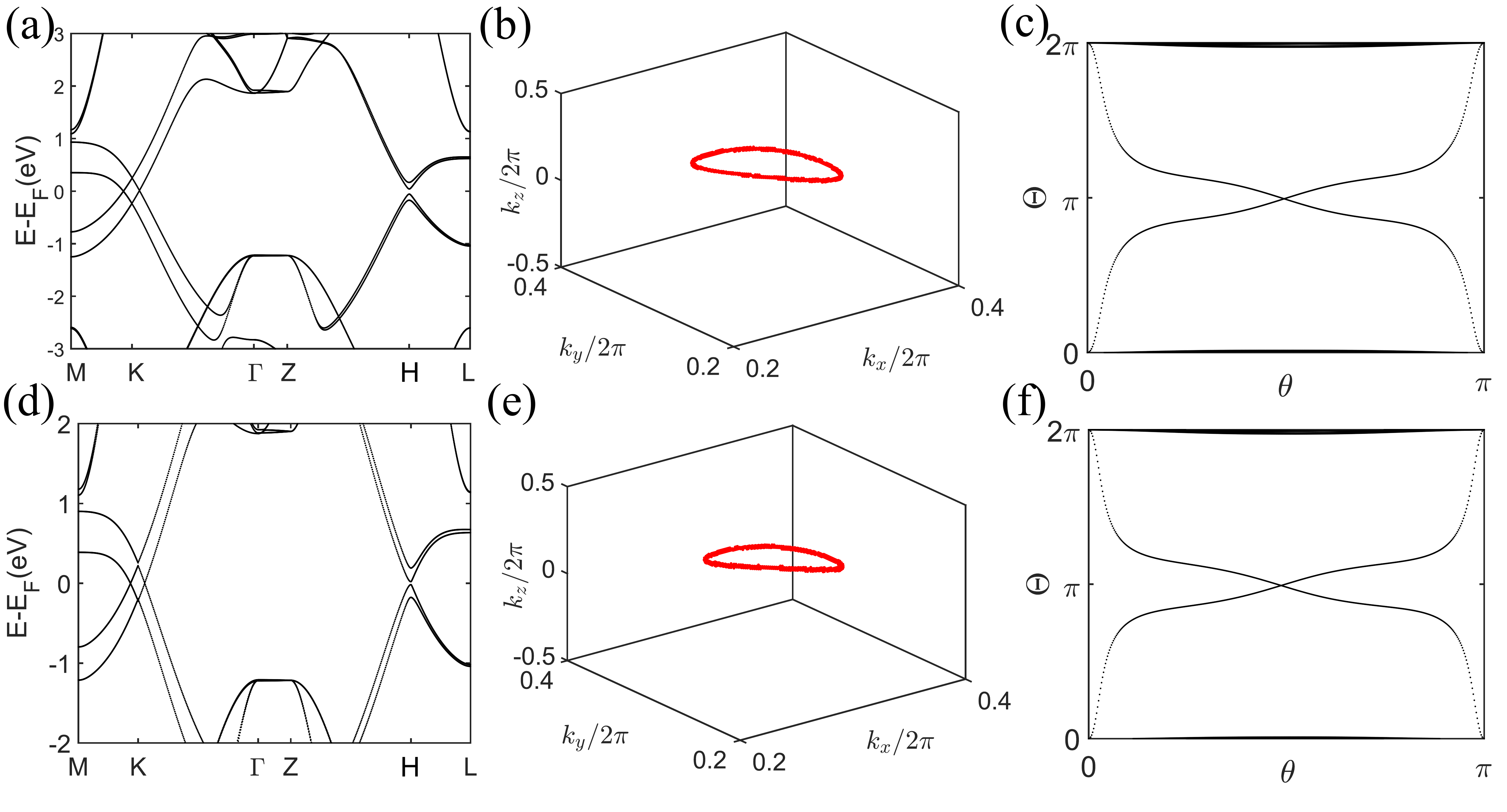}
\caption{\label{si1} Crystal 2: (a) The band structure. (b) The position of the nodal line around K. (c) Wilson loop on the sphere enclosing one of the nodal lines, indicating that the nodal line carries a unit $\mathbb{Z}_2$ monopole charge. (d)-(f) Same as (a)-(c) but for crystal 3.}
\end{figure}
\begin{figure}[h]
\includegraphics[width=1\textwidth]{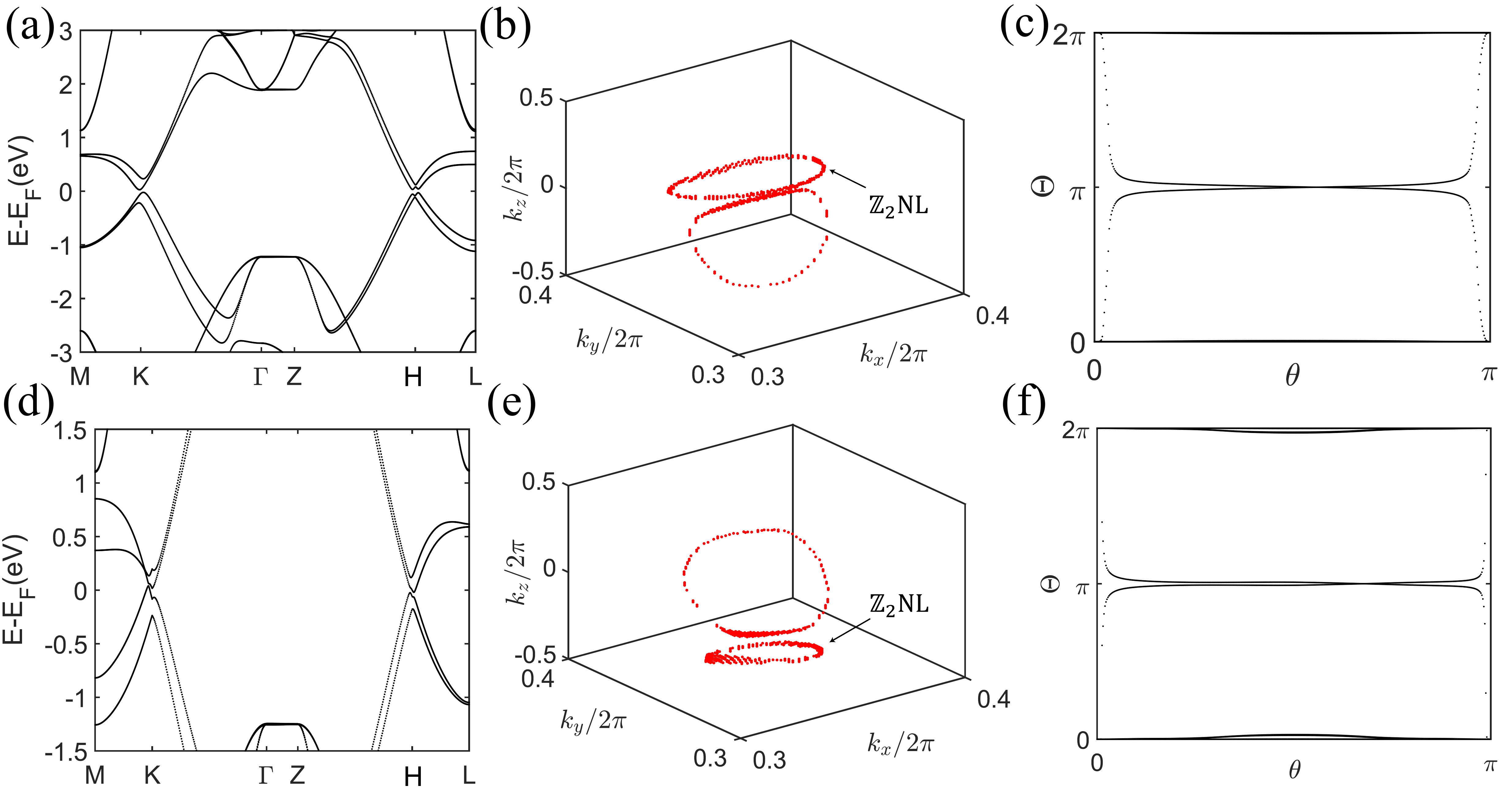}
\caption{\label{si2} Crystal 4: (a) The band structure. (b) The position of the nodal lines around K. Only one of them is $\mathbb{Z}_2$NL. (c) Wilson loop on the sphere enclosing the $\mathbb{Z}_2$NL, indicating that the nodal line carries a unit $\mathbb{Z}_2$ monopole charge. (d)-(f) Same as (a)-(c) but for crystal 5.}
\end{figure}
\begin{figure}[h]
\includegraphics[width=1\textwidth]{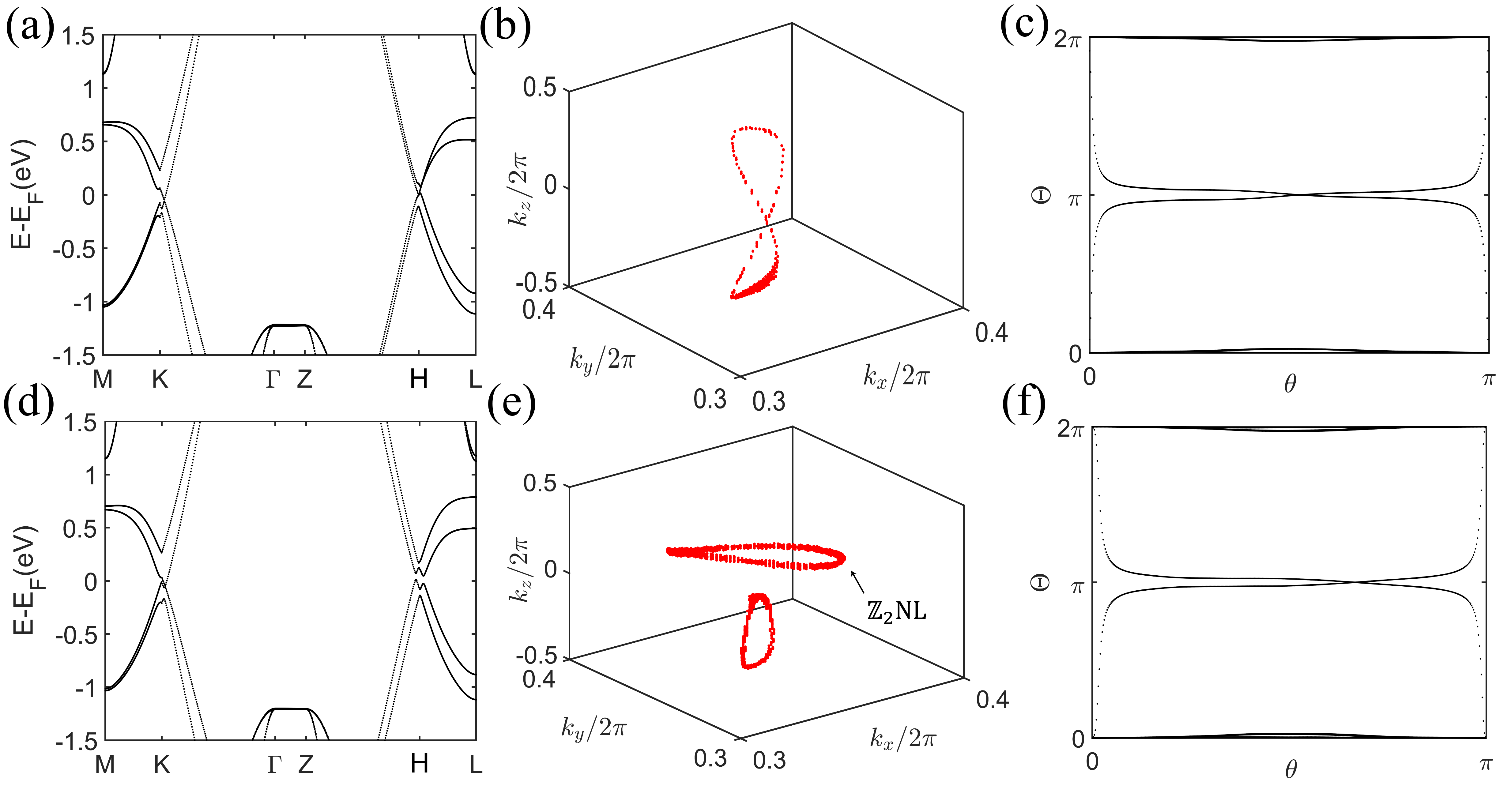}
\caption{\label{si3} Crystal 6: (a) The band structure. (b) The position of the nodal lines around K. (c) Wilson loop on the sphere enclosing the nodal line, indicating that the nodal line carries a unit $\mathbb{Z}_2$ monopole charge. Crystal 7: (d) The band structure. (e) The position of the nodal lines around K. Only one of them is a $\mathbb{Z}_2$NL. (f) Wilson loop on the sphere enclosing the $\mathbb{Z}_2$NL, indicating that the nodal line carries a unit $\mathbb{Z}_2$ monopole charge.}
\end{figure}
\clearpage

\bibliography{1reference}